\documentclass[twocolumn, times]{aastex62}

\usepackage{graphicx}
\usepackage{epstopdf}
\usepackage[intlimits,sumlimits]{amsmath}
\usepackage{epsfig} 
\usepackage{amssymb}
\usepackage{mathrsfs}  
\usepackage{gensymb}

\usepackage{natbib}

\usepackage{amsmath}
\definecolor{dark-red}{rgb}{0.4,0.15,0.15}
\definecolor{dark-blue}{rgb}{0.15,0.15,0.4}
\definecolor{medium-blue}{rgb}{0,0,0.5}
\hypersetup{
    colorlinks, linkcolor={dark-red},
    citecolor={dark-blue}, urlcolor={medium-blue}
}

\newcommand{\beqa}{\begin{eqnarray}} 
\newcommand{\eeqa}{\end{eqnarray}}

\newcommand{\bsub}{\begin{subequations}}
\newcommand{\esub}{\end{subequations}}
\newcommand{\beal}{\begin{align}}
\newcommand{\ealn}{\end{align}}

\received{\today}
\revised{}
\accepted{}
\submitjournal{ApJ}

\shorttitle{ZTF BTS}
\shortauthors{Fremling et al.}
\newcommand{\nBTS}{761}
\newcommand{\nIa}{547}

\newcommand{\nII}{155} % includes 6 SLSN-II
\newcommand{\nIbc}{40} % includes 12 SLSN-I
\newcommand{\nslsne}{19} % includes 12 SLSN-I
\newcommand{\rztf}{$r_\mathrm{ZTF}$}
\newcommand{\gztf}{$g_\mathrm{ZTF}$}

\begin{document}

\title{The Zwicky Transient Facility Bright Transient Survey I: \\ Spectroscopic Classification and the Redshift Completeness of Local Galaxy Catalogs}

\correspondingauthor{U.~C.~Fremling}
\email{fremling@caltech.edu}

\author{U.~C.~Fremling}
\affiliation{Division of Physics, Mathematics and Astronomy, California Institute of Technology, Pasadena, CA 91125, USA}

\author[0000-0001-9515-478X]{A.~A.~Miller}
\affiliation{Center for Interdisciplinary Exploration and Research in Astrophysics and Department of Physics and Astronomy, Northwestern University, 2145 Sheridan Road, Evanston, IL 60208, USA}
\affiliation{The Adler Planetarium, Chicago, IL 60605, USA}

\author{Y.~Sharma}
\affiliation{Division of Physics, Mathematics and Astronomy, California Institute of Technology, Pasadena, CA 91125, USA}

\author{A.~Dugas}
\affiliation{Division of Physics, Mathematics and Astronomy, California Institute of Technology, Pasadena, CA 91125, USA}

\author{D.~A.~Perley}
\affiliation{Astrophysics Research Institute, Liverpool John Moores University, Liverpool Science Park, 146 Brownlow Hill, Liverpool L35RF, UK}

\author[0000-0002-5748-4558]{K.~Taggart}
\affiliation{Astrophysics Research Institute, Liverpool John Moores University, Liverpool Science Park, 146 Brownlow Hill, Liverpool L35RF, UK}

\author{J.~Sollerman}
\affiliation{Department of Astronomy, The Oskar Klein Center, Stockholm University, AlbaNova, 10691 Stockholm, Sweden}

\author[0000-0002-4163-4996]{A.~Goobar}
\affiliation{Department of Physics, The Oskar Klein Center, Stockholm University, AlbaNova, 10691 Stockholm, Sweden}

\author{M.~L.~Graham}
\affiliation{Department of Astronomy, University of Washington, Box 351580, U.W., Seattle, WA 98195, USA}

\author{J.~D.~Neill}
\affiliation{Division of Physics, Mathematics and Astronomy, California Institute of Technology, Pasadena, CA 91125, USA}

\author{J.~Nordin}
\affiliation{Institut fur Physik, Humboldt-Universitat zu Berlin, Newtonstr. 15, 12489, Berlin, Germany}

\author[0000-0002-8121-2560]{M.~Rigault}
\affiliation{Universit\'e Clermont Auvergne, CNRS/IN2P3, Laboratoire de Physique de Clermont, F-63000 Clermont-Ferrand, France.}

\author{R.~Walters}
\affiliation{Division of Physics, Mathematics and Astronomy, California Institute of Technology, Pasadena, CA 91125, USA}
\affiliation{Caltech Optical Observatories, California Institute of Technology, Pasadena, CA 91125, USA}

\author{I.~Andreoni}
\affiliation{Division of Physics, Mathematics and Astronomy, California Institute of Technology, Pasadena, CA 91125, USA}

\author{A.~Bagdasaryan}
\affiliation{Division of Physics, Mathematics and Astronomy, California Institute of Technology, Pasadena, CA 91125, USA}

\author{J.~Belicki}
\affiliation{Caltech Optical Observatories, California Institute of Technology, Pasadena, CA 91125, USA}

\author{C.~Cannella}
\affiliation{Department of Electrical and Computer Engineering, Duke University, Durham, North Carolina 27708, United States}

\author[0000-0001-8018-5348]{Eric C. Bellm}
\affiliation{DIRAC Institute, Department of Astronomy, University of Washington, 3910 15th Avenue NE, Seattle, WA 98195, USA} 
\author{S.~B.~Cenko}
\affiliation{Astrophysics Science Division, NASA Goddard Space Flight Center, 8800 Greenbelt Road, Greenbelt, MD 20771, USA}

\author{K.~De}
\affiliation{Division of Physics, Mathematics and Astronomy, California Institute of Technology, Pasadena, CA 91125, USA}

\author{R.~Dekany}
\affiliation{Caltech Optical Observatories, California Institute of Technology, Pasadena, CA 91125, USA}

\author{S.~Frederick}
\affiliation{Department of Astronomy, University of Maryland, College Park, MD 20742, USA}

\author[0000-0001-8205-2506]{V. Zach Golkhou}
\affiliation{DIRAC Institute, Department of Astronomy, University of Washington, 3910 15th Avenue NE, Seattle, WA 98195, USA} 
\affiliation{The eScience Institute, University of Washington, Seattle, WA 98195, USA}
\altaffiliation{Moore-Sloan, WRF Innovation in Data Science, and DIRAC Fellow}

\author{M.~Graham}
\affiliation{Division of Physics, Mathematics and Astronomy, California Institute of Technology, Pasadena, CA 91125, USA}

\author{G.~Helou}
\affiliation{IPAC, California Institute of Technology, 1200 E. California
             Blvd, Pasadena, CA 91125, USA}

\author{A.~Y.~Q.~Ho}
\affiliation{Division of Physics, Mathematics and Astronomy, California Institute of Technology, Pasadena, CA 91125, USA}

\author{M.~Kasliwal}
\affiliation{Division of Physics, Mathematics and Astronomy, California Institute of Technology, Pasadena, CA 91125, USA}

\author[0000-0002-6540-1484]{T.~Kupfer}
\affiliation{Kavli Institute for Theoretical Physics, University of California, Santa Barbara, CA 93106, USA}

\author[0000-0003-2451-5482]{Russ R. Laher}
\affiliation{IPAC, California Institute of Technology, 1200 E. California
             Blvd, Pasadena, CA 91125, USA}

\author{A.~Mahabal}
\affiliation{Division of Physics, Mathematics and Astronomy, California Institute of Technology, Pasadena, CA 91125, USA}
\affiliation{Center for Data Driven Discovery, California Institute of Technology, Pasadena, CA 91125, USA}

\author[0000-0002-8532-9395]{Frank J. Masci}
\affiliation{IPAC, California Institute of Technology, 1200 E. California
             Blvd, Pasadena, CA 91125, USA}

\author{R.~Riddle}
\affiliation{Caltech Optical Observatories, California Institute of Technology, Pasadena, CA 91125, USA}

\author[0000-0001-7648-4142]{Ben Rusholme}
\affiliation{IPAC, California Institute of Technology, 1200 E. California
             Blvd, Pasadena, CA 91125, USA}

\author{S.~Schulze}
\affiliation{Benoziyo Center for Astrophysics, The Weizmann Institute of Science, Rehovot 76100, Israel}

\author[0000-0003-4401-0430]{David L. Shupe}
\affiliation{IPAC, California Institute of Technology, 1200 E. California
             Blvd, Pasadena, CA 91125, USA}

\author{R.~M.~Smith}
\affiliation{Caltech Optical Observatories, California Institute of Technology, Pasadena, CA 91125, USA}

\author{Lin~Yan}
\affiliation{Caltech Optical Observatories, California Institute of Technology, Pasadena, CA 91125, USA}

\author[0000-0001-6747-8509]{Y.~Yao}
\affiliation{Division of Physics, Mathematics and Astronomy, California Institute of Technology, Pasadena, CA 91125, USA}

\author{Z.~Zhuang}
\affiliation{Division of Physics, Mathematics and Astronomy, California Institute of Technology, Pasadena, CA 91125, USA}

\author{S.~R.~Kulkarni}
\affiliation{Division of Physics, Mathematics and Astronomy, California Institute of Technology, Pasadena, CA 91125, USA}

\begin{abstract}
The Zwicky Transient Facility (ZTF) is performing a three-day cadence survey of the visible Northern sky ($\sim$3$\pi$\,steradian). The transient candidates found in this survey are announced via public alerts. As a supplementary product ZTF is also conducting a large spectroscopic campaign: the ZTF Bright Transient Survey (BTS). The goal of the BTS is to spectroscopically classify all extragalactic transients brighter than $18.5$\,mag in either the \gztf\ or \rztf-filters at peak brightness and immediately announce those classifications to the public. Extragalactic discoveries from ZTF are predominantly Supernovae (SNe). The BTS is the largest flux-limited SN survey to date. Here we present a catalog of the \nBTS~BTS SNe that were classified during the first nine months of the survey (2018 Apr.~1 to 2018 Dec.~31). The BTS SN catalog contains redshifts based on SN template matching and spectroscopic host galaxy redshifts when available. Based on this data we perform an analysis of the redshift completeness of local galaxy catalogs, dubbed as the Redshift Completeness Fraction (RCF; the number of SN host galaxies with known spectroscopic redshift prior to SN discovery divided by the total number of SN hosts). In total, we identify the host galaxies of 512 Type Ia supernovae, 227 of which have known spectroscopic redshifts, yielding an RCF estimate of $44\% \pm 1\%$ (90\% confidence interval). We find a steady decrease in the RCF with increasing distance in the local universe. For $z\lesssim0.05$, or $\sim200$~Mpc, we find $\mathrm{RCF}\approx0.6$, which has important ramifications when searching for multimessenger astronomical events. Prospects for dramatically increasing the RCF are limited to new multi-fiber spectroscopic instruments that can catalog $\ga$10 million galaxies in the local universe, or wide-field narrowband surveys. We find that existing galaxy redshift catalogs are only 50\% complete at $r \approx 16.9$\,mag (AB). Pushing this limit several magnitudes deeper will pay huge dividends when searching for electromagnetic counterparts to gravitational wave events or sources of ultra high energy cosmic rays or neutrinos.
\end{abstract}


\section{Introduction}
Fritz Zwicky and Walter Baade first hypothesized that supernovae (SNe) were the transition of normal stars into neutron stars \citep{Baade1934}. To test this hypothesis, Zwicky used the 18-inch Schimdt telescope commissioned on Palomar mountain in 1936, to carry out the first systematic SN survey \citep{Zwicky1938PASP,Zwicky1938ApJ,Zwicky1942}. This survey was carried out by visually inspecting photographic plates of nebulae,\footnote{At the time, the term nebulae encompassed any diffuse astronomical object, including galaxies.} and identifying new point-sources. Twelve SNe were identified by Zwicky between 1936 Sept. 5 to 1940 Jan. 1. %These discoveries were based on the excessive brightness of these SNe compared to regular novae (Zwicky estimated SNe to be $>1000$ times brighter). %Comment on the two-fold nature of SN discovery. Just an image is not a discovery; the true discovery is the spectroscopic classification. Zwicky did not obtain spectra
%(http://adsabs.harvard.edu/abs/1934PhRv...46...76B)

Since the pioneering efforts by Zwicky, a variety of SN types have been identified through spectroscopy (see e.g., \citealp{Filippenko1997}). Thermonuclear SNe (SNe Ia) in particular have proven to be invaluable tools in order to measure cosmological distances (e.g., \citealp{Goobar2011}), and the study of SNe Ia eventually led to the remarkable discovery of the accelerating expansion of the universe \citep{Riess1998,Perlmutter1999}. Studies of core-collapse (CC) SNe have led to considerable insights in massive star evolution; extragalactic neutrinos were detected in SN\,1987A \citep{Hirata1987}, a $\gamma$-ray burst was associated with SN\,1998bw \citep{Galama1998}, direct evidence for binary-star driven mass loss was seen in SN\,1993J (e.g., \citealp{Schmidt1993,Fox2014}).
%\citealp{Schmidt1993,Filippenko1993,Nomoto1993,Fox2014}

In order to constrain cosmological models and to characterize both SNe in general and the various SN types and their host galaxies, a large number of SN surveys have been carried out since Zwicky's time. The scope of these surveys largely traces the progress made in both automation and detector technology during the last few decades. %Noteworthy examples are: the Center for Astrophysics SN Program (CfA; \citealp{Riess1999,Jha2006,Hicken2009,Hicken2012}), the Lick Observatory Supernova Search (LOSS; \citealp{Li2000}), the Nearby Supernova Factory (Snfactoty; \citealp{Aldering2002}), the Carnegie Supernova Project (CSP; \citealp{Hamuy2006}), the Sloan Digital Sky Survey SN Survey \citep{Frieman2008}, the Palomar Transient Factory (PTF; \citealp{Law2009}), the Catalina Real-Time Transient Survey (CRTS; \citealp{Drake2009}), the All-Sky Automated Survey for SuperNovae (ASAS-SN; \citealp{Shappee2014}), the Asteroid Terrestrial-impact Last Alert System (ATLAS; \citealp{Tonry2018}), and the Foundation Supernova Survey \citep{Foley2018}.
%
%, Foundation SN survey uses the Pan-STARRS telescope \citep{Kaiser2002}.
% texas supernova search (quimby2006), OGLE, ESSENCE, SLNS etc,. %higher z Pan-STARRS (Rest et al. 2014; Scolnic et al. 2014), 
% SDSS first survey paper (SDSS; \citealp{York2000})
%
%CHilean Automatic Supernova sEarch (CHASE; Pignata et al. 2009), the Gaia transient survey (Hodgkin et al. 2013), the La Silla-QUEST (LSQ) Low-Redshift Supernova Survey (Baltay et al. 2013), the Mo- bile Astronomical System of TElescope Robots (MASTER; Gor- bovskoy et al. 2013) survey, and the Optical Gravitational Lensing Experiment-IV (OGLE-IV; Wyrzykowski et al. 2014),
%
The first systematic search for SNe using a charge-coupled device (CCD) was performed on the 1.5-m telescope at La Silla \citep{Norgaard89}. The Field-of-View (FoV) of this telescope and CCD was $2.5\arcmin\times4\arcmin$, and the survey was designed to find a thermonuclear supernova at high-redshift. Two SNe, one SN Ia and one probable SN II, were found in two years.
%%% They found 2 SNe in 2 years!! their FoV was 2.5x4.0 arcminutes 17000 times smaller than ZTF!
More recent examples of SN surveys that have also been able to systematically classify their supernova candidates using spectroscopy include for example: the Lick Observatory Supernova Search (LOSS; \citealp{Li2000}), the Nearby Supernova Factory (SNfactory; \citealp{Aldering2002}), the Sloan Digital Sky Survey-II (SDSS-II) SN Survey \citep{Frieman2008}, and the Supernova Legacy Survey (SNLS; \citealp{Astier2006}). In the last few years, based on statistics on the Transient Name Server (TNS\footnote{https://wis-tns.weizmann.ac.il/stats-maps}),
several surveys are discovering hundreds of SNe that are also being spectroscopically classified, including
the Palomar Transient Factory (PTF; \citealp{Law2009}), the Asteroid Terrestrial-impact Last Alert System (ATLAS; \citealp{Tonry2018}), the All-Sky Automated Survey for SuperNovae (ASAS-SN; \citealp{Shappee2014}), and the Panoramic Survey Telescope and Rapid Response System (Pan-STARRS1, hereafter PS1; \citealp{Chambers2016}) Medium Deep Survey.
% , and the Foundation Supernova Survey \citep{Foley2018}, which uses the Panoramic Survey Telescope and Rapid Response System (Pan-STARRS1, hereafter PS1; \citealp{Kaiser2002}).

%\cf{I (AAM) re-worded this, let me know if you think it's more clear}. CF: Looks good to me, did a few minor edits. Certainly less complicated now.   
The past few decades have seen a growing complexity in SN search surveys, with the general trend being an increase in volumetric survey speed (e.g., \citealt{Bellm16a}) and consequently the number of SN discoveries. Given the scarcity of spectroscopic resources for SN follow-up observations, the increase in SN discoveries has resulted in a smaller fraction of the SNe being classified with time. Of the on-going surveys, only ASAS-SN is able to maintain close to complete spectroscopic coverage ($95\pm3\%$ for $m_\mathrm{peak}<16.5$; \citealp{Holoien19}), largely since ASAS-SN only detects very bright SNe. Otherwise, the typical strategies are to either: (i) focus entirely on the most nearby galaxies (LOSS employed this strategy and maintained a nearly complete survey for $\sim$10\,yr), (ii) focus observations on likely SNe Ia to study cosmology (e.g., SDSS-II, SNLS), or (iii) target only a subset of SN candidates (e.g., PTF, ATLAS). Any of these choices result in major systematic ambiguities underlying any attempt to derive SN rates and demographics, or to use SNe from these surveys as population probes of galaxies. Nevertheless, these compromises have been necessary given the resources at hand.

% In general, the effective survey volume\footnote{Survey volume refers to a combination of survey speed and depth.} and spectroscopic resources available have forced either targeting only a subset of SN candidates for spectroscopic followup (PTF, ATLAS), targeting only very bright SNe (ASAS-SN), targeting only a list of very nearby galaxies (as did LOSS), or targeting only a smaller portion of the sky, as did for example SDSS with stripe 82, and SNLS. These compromises result in major systematic ambiguities underlying any attempt to derive SN rates and demographics, or to use SNe from these surveys as population probes of galaxies. However, they have been necessary in order to (1) achieve an adequate cadence so as to not lose completeness in terms of detecting SN candidates\footnote{Faint SNe and SNe that evolve quickly (e.g., SNe Ibn) cannot be detected if there are no observations when the SNe are at peak brightness.}, and (2) to achieve high completeness in terms of spectroscopic classification of SN candidates within the boundaries defined by the survey (when applicable).  %\footnote{Surveys targeting a subset of SNe for classification based on human selection, as was the case with for example PTF, are particularly prone to unknown bias. However, it is still possible that strict data quality and observing condition cuts can be applied in order to identify periods where the survey was complete in terms of both SN candidate detection and classification (see e.g., \citealp{Frohmaier2019})}.

With the Zwicky Transient Facility (ZTF; \citealp{Bellm2019,Bellm2019b,Graham2019}) in combination with the fully automated Spectral Energy Distribution Machine (SEDM; \citealp{Ben-Ami12,Blagorodnova2018,Rigault2019}), a low-resoultion (R$\sim100$) Integral-Field-Unit (IFU) spectrograph mounted on the robotic Palomar 60-inch telescope (P60; \citealp{Cenko2006}), we have set out to address the lack of spectroscopic completeness described above.
% both points (1) and (2) above for bright SNe in the northern sky. 
We aim to monitor the entire visible sky at moderate cadence while being complete in terms of spectroscopic classification. The 47\,$\deg^2$ field of view of the ZTF camera, along with upgrades to the Palomar 48-inch (P48) telescope and dome, achieves a survey speed of 3750 \,$\deg^2$\,hr$^{-1}$, to a $5\sigma$ depth of $\sim$20.5\,mag in $r_\mathrm{ZTF}$ using $30$\,s exposures. This allows most of the sky visible from Palomar to be imaged at a 3-day cadence (see \S\ref{sec:observations} for details). Furthermore, SEDM is capable of classifying $>10$ SNe in the $18.5-19$~magnitude range every night. A significant amount of time is also allocated to this project on the Palomar 200-inch telescope (P200), Keck I, the Liverpool Telescope (LT), Apache Point Observatory (APO) and the Nordic Optical Telescope (NOT). These resources are used to supplement our SEDM observations when SEDM classification is not possible, and the combination enables the ZTF Bright Transient Survey (BTS): a SN survey of unprecedented scale and spectroscopic completeness in the local universe.

The primary goal of the BTS is to spectroscopically classify and publicly report every extragalactic $r<18.5$\,mag transient in the Northern sky covered by the public ZTF surveys,\footnote{Excluding the galactic plane ($\pm7\degree$).} producing the first large, fully magnitude-complete sample to $r_{\rm peak}<18.5$~mag.\footnote{The saturation limit of ZTF is $\sim14$~mag.} %\footnote{Even this choice requires compromise as fainter SNe are not classified as part of the BTS.} %CF: I removed this footnote since I think it should be clear that this is a coice we had to make. Also,  we do classify down to 19 but with less completeness. I think these details should be explained later
In this paper we will focus on SNe, but the BTS also finds and classifies Tidal Disruption Events (TDEs), which will be analysed separately, and other extragalactic phenomena such as massive active galactic nucleus (AGN) flares, and fast and highly energetic transients such as AT\,2018cow\footnote{The internal ZTF designation for AT\,2018cow is ZTF18abcfcoo.} \citep{Prentice2018,Perley2019b,Ho2019}, whose nature remains mysterious. Here we present a catalog of the \nBTS~BTS SNe classified during the first 9 months of the survey (2018 Apr.~1 to 2018 Dec.~31; Table~\ref{tab:sample}). Our catalog contains redshifts based on SN template matching (\citealp{Blondin2007}) and spectroscopic host galaxy redshifts when available. WISE \citep{Wright10} $W1$-band and PS1 \citep{Chambers2016} $i$-band host galaxy magnitudes are also reported. 

We expect that this sample, and its ongoing extension through 2019 and 2020, will be useful for a wide variety of topics within supernova astrophysics, some of which will be the focus of follow-up papers. In this paper, we focus on an analysis of the redshift completeness of local galaxy catalogs (\S\ref{sec:rcf}), dubbed as the Redshift Completeness Fraction (RCF; the number of SN host galaxies with known redshift prior to SN discovery divided by the total number of SNe). The methodology for this analysis closely follows that of \cite{Kulkarni2018}.

\begin{figure*}
    \centering
    \includegraphics[width=0.5\linewidth]{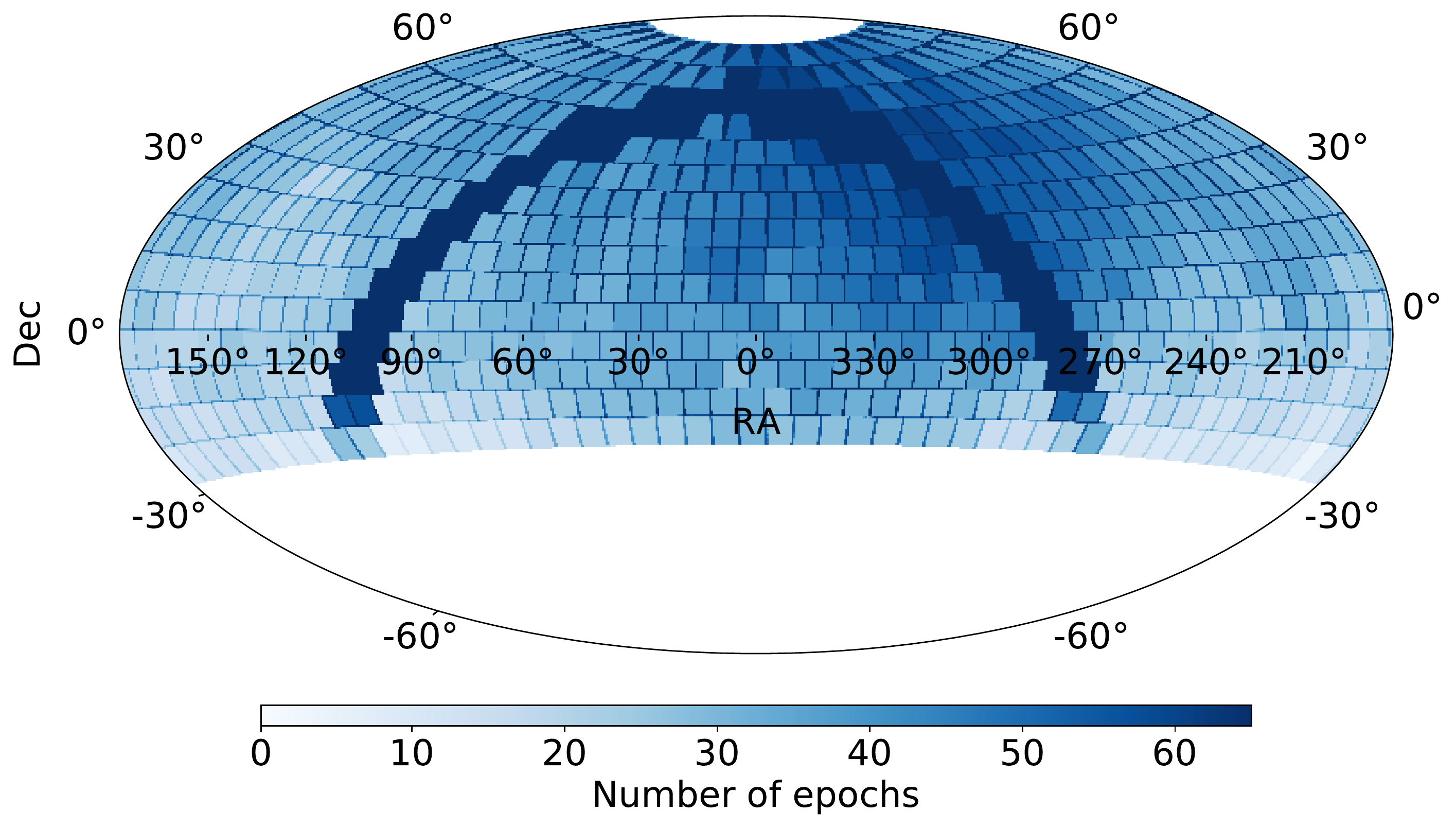}\includegraphics[width=0.5\linewidth]{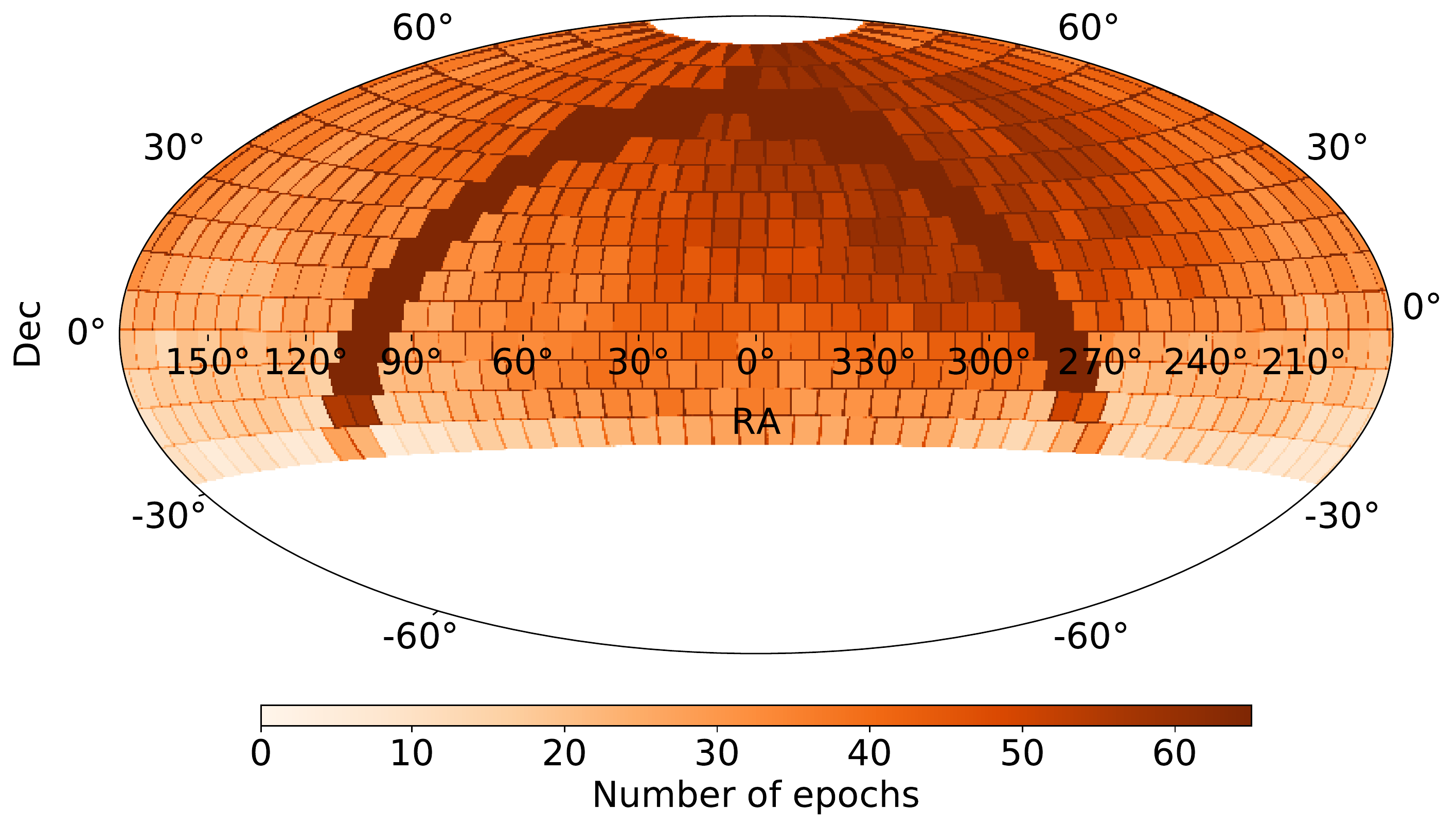}
    \caption{Coverage maps for the ZTF MSIP surveys, in the $g_\mathrm{ZTF}$- (left panel) and $r_\mathrm{ZTF}$-band (right panel) between 2018 Apr.~1 and 2018 Dec.~31. The colored rectangles represent the fixed ZTF main field grid. The color intensity indicates the number of observations during this time period, truncated to a maximum of 65.} 
    \label{fig:coverage}
\end{figure*}
%Note: The number of observations on the galactic plane greatly exceeds 60 since the galactic plane is observed at a 1-day cadence. Should we remove the galactic plane fields?

\section{Survey Design}\label{sec:observations}
Transient candidates for the BTS are provided by the public ZTF surveys: the Northern Sky Survey (NSS) and the Galactic Plane Survey (GPS). These surveys are made possible by an award from the NSF Mid-Scale Innovations Program (MSIP), and we henceforth refer to them as the ZTF MSIP surveys (see \citealp{Bellm2019b} for details). The NSS covers $\approx13{,}000\,\deg^2$ of the Northern sky at a 3-day cadence in the $g_\mathrm{ZTF}$ and $r_\mathrm{ZTF}$ filters, while using $34\%$ of the P48 telescope time. The GPS covers $\approx1{,}500\,\deg^2$ of the galactic plane at a 1-day cadence, also in the $g_\mathrm{ZTF}$ and $r_\mathrm{ZTF}$ filters, and uses $6\%$ of the P48 telescope time.

The BTS avoids low Galactic latitudes by design; we reject all transient candidates found within $7\degree$ of the Galactic plane (see \S\ref{sec:filter}). The combination of significant Galactic extinction and the BTS Galactic plane cut means that the vast majority of the transients in the BTS originate from the NSS. However, due to the large FoV and the fixed main field-grid used by ZTF (\citealp{Masci2019}), some of the GPS fields still allow transients to be found and monitored after a $7\degree$ galactic plane cut. During 2018 we classified two SNe within GPS fields. %These SNe are subject to the higher 1-day cadence of the GPS.

The images from the ZTF MSIP surveys are processed and analysed at IPAC by an automated pipeline \citep{Masci2019}, that uses the \citet{Zackay16} difference-imaging method. The pipeline produces transient alert packets from the difference images in the Apache Avro$^{\mathrm{TM}}$ format.\footnote{\url{https://avro.apache.org}} The Avro alert packets are distributed through the University of Washington (UW) as a Kafka data stream.\footnote{\url{https://kafka.apache.org}} The alert stream originating from the ZTF MSIP surveys is the data source for BTS transient candidates. The full ZTF Alert Distribution System (ZADS) is described in detail in \cite{Patterson2019}. 

\begin{figure}
    \centering
    \includegraphics[width=0.95\linewidth]{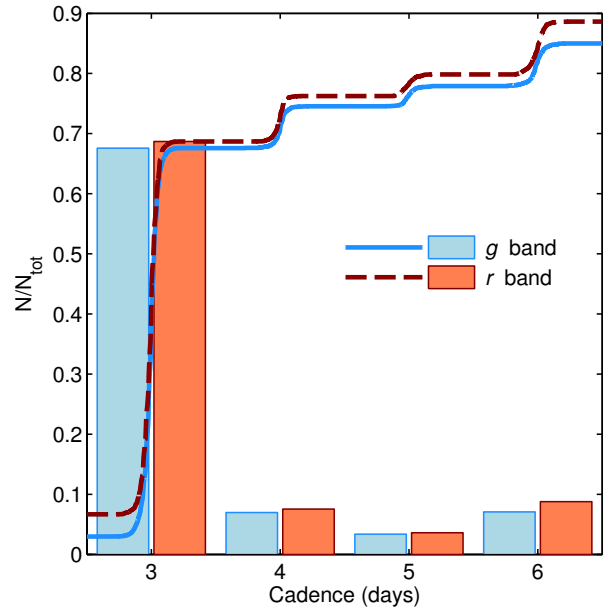}
    \caption{Cadence distribution for the ZTF NSS, in the $g_\mathrm{ZTF}$- (blue bars) and $r_\mathrm{ZTF}$-band (red bars), truncated at six days. Cumulative distributions are shown as a blue solid line for the $g$ band and a red dashed line for the $r$ band. $N/N_{\mathrm{tot}}$ is the fraction of observations at a specific cadence compared to the total number of observations between 2018 Apr.~1 and 2018 Dec.~31.}
    \label{fig:cadence}
\end{figure}

Figure~\ref{fig:coverage} shows the $g_\mathrm{ZTF}$ and $r_\mathrm{ZTF}$-band coverage maps of the ZTF MSIP surveys, between 2018 April 1 and 2018 December 31. The distribution of re-visit times (cadence) for the NSS for each field during the same time period, excluding 2018 September 29 to 2018 October 31 when the P48 was undergoing maintenance, is shown in Figure~\ref{fig:cadence}. Approximately $70\%$ of the NSS observations were carried out at the planned 3\,d cadence, and $\sim$90\% of re-visits occurred within $\leq6$\,d during 2018. For the GPS, around $80\%$ of the observations were carried out at a 1-day cadence. However, since only two BTS SNe were found and classified during 2018 in the GPS fields, the GPS cadence is not representative for the BTS.

In order to identify bright SN candidates within the raw ZTF alert stream produced by the MSIP surveys we utilize the filtering capability within the GROWTH \texttt{Marshal} framework \citep{Kasliwal2019} to apply the candidate filter described below (\S\ref{sec:filter}). The GROWTH \texttt{Marshal} is also used to organize BTS sources and the corresponding spectroscopic follow-up efforts (\S\S\ref{sec:scanning},~\ref{sec:followup}).

\subsection{Supernova Candidate Filter}
\label{sec:filter}

The BTS filter used between 2018 Apr.~1st and 2018 Dec.~31st was deliberately designed to be simple, in order to minimize the risk of false negatives (i.e., real transients that are not saved as part of the program). Using the GROWTH \texttt{Marshal} alert filtering system, we applied the following cuts to the raw ZTF alert stream in order to identify bright SN candidates: %This filter ran from Jan 2018 to Jun 14 2019. Then significant changes made
\begin{itemize}
\item Alerts at low Galacitc latitudes ($\left|b\right| \le 7\degree$) are rejected.
\item Alerts with a random forest based machine learning real-bogus score (\texttt{rbscore}; \citealp{Mahabal2019}) of less than 0.2 are rejected. This choice results in a completeness (i.e., 1 - false negative rate) of $>99\%$ (figure 16 in \citealp{Duev2019}). %\cf{I'm worried we are mixing results here - we were using Umaa's RB, and the results for Dimitri's RB are different... CF: I have clarified. We are indeed using random forest RB, but the statistics on that (and also braai) are shown in Dimas paper figure 16.}
\item Alerts produced at the position of known stars, as identified in the catalog created by \citet{Tachibana2018}, are rejected. A small fraction of galaxies (estimated to be $<$0.5\% in \citealt{Tachibana2018}), and thus nuclear SNe will be missed as a result of this cut.
\item Alerts produced close to very bright stars have been rejected ($<20$\arcsec\, for $<15$\,mag stars; $\sim1\%$ loss of survey area\footnote{Based on randomly injecting $10,000$ SNe within the survey footprint and matching against PS1 stars.}).
\item Alerts that do not include at least two detections separated by a minimum of $30$\,min are rejected (moving object filter).
\item Alerts with negative flux relative to the reference image are rejected.
\item The alert must include at least one epoch with $m>19$\,mag (in $g_\mathrm{ZTF}$- or $r_\mathrm{ZTF}$-band), otherwise it is rejected.
\end{itemize}
The BTS filter effectively passes all alerts brighter than $19$\,mag that are not consistent with stellar events (star detected in PS1) or moving objects for human vetting (scanning; \S\ref{sec:scanning}).

\subsection{Human candidate vetting}
\label{sec:scanning}
On a typical night in 2018 a few hundred alerts passed the BTS filter. These BTS transient candidates were visually inspected by a team of scanners on a daily basis. To organize this effort we use the GROWTH \texttt{Marshal} \citep{Kasliwal2019}, where the light curves and image cutouts (science, reference, and subtraction) contained in the Avro packets of the passing alerts are collected on a scanning page for each night. Supplementary information is also displayed, such as: PS1 and SDSS color composite cutouts centered on the position of the transient, the star-galaxy separation score (\texttt{sgscore}), which gives a probability that the closest PS1 counterpart is an extended galaxy or a point-like star (\citealp{Tachibana2018}), multi-band photometry of this PS1 counterpart, a cross-check for known near earth objects (NEOs), and information about if and when there have been previous ZTF alerts at the same position that are not part of the $30$-day history contained in the alert packet itself. External catalog cross-matches (e.g., NED, TNS, SIMBAD, VizieR) are also linked through the GROWTH \texttt{Marshal} to provide additional contextual information for each potential SN candidate.

The human vetting process essentially consisted of inspecting the information contained in the Avro packets for each alert that passed our filter and also taking into account any relevant supplementary information available in order to rule out variability from a stellar counterpart, and to reject alerts produced by known AGN.

Among the passing alerts, 5--15 SN candidates are typically identified by the human scanners as good SN candidates and assigned for spectroscopic followup, per night.\footnote{This number strongly depends on the weather; after a period of bad weather a large number of SNe will be recovered when observations are resumed. Good weather periods produce a more constant number each night.} The two main contaminants in our scanning process are cataclysmic variables (CVs) that are too faint to be seen in PS1 in their quiescent phase, and therefore lack an \texttt{sgscore}, and AGN. Both of these must be avoided, given our limited spectroscopic resources. We have found that the vast majority of CVs can be avoided by monitoring the lightcurve behavior until $\sim1$~week after the initial outburst and comparing the evolution with typical CV lightcurves.\footnote{CVs feature a fast evolution (rise time of $\sim$1--2\,d, decline of $\sim$7--10\,d) and their \gztf $-$ \rztf\ colors are persistently blue. We do not follow-up events with these characteristics that also lack an obvious host-galaxy counterpart.} This is especially effective since ZTF produces both $g_\mathrm{ZTF}$- and $r_\mathrm{ZTF}$-band photometry for virtually all transients that are detected.

Filtering AGN is more challenging: excluding all AGN from the BTS could inadvertently reject a SN that has exploded near the nucleus of a galaxy that harbors an AGN. Filtering on past variability is a very effective way of excluding AGN, but for the sample presented here the baseline of ZTF observations was only a few weeks or months, which is not always a sufficient amount of time for the AGN to change in flux enough to be recognized as a variable object. We have generally not triggered spectroscopic follow-up for alerts that are positionally coincident with known AGN (e.g., the ALLWISE mid-infrared AGN catalog; \citealp{Secrest2015}, the Million Quasar catalog; \citealp{Flesch2015}), unless the photometric evolution of the transient is very similar to that of a SN. Thus, the BTS is incomplete for SNe near AGN (a focused survey with the specific goal of discovering SNe in galaxies with AGN is needed to address this).

\subsubsection{Completeness of the BTS filter}
To assess the completeness of our BTS GROWTH \texttt{Marshal} filter (\S\ref{sec:filter}) and human scanning effort (\S\ref{sec:scanning}) we have re-processed and re-filtered all public ZTF alerts between 2018 Apr.~1 to 2018 Dec.~31 using the \texttt{AMPEL} system \citep{Nordin2019,Soumagnac2018}. Two filters were applied for this exercise: a variant of our BTS GROWTH \texttt{Marshal} filter converted to work within \texttt{AMPEL}\footnote{We have confirmed that this filter passes all candidates we found and classified in 2018}, and the \texttt{AMPEL} TNS channel filter described in \cite{Nordin2019}.

After the filtering step, we combined the passing candidates from both filters and imposed a cut requiring: at least 5 detections, a peak mag $<18.5$, no more than one negative detection, and a time between the first and last detection more than 5 days and less than 90 days\footnote{A subset of the $>$90 day events were also vetted and no SNe were found.}. Finally, we also required that the candidate passes a version of the GROWTH \texttt{Marshal} filter that checks all associated Avro packets for that candidate. All of the remaining objects that were not saved on the GROWTH \texttt{Marshal} were then vetted individually to remove any remaining CVs, AGNs, classical novae, and artifacts.
%Then, instead of human vetting of the candidates, for those candidates where the data allows the rise and fade time from and to $0.75$~mag below peak to be determined we employed the following lightcurve cuts in order to identify possible SNe: rise-time to peak within $4-100$~d, fade-time between $10-200$~d.

This singled out 17 likely SNe that our BTS filter scanning efforts had not picked up. Among these 9 were saved by other ZTF science programs or \texttt{AMPEL}, and were spectroscopically classified. The remaining 8 are unclassified. The BTS sample contains 520 SNe peaking at $<18.5\pm0.05$~mag. This implies a completeness in our scanning of the BTS filter in 2018 of 97\% for candidates with peak mag $\lesssim18.5$. The 17 objects identified here are not included in our sample or analysis but will be included and analysed in our next data-release. We also note that \cite{Nordin2019} showed that ZTF has been complete with respect to detecting SNe reported to TNS by other groups when they fall on active ZTF CCD regions.
%\\
%\\
%(1) More than four high quality detections\\
%(2) More than two detections brighter than 19 mag\\
%(3) More than four detections brighter than 19.5 mag\\
%(4) More than seven detections brighter than 19.5 mag\\
%(5) Transient duration less than 50 days\\
%(6) Transient duration less than 100 days\\
%(7) rms(Detection JD) $\leq20$~days\\
%(8) mean(Detection JD) $-$ median(Detection JD) $<5$~days\\
%(9) Distance to closest reference position larger than 1\arcsec\\

%Nearby SNe should fullfill \textit{most} of these criteria, with the max rank being 9. Visual inspection showed that candidates with rank $<7$ appear to be predominantly CVs and AGN. Candidates with ranks 7, 8, and 9 were visually inspected. Among these we have identified 3 probable SNe that were either missed by our human scanners, or never shown to our human scanners due to technical issues (such as GROWTH Marshal database maintenance, or temporary communication issues with the UW alert stream). Based on this procedure we estimate that the alert completeness for the BTS filter and human scanning for 2018 was $99.6\%$ for alerts with lightcurves peaking at $18.5$~mag or brighter.

%add that Nordin et al has shown that our completeness with resp. to TNS is very high (chip gaps explain any misses, see the Nordin paper)

%\textcolor{red}{to-do: Add the final numbers.} 

\subsection{Spectroscopic followup assignment}
\label{sec:followup}
The primary classification instrument of BTS candidates is the SEDM \citep{Blagorodnova2018,Rigault2019}. For candidates that pass the BTS filter (\S\ref{sec:filter}) and visual inspection (\S\ref{sec:scanning}), we assign spectroscopic SEDM observations with the following priorities: transients that are, or are likely to become, brighter than 18.5\,mag are scheduled with the highest priority (P3). Transients that appear likely to peak between $18.5$ and $18.75$\,mag are assigned a lower priority (P2), and transients expected to peak fainter than  $18.75$\,mag are triggered at the lowest priority (P1). For a typical BTS source ($m \approx 18.5$\,mag) observed under typical observing conditions for 1800\,s, SEDM obtains $\mathrm{SNR} \approx 12$ per resolution element, and an integrated $\mathrm{SNR} \approx 50$ in the region covered by the \rztf\ filter.
%\citep{Schlegel90}

%To organize spectroscopic follow-up of the candidates that passed our candidate filter (\S\ref{sec:filter}) and human vetting (\S\ref{sec:scanning}), our first line of attack is SEDM, for which we employ a prioritization system as follows: when a transient appears likely to become (or already is) brighter than 18.5~mag, SEDM-based follow-up spectroscopy is triggered at our highest priority (P3). Transients that appear likely to peak in the range $18.5-18.75$~mag range are assigned a lower priority (P2), and transients expected to fall in the $18.75-19.0$~mag range are triggered at the lowest priority (P1).

The SEDM queue is designed so that lower priority targets (e.g., P1, P2) are only observed if no higher priority targets are present in the queue that could be observed within the same observation-time window. The purpose of this priority scheme is two-fold. First, it ensures that we can reach a high level of completeness for $m_\mathrm{peak} < 18.5$\,mag sources, by allowing some margin for error on the fainter end of $18.5$\,mag. Second, when the queue allows (e.g., periods of consistently good weather), significantly fainter targets, including as faint as $\sim$19\,mag, can also be observed in addition to our high priority (P3) targets.

By default, triggers that enter the SEDM queue remain active for $7$\,d. If the transient has not been observed at this time, the candidate is re-assigned to SEDM, or to larger telescopes if the candidate has become too faint for SEDM ($m \ga 19$\,mag). Larger telescopes are also used if classification with SEDM is unsuccessful, which typically happens only if the candidate was observed in poor sky conditions, there is strong host galaxy contamination, or if higher resolution is needed for a secure classification. For this purpose we have primarily used P200 and Keck I, but supporting observing programs at LT, NOT and APO have contributed significantly as well. Community efforts (e.g., ePESSTO; \citealp{Smartt15}) have also contributed through TNS (see \S\ref{sec:classifications} for details).%, or if weather losses completely prevent SEDM spectra from being obtained before the transient fades past 19~mag
%Talk about re-triggering of failed attempts; periodically we check older targets and re-assign to SEDM; if too faint P200, Keck, Lick3m, APO, LT

%ePESSTO; Smartt et al. 2015, A&A, 579, 40 http://www.pessto.org 

%If classification with SEDM is unsuccessful (e.g., target observed under less than ideal conditions), or if weather losses completely prevent SEDM spectra from being obtained before the transient fades past 19~mag, larger telescopes are used in order to classify the transient.

\subsubsection{Spectroscopic completeness}
A key goal for the BTS is to obtain high spectroscopic completeness for all events passing our basic selection criteria (essentially, $m<18.5$\,mag and extragalactic, see \S\ref{sec:filter}).  While we made every effort to spectroscopically classify every transient saved to the BTS program, our efforts were inevitably imperfect. Following the conclusion of 2018 (i.e., the period covered in this early-release paper), we conducted two independent tests of our completeness as described below. 

As an initial test of the completeness of BTS spectroscopic follow-up, we compiled every object that was saved following visual inspection (\S\ref{sec:scanning}) and applied additional filters designed to remove most variables and AGN using a more sophisticated point-source coincidence check and the long-term light-curve history.\footnote{All but 3 of the early-release SNe passed these additional filters. These 3 were rejected due to their proximity to bright stars, meaning the additional filters still find genuine SNe with high fidelity.}  We restrict these candidates to those brighter than $m<18.5$ in at least one observation. Every such object without a formal spectroscopic classification (74 in total) was then visually examined by our team of scanners.  Most of these events are clearly not transients (e.g., subtraction artifacts, AGN, stars) based on their full light curves. However, 31 events had properties consistent with SNe and were unclassified. If each of these events is a genuine SN, this would suggest a completeness of $\sim$96\%.%\todo{Should ideally recheck that these numbers are still accurate following any final re-classifications}

We separately examined the filtered subset of alerts described above and estimated the characteristic rise and fade times (from, and to, 0.75\,mag below peak, respectively) for every event using an automated procedure. For events with sufficient P48 data around peak to accurately constrain the rise and fade times, and with rise times between 4--100\,d and fade times between 10--200\,d (i.e., probable SNe), our classification completeness rate is 100\% to $m_{\rm peak}<$16.5\,mag, 98.8\% to $m_{\rm peak}<$17.5\,mag, 93.6\% to $m_{\rm peak}<$18.5\,mag, and 88.8\% to $m_{\rm peak}<$19.0\,mag. This method, which is easily automated (and could be applied to samples larger than what can be manually inspected), is consistent with the results from our visual inspection described above.

\section{BTS SN Classifications}

\subsection{Classification Method}\label{sec:classifications}

As previously noted, spectroscopic observations of BTS SN candidates are primarily obtained with SEDM. In cases where the candidates were too faint to be observed with SEDM, or scheduling conflicts prevented SEDM observations, or the SEDM spectra proved to be ambiguous, spectra were obtained with spectrographs on larger aperture telescopes: the Double Spectrograph (DBSP; \citealp{Oke1982}) on P200, the Low Resolution Imaging Spectrometer (LRIS; \citealp{Oke1995}) on the Keck 1 telescope, the Dual Imaging Spectrograph (DIS) on the APO 3.5\,m telescope, the Spectrograph for the Rapid Acquisition of Transients (SPRAT; \citealp{SPRAT}) on the 2.0\,m LT, and the Alhambra Faint Object Spectrograph and Camera (ALFOSC) on the 2.56\,m NOT.

SEDM data were reduced by the pipeline described in \cite{Blagorodnova2018}, for data until August 2018, and the automatic \texttt{PYSEDM} pipeline described in \cite{Rigault2019} for data after August 2018. DIS data were reduced with the \texttt{pyDIS} package \citep{pydis}. DBSP data were reduced using the \texttt{PyRAF} \citep{pyraf} based pipeline \texttt{pyraf-dbsp} \citep{Bellm2016}. ALFOSC data were reduced using standard procedures and tools based on \texttt{IRAF} \citep{Tody1986}. LRIS data were reduced using the LRIS automated reduction pipeline (\texttt{LPipe}; \citealp{Perley2019}). 

For SN candidates where we could not obtain spectroscopic observations from any of the above mentioned telescopes, public TNS spectra were analysed from the following instruments: the Asiago Faint Object Spectrograph and Camera (AFOSC) on the Asiago Ekar 182\,cm telescope, FLOYDS on the Faulkes Telescope South (FTS) in Australia operated by Las Cumbres Observatory (LCO), the Wide Field Reimaging CCD (WFCCD) on LCO's duPont telescope, the ESO Faint Object Spectrograph and Camera v.2 (EFOSC2) on the New Technology Telescope (NTT), the Dolores (Device Optimized for the LOw RESolution) on the Telescopio Nazionale Galileo (TNG), the Intermediate-dispersion Spectrograph and Imaging System (ISIS) on the William Herschel Telescope (WHT), and the DeVeny spectrograph on the Discovery Channel Telescope (DCT).

%EFOSC ref is \citealp{Buzzoni1984}

Preliminary classifications are made via \texttt{SuperNova} \texttt{IDentification} (\texttt{SNID}; \citealt{Blondin2007}) template matching \textit{and} visual inspection. \texttt{SNID} is automatically applied to all SEDM spectra, whereas for all other instruments \texttt{SNID} is applied to the spectra by the observer. These preliminary classifications are annotated and recorded within the BTS program on the GROWTH \texttt{Marshal}, and subsequently sent to TNS within one to two days.\footnote{The BTS classifications on TNS should also be considered preliminary. They reflect the initial classifications on the GROWTH \texttt{Marshal}, since turnaround speed is important, but they are subsumed by the efforts described in this paper.}

We revisit each of the preliminary classifications for this study in order to develop a homogeneous classification scheme. For this purpose we developed a custom process to spectroscopically classify the \nBTS~SNe in the BTS sample. For each BTS SN, we identified the top 15 spectral matches ($r$lap$_\mathrm{min}\geq5$) from \texttt{SNID} to the most recent spectrum available on the GROWTH \texttt{Marshal}. We used the latest spectrum from the Marshal under the assumption that BTS targets only received additional spectroscopic observations when the initial classification was inconclusive. The \texttt{SNID} templates used for this process include the developer defaults,\footnote{https://people.lam.fr/blondin.stephane/software/snid/\#Download} as well as SNe Ia and a few non SN templates from the Berkeley SN Ia program (BSNIP; \citealt{bsnip}), SN Ib/c templates from \citet{modjaz2014,modjaz2016,liu2016}, and  \citet{williamson}, and SN IIP templates from \citet{gutirrez}.

Following the SNID matching, we produced plots showing a comparison between the observed BTS spectrum and the (redshift-corrected) template spectrum from SNID. These plots were visually inspected to identify the best matching template. In practice, the sample was split into 6 groups, and each group was inspected by a member of our team (CF, AAM, AD, YS, KT, AG). While identifying the best-matching template, we recorded the SN type and redshift, as well as the name and phase of the template SN spectrum. In cases where the same classification was reported for all 15 matches, we recorded the type, redshift, and phase from the top match from the SNID output. We otherwise selected the best match based on common prominent SN spectral features (H, He, Si, Ca, Fe, etc.). If the top 15 matches from SNID proved ambiguous, we used either the ZTF light curve or alternative spectra to remove the ambiguity. For example, in cases with multiple matches to both SNe Ia and SNe Ic, the telltale secondary near-infrared (nIR) peak of SNe Ia can typically be seen in ZTF $r_\mathrm{ZTF}$-band light curves. The secondary nIR peak is unique to SNe Ia, as explained in \cite{2006ApJ...649..939K}. It occurs following a recombination transition of the iron-group elements in the ejecta, whereby the strength of the \ion{Fe}{3} and \ion{Co}{3} lines decreases and there is a corresponding strengthening of the \ion{Fe}{2} and \ion{Co}{2} lines \citep[see also][]{2015MNRAS.448.2766B}. While these papers have mostly considered $\lambda \ga 7500$\,\AA, a related "shoulder" in the $r$-band lightcurve is also observed at approximately the same time \citep{2019MNRAS.485.2343P,2019MNRAS.483.5045P}. 

If at this stage a classification still proved ambiguous, then the SN was examined by another member of the team. For consistency, a final check of all ambiguous classifications was performed by two members of the team (CF, AAM). Ultimately, we have classified \nBTS~SNe via their spectra and light curves (Table~\ref{tab:sample}). Out of these, 503 were classified using SEDM spectra, 86 using P200-DBSP, 76 using Keck 1-LRIS, 20 using LT-SPRAT, 11 using APO-DIS, and 9 using NOT-ALFOSC. Finally, 56 were classified based on publicly available spectra on TNS. 

We note that the positions reported in Table~\ref{tab:sample} are obtained by taking the weighted average of the position of the SN in every image in which the SN is detected (i.e., for every alert associated with the SN). The updated positions are more accurate than those reported to TNS, which typically only include a single low signal-to-noise ratio (SNR) detection of the SN.

\begin{deluxetable*}{lrrlllccrrrrrrr}
\tabletypesize{\scriptsize}
\tablewidth{0pt}
\tablecaption{ZTF BTS SNe\label{tab:sample}}
\tablehead{
\colhead{}
& \colhead{}
& \colhead{}
& \colhead{}
& \colhead{}
& \colhead{}
& \colhead{}
& \colhead{}
& \colhead{}
& \multicolumn{6}{c}{Observed time and mag at maximum brightness\tablenotemark{a}} \\
\cline{10-15}
\colhead{ZTF}
& \colhead{$\alpha_\mathrm{SN}$}
& \colhead{$\delta_\mathrm{SN}$}
& \colhead{IAU}
& \colhead{TNS Internal}
& \colhead{Discovered}
& \colhead{$E(\bv)$\tablenotemark{b}}
& \colhead{SN}
& \colhead{$z_\mathrm{SN}$\tablenotemark{c}}
& \colhead{$t_{g}$}
& \colhead{$m_{g}$}
& \colhead{$\sigma_{m_g}$}
& \colhead{$t_{r}$}
& \colhead{$m_{r}$}
& \colhead{$\sigma_{m_r}$} \\
\colhead{Name}
& \colhead{(J2000.0)}
& \colhead{(J2000.0)}
& \colhead{Name}
& \colhead{Name}
& \colhead{by}
& \colhead{(mag)}
& \colhead{Type}
& \colhead{}
& \colhead{(JD$^\prime$)\tablenotemark{d}}
& \colhead{(mag)}
& \colhead{(mag)}
& \colhead{(JD$^\prime$)\tablenotemark{d}}
& \colhead{(mag)}
& \colhead{(mag)}
}
\startdata
ZTF18aabssth & $11{:}00{:}45.38$ & $+22{:}17{:}15.0$ & SN\,2018aex & ZTF18aabssth & ZTF & 0.015 & II & 0.026 & 494.05 & 20.44 & 0.19 & 218.71 & 18.74 & 0.05 \\
ZTF18aabxlsv & $10{:}29{:}51.62$ & $+09{:}00{:}46.6$ & SN\,2018aks & ASASSN-18ga & ASAS-SN & 0.025 & Ib & 0.055 & 605.71 & 20.33 & 0.19 & 224.68 & 18.64 & 0.04 \\
ZTF18aaemivw & $10{:}33{:}42.69$ & $+39{:}29{:}26.6$ & SN\,2018hus & ZTF18aaemivw & ZTF & 0.012 & Ia & 0.065 & 423.96 & 19.00 & 0.16 & 534.87 & 17.95 & 0.15 \\
ZTF18aagpzjk & $07{:}59{:}25.01$ & $+16{:}25{:}34.5$ & SN\,2018afm & $\ldots$ & POSS & 0.031 & II & 0.013 & \ldots & \ldots & \ldots & 217.66 & 17.46 & 0.02 \\
ZTF18aagrdcs & $14{:}33{:}19.98$ & $+41{:}16{:}02.3$ & SN\,2018alc & ASASSN-18ge & ASAS-SN & 0.012 & Ib & 0.024 & 547.91 & 17.37 & 0.05 & 217.90 & 16.67 & 0.01 \\
ZTF18aagrtxs & $13{:}14{:}25.46$ & $+50{:}58{:}39.7$ & SN\,2018amo & ASASSN-18gi & ASAS-SN & 0.010 & Ia & 0.018 & 214.76 & 16.52 & 0.02 & 214.73 & 16.46 & 0.02 \\
ZTF18aagstdc & $15{:}50{:}03.56$ & $+42{:}05{:}18.5$ & SN\,2018apn & ASASSN-18gs & ASAS-SN & 0.018 & Ia & 0.038 & 210.86 & 17.24 & 0.02 & 539.01 & 17.27 & 0.06 \\
ZTF18aagtcxj & $16{:}32{:}11.55$ & $+42{:}42{:}48.3$ & SN\,2018aqm & $\ldots$ & TNTS & 0.011 & Ia & 0.033 & 539.03 & 19.01 & 0.19 & 219.96 & 18.09 & 0.03 \\
ZTF18aahesrp & $08{:}35{:}45.43$ & $+28{:}16{:}12.9$ & SN\,2018aqy & ATLAS18mzs & ATLAS & 0.036 & Ia & 0.051 & \ldots & \ldots & \ldots & 217.69 & 18.54 & 0.07 \\
ZTF18aahfeiy & $10{:}17{:}15.57$ & $+43{:}31{:}24.2$ & SN\,2018loy & ZTF18aahfeiy & ZTF & 0.010 & Ia & 0.071 & 429.99 & 20.53 & 0.21 & 214.71 & 18.15 & 0.05 \\
\enddata
\tablecomments{
This table is available in its entirety in a machine-readable 
form in the online journal. A portion is shown here for guidance 
regarding its form and content. \\}
References for SNe that were recovered by ZTF and discovered elsewhere (``Discovered by'' in the table) are as follows: 
ASAS-SN \citep{Shappee2014}, ATLAS \citep{Tonry2018}, 
the Corona Borealis Observatory Supernova Survey (CSNS; \citealt{Sun18}),
G.~Cortini \citep{Cortini18},
DLT40 \citep{Tartaglia18}, \textit{Gaia} \citep{Hodgkin13},
K.~Itagaki \citep{Itagaki18, Itagaki18a},
the Italian Supernovae Search Project (ISSP; \url{http://italiansupernovae.org/}), 
LOSS \citep{Filippenko01}, 
the Mobile Astronomical System of TElescope Robots (MASTER; \citealt{Gorbovskoy13}),
the Puckett Observatory Supernova Search (POSS; \url{http://www.posssupernova.com/}, 
PS1 \citep{Chambers2016},
J.~Grzegorzek \citep{Grzegorzek18}, 
the PMO-Tsinghua Supernova Survey (PTSS, \url{http://www.cneost.org/ptss/}),
the Great Supernova Hunt (SNhunt; \url{http://nesssi.cacr.caltech.edu/SNhunt/}),
the Tsinghua University-National  Astronomical  Observatories,  Chinese  Academy  of  Sciences Transient Survey (TNTS; \citealt{Zhang15}),
the Xinming Observatory Supernova Survey (XOSS; \citep{Zhang18}),
Y.~Tanaka \citep{Tanaka18}
\tablenotetext{a}{Time and magnitude of maximum brightness are 
determined directly from the observations, as available in the AVRO alert packets. 
``$\ldots$'' is used when the SN was not detected in either the $g_\mathrm{ZTF}$ or $r_\mathrm{ZTF}$ filter. 
No correction for extinction has been applied.}
\tablenotetext{b}{$E(\bv)$ is determined using the \citet{Schlafly11} updates to the 
\citet{Schlegel98} maps.}
\tablenotetext{c}{Determined from \texttt{SNID} (see text).}
\tablenotetext{d}{JD$^\prime$ = JD $-$ 2,458,000.}
\end{deluxetable*}

\subsection{Classifications}

We broadly classify all BTS SNe as belonging to one of 4 different classes: SNe Ia, SNe II, SNe Ib/c, and super luminous SNe (SLSNe). As detailed above, these classifications are primarily made via the SN spectra, however, in some cases the photometric evolution of the SN also informs the classification. This is especially true of the SLSNe, which are defined by their luminosity (typically $M < -21$\,mag, \citealt{Gal-Yam12}; although ZTF adopts $M < -20$\,mag).

\begin{figure}
    \centering
    \includegraphics[width=1.0\linewidth]{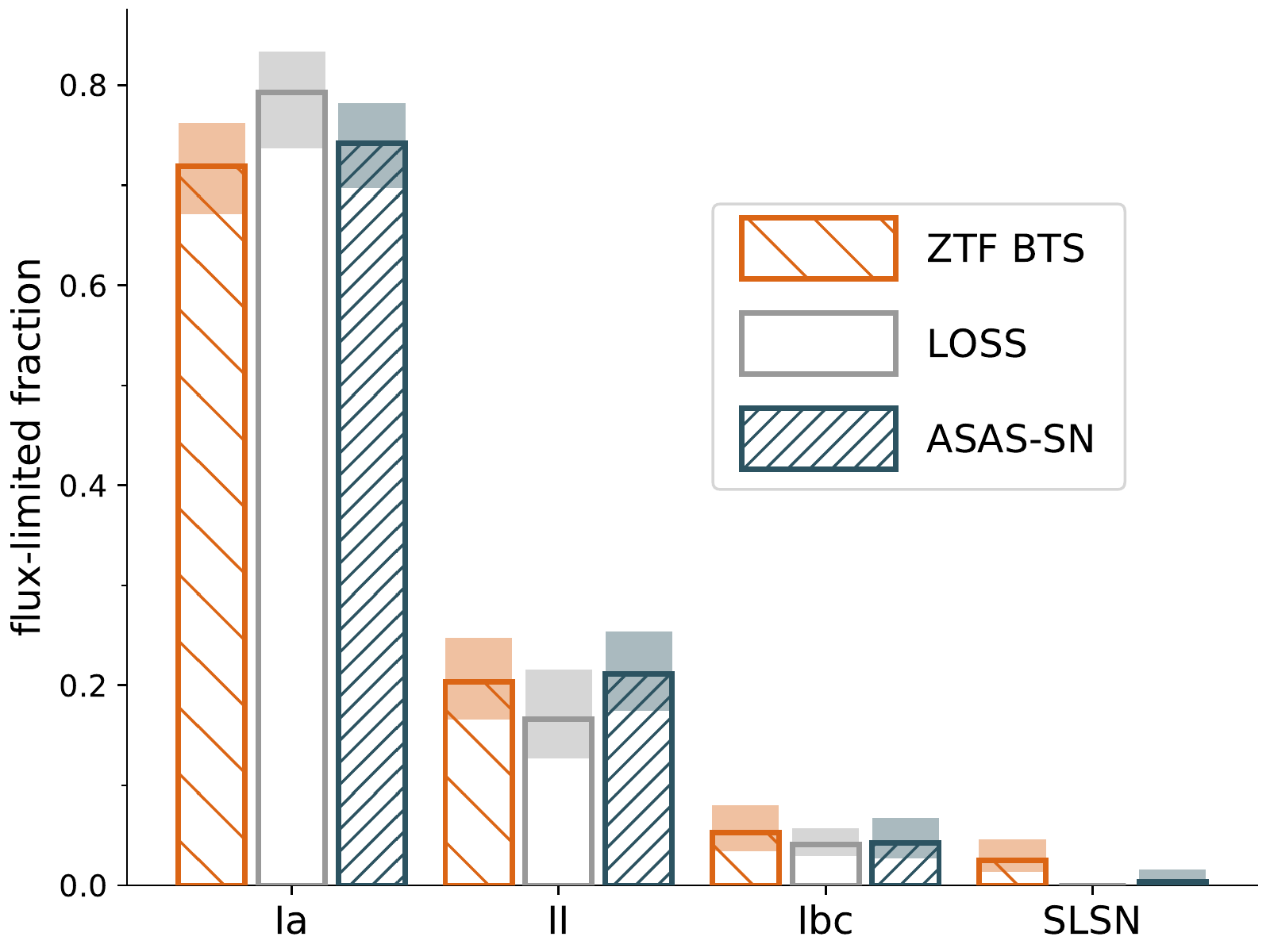}
    \caption{Relative fractions of SNe Ia, II, Ib/c, and SLSNe in the flux-limited ZTF BTS survey. The orange, hatched bars show the results for the BTS, while the light grey, open bars show results from LOSS \citep{Li11}, and the blue double-hatched bars show the results from ASAS-SN \citep{Holoien19}. The lightly shaded regions show the uncertainty on each estimate (see text for further details). To within the uncertainties, the results agree for all 3 surveys (\citealt{Li11} did not report results for the SLSN class -- see text).
    }
    \label{fig:mag_rates}
\end{figure}

Of the \nBTS~SNe, we find that \nIa~are SNe Ia, \nII~are SNe II, \nIbc~are SNe Ib/c, and \nslsne~are SLSNe. The fraction of discoveries belonging to each of these classes is in agreement with the results from \citet{Li11} for a magnitude-limited survey, as shown in Figure~\ref{fig:mag_rates}.\footnote{The ZTF BTS utilizes a 3-d cadence (see \S\ref{sec:observations}), whereas Table~7 in \citet{Li11} reports results for surveys with a 1-d and 5-d cadence. The results in \citet{Li11} are identical for 1-d and 5-d cadences, and thus we assume an intermediate 3-d cadence would also yield identical relative fractions of SNe.} Figure~\ref{fig:mag_rates} also shows the relative rate of SNe found by ASAS-SN \citep{Holoien17,Holoien17a,Holoien17b,Holoien19}, which, like the ZTF BTS and unlike LOSS, does not target specific galaxies when searching for transients.\footnote{To calculate the relative rates of SNe found by ASAS-SN (and host-galaxy offsets, which are discussed below) we include both ASAS-SN discoveries, and SNe recovered by ASAS-SN, as all discovered and recovered SNe are included in our analysis of the ZTF BTS. Therefore, the numbers shown here differ slightly from what is shown in e.g., Figure~1 of \citet{Holoien19}, which only considers SNe discovered by ASAS-SN.} Both ZTF and ASAS-SN find a higher fraction of SNe II than LOSS, although these estimates all agree to within the uncertainties. By targeting massive galaxies, including a significant fraction of passive elliptical galaxies, LOSS may have been slightly biased against finding CC SNe (see \citealt{Taubenberger17} and references therein). The relative rate of SLSNe is somewhat higher for ZTF compared to ASAS-SN, but still consistent within the uncertainties (see \S\ref{sec:slsne}).

\begin{deluxetable*}{lrrrrrrrr}
\tabletypesize{\small}
\tablewidth{0pt}
\tablecaption{Relative SN rates\label{tab:rates}}
\tablehead{
\colhead{Survey}
& \colhead{$\mathcal{R}$(Ia)}
& \colhead{$N_\mathrm{Ia}$}
& \colhead{$\mathcal{R}$(II)}
& \colhead{$N_\mathrm{II}$}
& \colhead{$\mathcal{R}$(Ibc)}
& \colhead{$N_\mathrm{Ibc}$}
& \colhead{$\mathcal{R}$(SLSN)}
& \colhead{$N_\mathrm{SLSN}$}
}
\startdata
LOSS & 0.792$\pm^{0.044}_{0.055}$ & \ldots & 0.166$\pm^{0.050}_{0.039}$ & \ldots & 0.041$\pm^{0.016}_{0.013}$ & \ldots & \ldots & \ldots \\
ASAS-SN & 0.742$\pm^{0.040}_{0.045}$ & 607 & 0.211$\pm^{0.043}_{0.037}$ & 173 & 0.043$\pm^{0.024}_{0.016}$ & 35 & 0.004$\pm^{0.012}_{0.003}$ & 3\\
ZTF BTS & 0.719$\pm^{0.043}_{0.048}$ & 547 & 0.204$\pm^{0.043}_{0.038}$ & 155 & 0.053$\pm^{0.027}_{0.018}$ & 40 & 0.025$\pm^{0.021}_{0.012}$ & 19\\
\enddata
\tablecomments{
Relative rates $\mathcal{R}$ of SNe Ia, II, Ibc, and SLSN from a flux limited search. 
Results for LOSS are taken directly from Table~7 in \citet{Li11} for a 1-d cadence and are based on an assumed 
luminosity function and Monte Carlo simulations. Thus, the total number of each type of SN is not relevant 
and therefore not reported.
Results for ASAS-SN use all discovered and recovered SNe reported in \citet{Holoien17,Holoien17a,Holoien17b,Holoien19}.
Uncertainties for both ASAS-SN and the ZTF BTS include 90\% confidence intervals (see text).
}
\end{deluxetable*}

Table~\ref{tab:rates} summarizes the relative fraction of SNe in each class for the ZTF BTS, ASAS-SN, and LOSS. For LOSS we directly use the estimates from \citet{Li11}, while for the BTS and ASAS-SN we assume the observations are drawn from a  multinomial distribution and estimate 95\% confidence intervals on the true rate via the approximate method of \citet{Goodman65} as implemented in the \texttt{MultinomCI} \citep{Signorell19} package in \texttt{R}. The true uncertainties on these fractions require a detailed estimate of the completeness of the BTS, which is beyond the scope of this paper, and will be addressed in future work (Nordin et al., in prep.). 

As a test of our classification accuracy, we compare our final classifications to those made by ePESSTO \citep{Smartt15}.\footnote{We adopt ePESSTO for comparison because (i) it is the survey with the most overlap with the ZTF BTS, and (ii) all ePESSTO classifications are made with the 3.6\,m NTT, which, on average, will perform better than SEDM for SN classification.} There are 26 sources in common between our BTS classifications and those made by ePESSTO, and the classifications are in agreement for all but 2 sources: ZTF18abmrhom (SN\,2018ffi) and ZTF18abtswjk (SN\,2018gfx). We classify ZTF18abmrhom as a SN Ia, whereas the ePESSTO spectrum is classified as having just galaxy light. The ZTF spectrum of ZTF18abmrhom, which was obtained 2 nights prior to the ePESSTO spectrum, shows clear and strong \ion{Si}{2} absorption. The same broad feature can be seen in the ePESSTO spectrum, which is otherwise dominated by emission from the host galaxy. Furthermore, the \rztf-band light curve shows a ``shoulder'' a few weeks after maximum light. Taken together, it is very likely that ZTF18abmrhom is a SN Ia. The classification of ZTF18abtswjk is more challenging: the spectrum clearly shows narrow emission lines, and we have classified the event as a SN IIn, whereas ePESSTO classified the spectrum as an active galactic nucleus (AGN). The photometric evolution of ZTF18abtswjk is far more reminiscent of SNe than AGN: the transient exhibits a $\sim$30\,d rise, followed by a very slow, monotonic decline (the transient has exhibited a monotonic fade by $\sim$1.5\,mag over the period covering 2018 and 2019) as is characteristic of many SNe IIn (e.g., \citealt{turatto93}). Furthermore, the \texttt{WISE} colors for the host galaxy are not consistent with AGN \citep{Jarrett11}. Thus, based on our comparison to ePESSTO, we conclude that our classifications are of high fidelity.

There were 3 BTS sources for which we attempted spectroscopic classification, but the nature of the transients remains unresolved. Each of the 3 candidates are positionally coincident with the nucleus of their host galaxies. ZTF18aaqkdwf (SN\,2018fhd) exhibits a broad rise and decline over $\sim$200\,d and significant P-Cygni-like feature around H$\alpha$. However, variability at the location of ZTF18aaqkdwf was detected $\sim$3.5\,yr prior to the 2018 variability by iPTF, and the host galaxy is a bright point source in the radio \citep{Helfand15}. Thus the variability may be due to an AGN. The light curve of ZTF18abuqhje (SN\,2018gki) has poor coverage with only 4 $r_\mathrm{ZTF}$ detections that show a decline of $\sim$0.7\,mag over $\sim$15\,d. We obtained 2 spectra of ZTF18abuqhje, both of which show many narrow lines that could be consistent either with an AGN or a SN IIn. 
However, we detected variability from this nucleus $\sim$2.5\,yr prior to 2018 with iPTF, which would be consistent with an AGN. The general evolution of the ZTF18aarcchg (AT\,2018boa) $r_\mathrm{ZTF}$ light curve is consistent with a SN: the transient rises by $\sim$1.5\,mag over $\sim$20\,d, followed by a monotonic decline. Spectra of ZTF18aarcchg exhibit a blue continuum with superposed narrow emission from the Balmer series, [\ion{O}{3}], and [\ion{S}{2}], all of which are consistent with star formation (the host is classified as a star-forming galaxy; \citealt{Maraston13}). Without discernible SNe features in the spectrum, we cannot classify this transient. We exclude these 3 candidates from the subsequent analysis.

\subsection{Sub-type Classifications}

The primary purpose of this study is to improve the measurement of the RCF, and to extend the redshift coverage of this measurement beyond what is presented in \citet{Kulkarni2018}. To that end, we only need to separate SNe Ia from all other transients in the ZTF BTS. Nevertheless, we attempt to make sub-type classifications, based on the \texttt{SNID} matches described above, as detailed for each class below. We caution, however, that the vast majority of these classifications are made with very low resolution ($R \approx 100$) SEDM spectra, and as a result there are significant uncertainties on the subclass of any individual SN. Furthermore, any biases in the \texttt{SNID} template set can influence the final BTS classifications. The \texttt{SNID} templates were compiled from heterogeneous sources and do not perfectly reflect the discovery-space of an un-targeted transient survey, meaning the \texttt{SNID} templates themselves are a further source of uncertainty for the final sub-classifications presented in Table~\ref{tab:sample} (see \citealt{Blondin2007} for further details).

\subsubsection{SNe Ia}

The vast majority of SNe Ia discovered in magnitude-limited surveys are considered ``normal.'' It is argued in \citet{Li11} that the most common subclass is SN\,1991T-like (SN Ia-91T in Table~\ref{tab:sample}), which are slightly over-luminous relative to normal SNe Ia, followed by SN\,1991bg-like (SN Ia-91bg in Table~\ref{tab:sample}), which are under-luminous and decline faster than normal SNe Ia. These rates are derived from LOSS, which targeted relatively massive, high star-formation rate galaxies. SNe Ia-91T seem to prefer late-type galaxies and may be associated with young stellar populations (e.g., \citealp{Howell2001}). Thus, the relative rate of SNe Ia-91T may have been over-estimated (see e.g., \citealt{Taubenberger17,Silverman12}).

In the ZTF BTS we identify 504 normal SNe Ia, 31 91T-like SNe, and 6 91bg-like SNe. This represents significantly less 91T-like ($\sim$6\%) and 91bg-like ($\sim$1\%) SNe than one would expect based on the LOSS results for a magnitude limited survey ($\sim$18\% 91T-like and $\sim$3\% 91bg-like SNe; \citealp{Li11}), however, it does agree with what is found by ASAS-SN ($\sim$6\% 91T-like and $\sim$1\% 91bg-like SNe; \citealp{Holoien17,Holoien17a,Holoien17b, Holoien19}). %These differences could be due to the differing targeting strategies of LOSS and ZTF (see \citealt{TauMillerbenberger17} and references therein). 

We caution that sub-type classification can be difficult with a single low-resolution SEDM spectrum, as is the case for the majority of the SNe in our sample. For example, the most distinguishing feature of 91T-like SNe is weak \ion{Si}{2} and \ion{Ca}{2} absorption prior to maximum light (e.g., \citealt{Filippenko1997,Branch06}). At $R \approx 100$ even relatively strong absorption can be smeared out, and as a result \texttt{SNID} frequently identifies both normal and 91T-like SNe Ia as the best matches for the SNe Ia in our sample. Furthermore, it is difficult to separate normal and 91T-like SNe in post-maximum spectra. Thus, we conservatively label SNe Ia as normal, unless there is strong spectroscopic (very weak \ion{Si}{2} and \ion{Ca}{2}) or photometric (over-luminous \textit{and} a slow decline) evidence to support a 91T-like classification. Similarly for 91bg-like SNe, unless there is strong spectroscopic (weak \ion{Fe}{2}, strong \ion{Ti}{2}; \citealt{Filippenko1997}) and photometric (under-luminous \textit{and} a fast decline) evidence, we label the SN as normal.

In addition to the above subclasses, we additionally identify 3 SN\,2002cx-like SNe, 1 SN Ia that shows signs of CSM interaction (SN Ia-CSM), and 2 SNe that appear to be super-Chandrasekhar mass explosions. %\todo{question - do we need to define these classes in more detail?}

\subsubsection{SNe II}

We have identified 162 H-rich SNe in the ZTF BTS. Of these, we make no effort to distinguish between Type IIP and IIL SNe, which are photometrically defined subtypes. We classify SNe II as either ``normal'' (119 of the 162), IIb (15), IIn (19), SLSNe-II (7, see also \S\ref{sec:slsne}), or SN\,1987A-like (2; SN II-87A in Table~\ref{tab:sample}). The relative fraction of SNe IIb and SNe IIn is significantly smaller in the BTS than that reported from LOSS \citep{Li11}. This could be a direct consequence of the different targeting strategies, although there is one important caveat for SNe IIn: for SEDm spectra, the SNe IIn represent the most difficult subclass to positively identify because the narrow emission that is the hallmark of SNe IIn (see \citealt{Schlegel90}) can easily be confused with emission lines from the host galaxy. When SEDM spectra indicate the presence of strong narrow H emission, we have generally attempted to obtain higher resolution spectra. However, some SNe IIn may not have strong enough narrow lines to be noticed in a SEDM spectrum. These would be classified as SNe II. In conclusion, SNe identified as Type IIn in the BTS all have clear evidence for strong H$\alpha$ emission lines that are significantly broader than would be expected from a galaxy or \ion{H}{2} region.

SNe for which there were both Type II and IIb \texttt{SNID} matches have been manually inspected and classified by comparing the absolute and relative strengths of the H and He features (both emission and absorption) to those seen in typical hydrogen rich SN II spectra. The two SN II-87A events exhibit a nearly identical spectroscopic evolution as SN\,1987A itself, as well as the highly unusual light curve with an initial decline followed by a $\sim$100\,d long rise (see e.g., \citealt{Arnett89,McCray93}).

\subsubsection{SNe Ibc}

There are 40 H-poor core-collapse SNe in the BTS (excluding 12 SLSNe-I; see \S\ref{sec:slsne}). We classify these sources as either SNe Ib (11), Ic (18), Ic-BL (5), Ib/c (3), Ibn (2), or Ic-pec (1). The relative ratio of SNe Ib to Ic in the BTS sample is in agreement with that found by LOSS \citep{Li11}, though we note that for both surveys the total number of stripped-envelope SNe discovered is relatively small and thus the uncertainties on the relative rates are high.

The SNe Ib clearly show He in their spectra, whereas the SNe Ic do not. There are three events that clearly lack H emission, but where we cannot distinguish between either a Ib or Ic classification (designated as SN Ib/c). The five SNe Ic-BL show very broad absorption features, similar to SN\,1998bw \citep{Patat2001}, while the two SNe Ibn display the hallmark narrow He emission lines that define the subtype \citep{pastorello07-06jc,foley06jc}. Finally, there is a single event, ZTF18aceqrrs (SN\,2018ijp), that lacks both H and He, but has a highly unusual spectroscopic and photometric evolution (Tartaglia et al., in prep.). Thus, we refer to ZTF18aceqrrs as a ``peculiar'' SN Ic (Ic-pec in Table~\ref{tab:sample}).

\subsubsection{SLSNe}\label{sec:slsne}

A population of so-called ``superluminous'' supernovae, with peak optical luminosities ($M_V$ up to $-$23\,mag) greatly in excess of any known supernova at the time, was first identified in the late 2000s \citep{Quimby2007} and quickly recognized to occur in both hydrogen-rich (SLSN-II) and hydrogen-free (SLSN-I) varieties \citep{smith07-2006gy,Gal-yam2009,quimby11}. More recent surveys have shown that the luminosity distributions of both types of SLSNe overlap with ``ordinary'' Type IIn and Ic SNe \citep{DeCia+2018,Angus+2019}, and the spectral properties may also form a continuum. Thus, the identification of a particular luminous transient as a superluminous SN, versus merely a luminous SN Ic or IIn is a challenge. Work by \cite{Quimby2018} does indicate that SLSN-I can be classified with spectral features alone without any luminosity cut. \citet{Quimby2018} show several SLSN-I events with peak luminosity below the traditional $-21$~mag threshold. However it is still true that post-peak spectra for SLSNe-I and SNe Ic can be very similar \citep{Pastorello2010}.  For the purposes of this analysis we use spectroscopic matches to previous ``unambiguous'' SLSNe-I as the primary determinant, but also place any transient with $M_g < -20$\,mag in this category even if it is well-fit by ordinary SNe (as is the case for nearly all SLSNe-II, which are good matches to SNe IIn). 

In total, we identify 19 SLSNe in the BTS sample.  Of these, 12 are classified as H-poor (SLSN-I), and the remaining 7 have H emission lines (SLSN-II).  This represents $2.5\%\pm^{2.1}_{1.2}$ of the sample---which is, not surprisingly, a far higher fraction than what was found in galaxy-targeted surveys such as LOSS. LOSS found a single SLSN (SN\,2006gy; \citealt{Foley06}), which was not recognized as part of a separate class in \citet{Li11}. SLSNe are volumetrically rare, and best found via untargeted wide-area surveys \citep{quimby11}. The relative rate of SLSNe in ASAS-SN is $0.4\%\pm^{1.2}_{0.3}$. This is lower but still consistent with our BTS estimate within the uncertainties on both measurements. 
% The low number of SLSNe in ASAS-SN is due to the shallower magnitude limit compared to BTS. ASAS-SN can only detect the most nearby SLSNe.

The true fraction of SLSNe may be even higher than what we report here: over this early-survey period our selection methods were biased against SLSNe, because these very long-lived and slow-rising transients were often present in the reference image, such that at peak the subtraction of their own pre-max flux made them appear fainter than they really were and thus less likely to pass the BTS filter.

%This is likely the result of differences in strategy and depth. SLSNe are volumetrically rare, and best found via untargeted wide-area surveys \citep{quimby11}. 

The detailed analysis of the four SLSN-I discovered during the ZTF commission phase has been submitted for publication by \citep{Lunnan2019}, and a thorough investigation of the full ZTF SLSN sample is underway.

% For the purpose of this study, to get secure classifications of SNe in our sample, latest spectrum of each SN was used for SNID classification and the top 15 matches were plotted and saved. All these plots were analysed manually and closest classification by visual inspection was reported along with SNID best match redshift and error in redshift (upto 3 decimal places) , best match template name and phase (in days). If there were multiple close matching templates, phase information from light curves was used to decide the best fit. For SNe spectra which didn't have any matches in SNID, a different spectra was manually ran SNID on to decide the best match. \\

\section{SN Distances and Host Galaxies}

As in \citet{Kulkarni2018}, our aim is to measure the RCF of local galaxy catalogs, in this case using SNe from the ZTF BTS. In order to make this measurement we need to both identify the host galaxy for every SN, and measure the SN redshift (for cases where the host redshift is unknown). Using this information it is then possible to calculate the RCF.

\subsection{SN redshifts}

\begin{figure}
    \centering
    \includegraphics[width=1.0\linewidth]{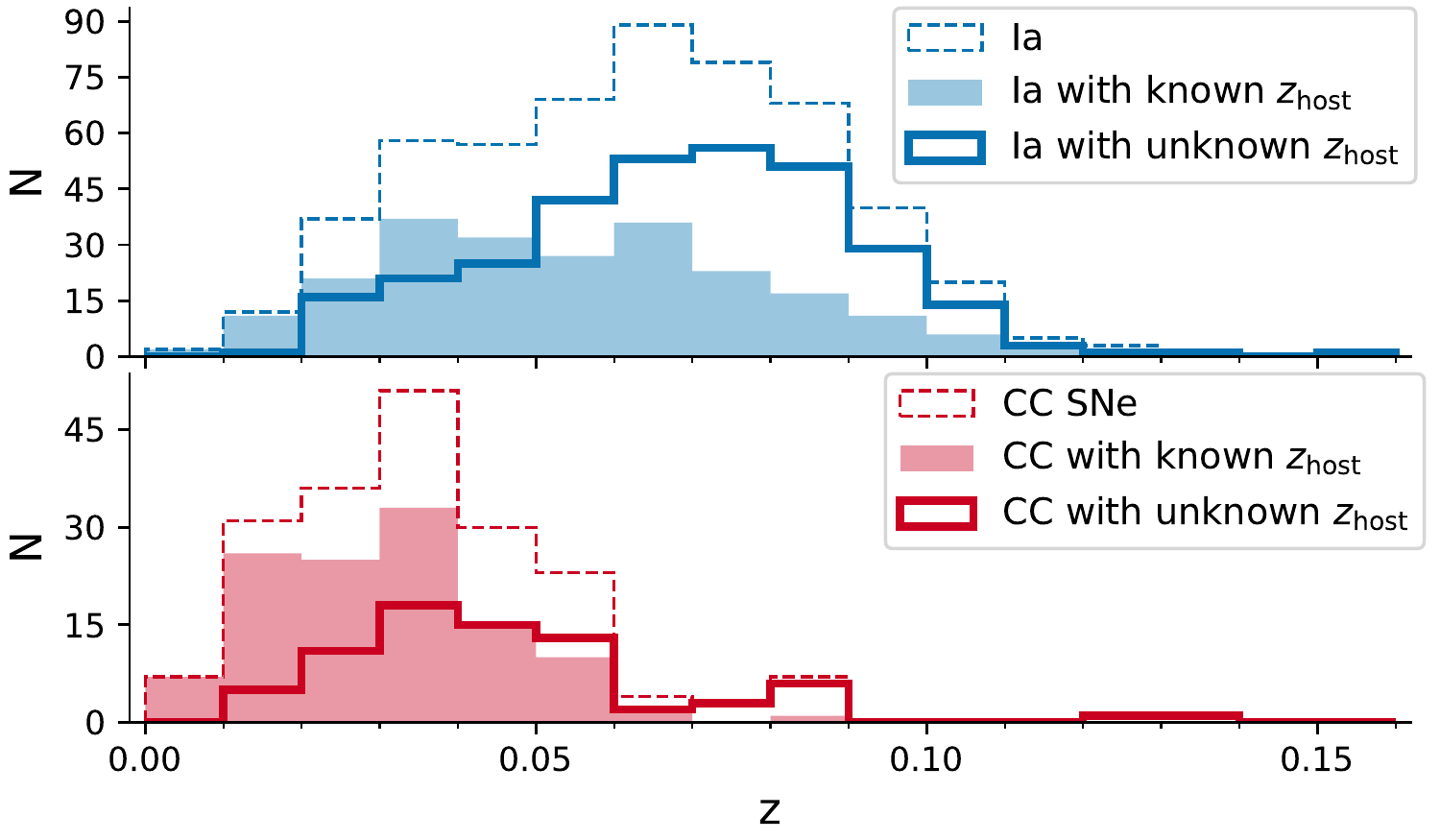}
    \caption{Redshift distribution of BTS SNe Ia (top) and CC SNe (bottom)  shown via thin dashed lines. SNe associated with host galaxies of known redshift are shown as solid histograms. If the host galaxy redshift is unknown we adopt $z_\mathrm{SN}$ as the host galaxy redshift, as shown by the thick solid lines. SLSNe are not shown.}
    \label{fig:redshift_distribution}
\end{figure}

In addition to providing SN spectral types, \texttt{SNID} estimates the redshift of the SN it is attempting to classify. We adopt the redshift of the best-matching \texttt{SNID} template as the redshift of the SN, $z_\mathrm{SN}$.\footnote{No corrections for Heliocentric, Galactocentric, or host-galaxy-rotation velocity are applied to $z_\mathrm{SN}$.} The redshift distribution for BTS SNe is shown in Figure~\ref{fig:redshift_distribution}, where we adopt the redshift of the host galaxy ($z_\mathrm{host}$) when known, otherwise we show $z_\mathrm{SN}$.

\begin{figure}
    \centering
    \includegraphics[width=1.0\linewidth]{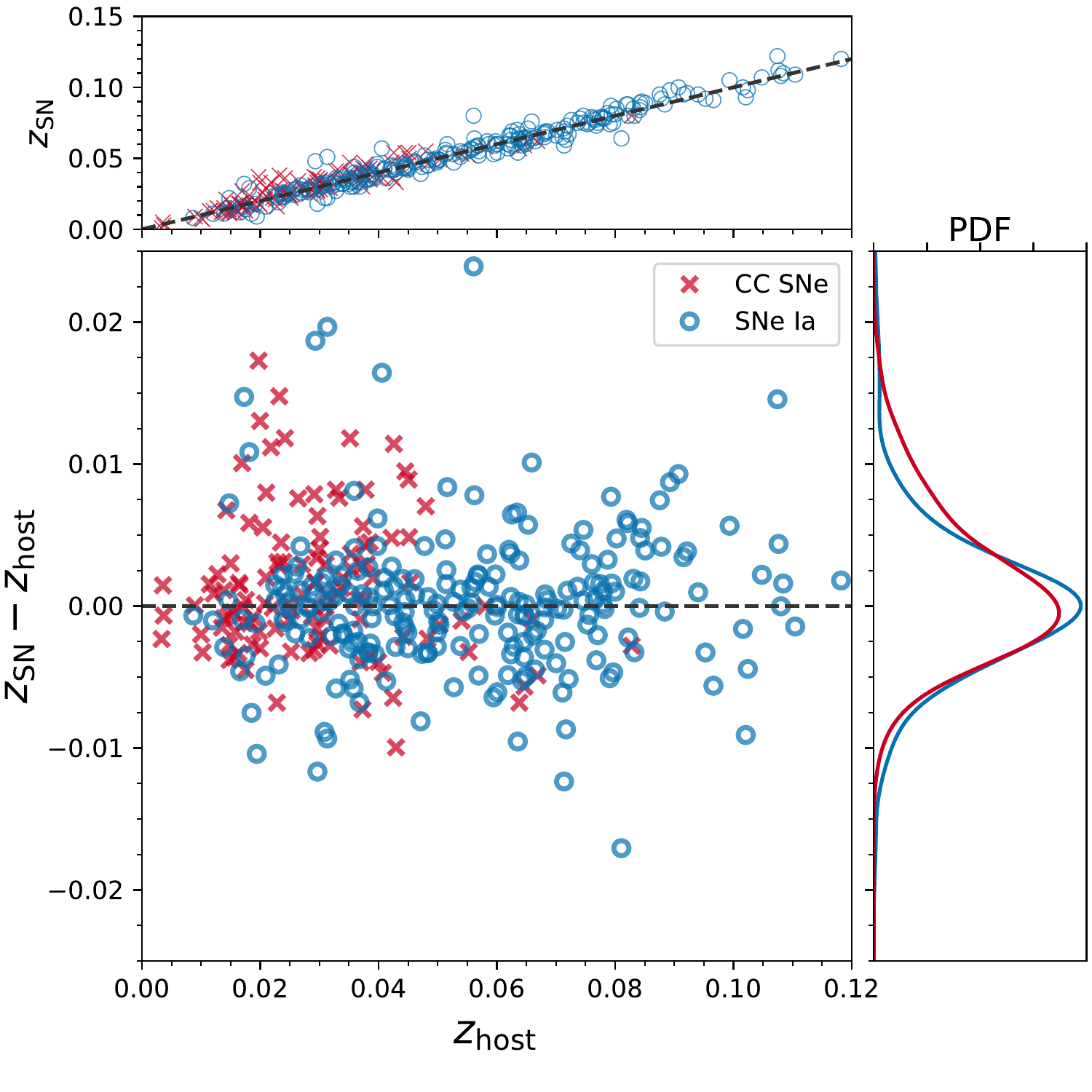}
    \caption{Difference between host-galaxy redshifts ($z_\mathrm{host}$) and SNID-derived redshifts ($z_\mathrm{SN}$) as a function of host-galaxy redshift. CC SNe are shown as crosses, while SNe Ia are shown as open circles. The scatter in the difference is nearly constant as a function of redshift. \textit{Top}: $z_\mathrm{SN}$ vs. $z_\mathrm{host}$. \textit{Right}: A Gaussian KDE of the PDF of the residuals for CC SNe (red) and SNe Ia (blue). The typical scatter is $\sim$0.004 (see text), while 95\% of all SNe Ia have $z_\mathrm{SN}$ within 0.01 of $z_\mathrm{host}$.}
    \label{fig:snid_host}
\end{figure}

We can estimate the accuracy of the \texttt{SNID} redshift measurements using the subset of BTS SNe that have host galaxies with known redshift (for more on the identification of BTS host galaxies, see \S\ref{sec:host_id}). We find that $z_\mathrm{SN}$ is a good estimator of the host galaxy redshift, $z_\mathrm{host}$, as summarized in Figure~\ref{fig:snid_host} for the 345 ZTF BTS host galaxies with known redshifts. 

The main panel in Figure~\ref{fig:snid_host} shows the difference between $z_\mathrm{SN}$ and $z_\mathrm{host}$ as a function of $z_\mathrm{host}$. The residuals show that for SNe Ia there is relatively small scatter ($\sigma \approx 0.0037$)\footnote{We use a robust estimate of scatter by taking half of the difference between the 84$^\mathrm{th}$ and 16$^\mathrm{th}$ percentiles. The sample standard deviation is $\sim$0.005. The sample standard deviation for the recession velocity is $\Delta\,v/v \approx 0.14$.} and virtually no bias in the estimates of $z_\mathrm{SN}$. The scatter is somewhat higher for CC SNe ($\sigma \approx 0.0047$), where \texttt{SNID} appears to systematically overestimate the true redshift, as is seen in the Gaussian Kernel Density Estimate (KDE) shown in the right panel of Figure~\ref{fig:snid_host}. We find some evidence for an increased scatter at higher redshifts. For SN Ia hosts with $z \leq 0.04$,  $\sigma \approx 0.0032$, while for hosts with $0.08 \leq z < 0.12$ $\sigma \approx 0.0045$ (though the 90$^\mathrm{th}$ percentile widths are nearly identical in these two regions). This increased scatter makes sense as the signal-to-noise ratio typically decreases for higher redshift SNe; for SNe outside the redshift range shown in Figure~\ref{fig:snid_host} the typical uncertainty on any individual redshift may be larger than $\sim$0.004. We also find that the scatter is not appreciably larger when restricting the sample to only those SNe that have been observed by SEDM. In the analysis that follows, we assume that $z_\mathrm{host} = z_\mathrm{SN}$ for normal SNe Ia, and that the typical uncertainty on $z_\mathrm{SN}$ is 0.004. We note that (i) this is very consistent with the uncertainty ($\sigma = 0.005$) reported in figure 19 in \cite{Blondin2007}, even though a very different redshift range was used ($z=0.1-0.8$), and (ii) our estimated redshift uncertainty on $z_\mathrm{SN}$ is much larger than the uncertainty in the wavelength calibration of SEDM (figure 13 in \citealp{Rigault2019}). These two findings indicate that the accuracy of our \texttt{SNID} based redshifts ($z_\mathrm{SN}$) is not limited by the low resultion of SEDM.

\subsection{Host Galaxy Identification}\label{sec:host_id}

Correctly associating a newly discovered transient with its host galaxy is a challenging problem, especially when the redshifts of the host candidates are unknown.  Simply identifying the closest galaxy (in angular offset) is likely to produce a significant number of misidentifications, especially in the case of nearby SNe for which angular offsets relative to the host nuclei may be quite large.

%The trivial solution to automatic identification is to assign whichever galaxy has the smallest angular offset from the transient as the host, however, this trivial solution leads to a significant number of misidentifications (especially for nearby galaxies where the angular offset may be quite large).  Alternative solutions include visual inspection, wherein someone examines the field of the transient and selects the most likely host (based on some combination of proximity and angular size), or automated approaches that simultaneously consider several galaxies and select the host via some optimal statistic (e.g., a brightness-weighted separation from the transient \dap{Didn't you do something like this for GRB host ID?}, or a galaxy orientation separation distance  \citealt{Sako18}).\dap{there is a rich literature about this - if you want to expand on what has been written, please do}

We use a combination of automated procedures and visual inspection to identify hosts for ZTF BTS SNe. As an initial pass, we query the NASA Extragalactic Database (NED)\footnote{\url{https://ned.ipac.caltech.edu/}} for all galaxies within 2\arcmin~of the SN position. Within this list, the galaxy with the smallest angular separation from the SN \textit{AND} a cataloged redshift in NED is automatically assigned as the host. In cases where there are no cataloged galaxy redshifts within 2\arcmin of the SN, the galaxy with the smallest angular separation from the SN is assigned as the host. In cases where the $z_\mathrm{host}$ and $z_\mathrm{SN}$ significantly differ ($\left|{z_\mathrm{SN} - z_\mathrm{host}}\right| > 0.05$), the galaxy with the smallest angular separation from the SN is assigned as the host. From here we calculated the projected separation, $d_\mathrm{p}$, between the SN and the host galaxy using the redshift of the SN (see above). We visually inspect host candidates with a projected separation $d_\mathrm{p} \geq 19$\,kpc. In most of these cases it is clear that the automated procedure identified a background galaxy that is clearly not the host, in which case we update the host with the NED galaxy with the smallest angular separation from the SN. Following this procedure, there were a total of 12 SN host identifications with separations $d_\mathrm{p} \geq 19$\,kpc (9 SNe Ia and 3 SNe II). In 11 of these 12, the host redshift is known \textit{and} that redshift matches that of the SN, providing confidence in these associations. We cannot rule out the possibility that these SNe occurred in faint dwarf galaxies that are associated with the bright galaxy that has been identified as the host. For the purposes of the RCF calculation below, we assume each of these identifications to be correct. For the remaining SN, ZTF18acrcetn (SN\,2018jag), there is a very bright ($r' = 14.3$\,mag) elliptical galaxy, PSO\,J015.9596+10.5902, in the field of the SN. The SN is $\sim$3.2 Petrosian radii from the galaxy which has an SDSS photometric redshift, $0.052 \pm 0.012$ \citep{Abolfathi18}, that agrees with the SN redshift determined by SNID, 0.053 (which gives $d_\mathrm{p}\approx24$~kpc). For the calculations below we assume that this identification is correct.

Following this procedure, we use deep $i$-band stack images from PS1 to visualize the position of each SN relative to its host galaxy. In the vast majority of cases these images confirmed a clear association between the SN and host galaxy. In some cases, the putative host was extremely faint and a significantly brighter galaxy with only slightly larger angular separation was selected at the likely host. Finally, there were a handful of cases where the association was ambiguous or with a low signal-to-noise ratio PS1 detection. We find that for 40 SNe in the sample, the host identification is ambiguous. We exclude these SNe from the host galaxy analysis below. We thus identify host galaxies for 721 of the \nBTS~SNe in the ZTF BTS sample. Properties of the BTS SNe host galaxies are summarized in Table~\ref{tab:hosts}, including notes on each of the SNe identified as having ambiguous hosts. For the 40 ambiguous cases, we find that 29 SNe have no discernible host, including 17 SLSNe, for which host galaxies are typically not found in imaging at the depth of PS1 in this redshift range (e.g., \citealt{quimby11}). The remaining 11 are either roughly equidistant between multiple galaxies of the same brightness, or very close to a faint galaxy, with a significantly brighter galaxy at a similar redshift of the SN residing at much larger angular separation.

\begin{figure}
    \centering
    \includegraphics[width=1.0\linewidth]{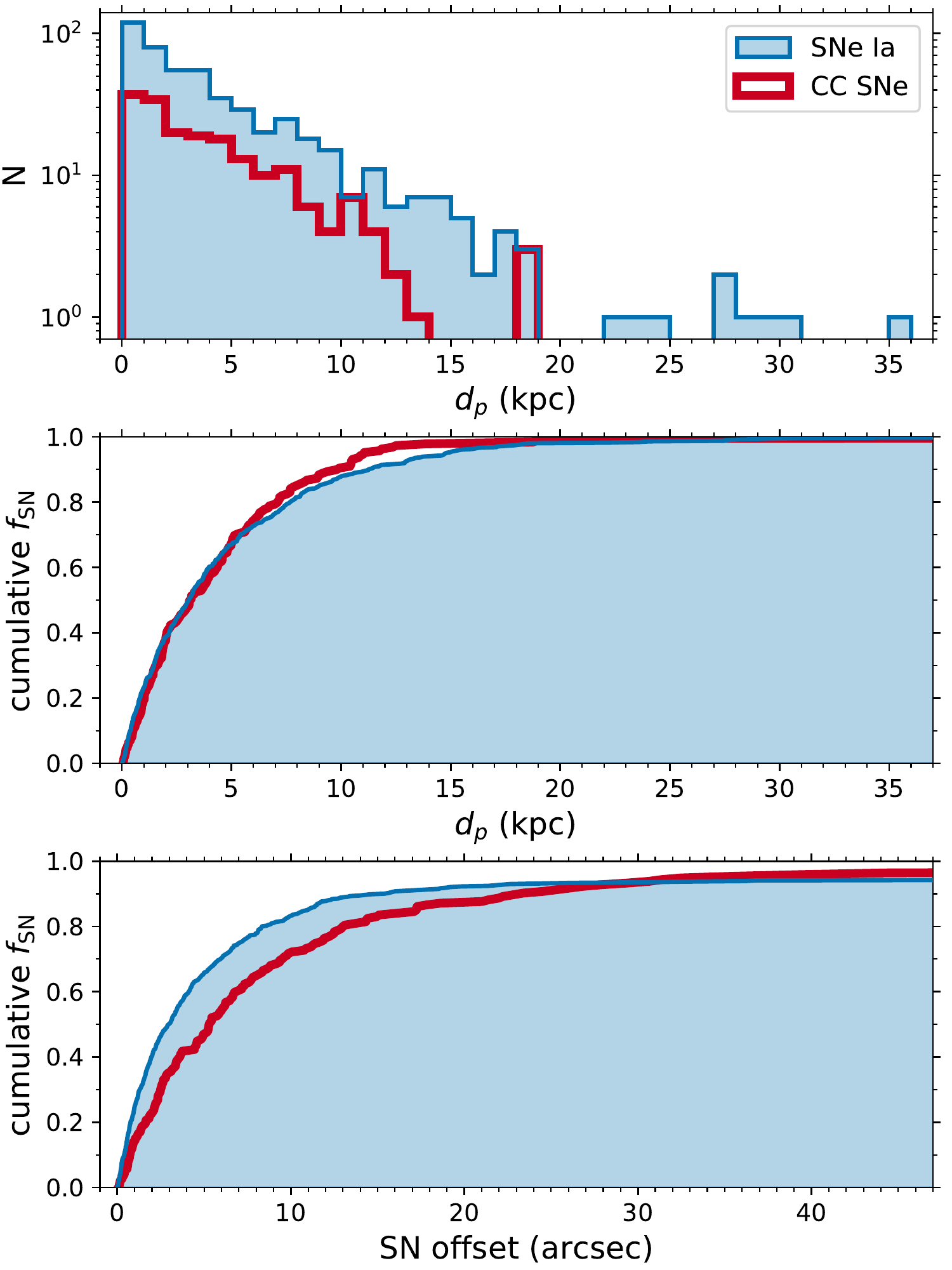}
    \caption{\textit{Top}: Projected physical separation $d_p$, in kpc, between ZTF BTS SNe and their respective host galaxies. SNe Ia are shown via a solid histogram, while CC SNe are shown via a thick, crimson line. The general shape of these distributions are similar to what was found during PTF, with roughly an order of magnitude fewer SNe at $d_p \approx 10$\,kpc, as there are with $d_p < 1$\,kpc \citep{Kasliwal12}.
    \textit{Middle}: Cumulative distribution of $d_p$ for BTS SNe and their hosts. The distribution for SNe Ia and CC SNe is nearly identical; the small discrepancies observed between $\sim$10--15\,kpc are likely due to small number statistics. \textit{Bottom}: Cumulative distribution of the angular offset, in arcsec, between BTS SNe and their hosts. Given that the average SN Ia is at higher redshift than the average CC SN in the BTS (Figure~\ref{fig:redshift_distribution}), but the physical separations are the same, it makes sense that SNe Ia, on average, have a smaller angular offset than CC SNe.}
    \label{fig:host_separation}
\end{figure}

The host galaxy coordinates available via NED come from a heterogeneous set of catalogs and surveys, resulting in an astrometric offset between NED and ZTF SN positions, which are measured relative to \textit{Gaia} \citep{Gaia-Collaboration16}. We crossmatch the host positions against the PS1 catalog, which is also calibrated against \textit{Gaia}, in order to place the BTS SNe and host galaxies on the same relative astrometric system. 715 of our initial host positions have counterparts within 2 arcsec in the PS1 DR1 \texttt{MeanObject} table, which is astrometrically calibrated against \textit{Gaia}. Four of the hosts, those associated with ZTF18aapgrxo (SN\,2018bym), ZTF18aayjyub (SN\,2018cod), ZTF18acaeous (SN\,2018hbu), and ZTF18acrknyn (SN\,2018jef), are too faint to be included in the PS1 \texttt{MeanObject} table, and instead we use PS1 positions from the \texttt{StackObjectThin} table. The last two hosts, associated with ZTF18acbzoyh (SN\,2018hqu) and ZTF18acdwohd (SN\,2018ids) are not detected in the PS1 catalog, and we instead use their positions from SDSS and the \textit{Galaxy Evolution Explorer} (\textit{GALEX}; \citealt{Martin05}), respectively, in Table~\ref{tab:hosts}. Following this update of the positions, we recalculate the distribution of host-galaxy separations, which is shown in Figure~\ref{fig:host_separation}. We find a nearly identical distribution in projected separation for CC SNe and SNe Ia (middle panel of Figure~\ref{fig:host_separation}). We use a two-sample Kolmogorov–Smirnov (KS) test and a $\chi^2$ test for independence to determine the statistical difference between the two distributions and find no significant difference between the projected offsets of CC SNe and SNe Ia.

\begin{deluxetable*}{lrrllrrrccrcccc}
\tabletypesize{\scriptsize}
\tablewidth{0pt}
\tablecaption{BTS Host Galaxies\label{tab:hosts}}
\tablehead{
\colhead{ZTF}
& \colhead{Host}
& \colhead{$\alpha_\mathrm{host}$}
& \colhead{$\delta_\mathrm{host}$}
& \colhead{$z_\mathrm{host}$}
& \colhead{SN offset}
& \colhead{$d_p$}
& \colhead{$m_g$}
& \colhead{$m_r$}
& \colhead{$m_i$}
& \colhead{$m_z$}
& \colhead{$m_y$}
& \colhead{$m_{W1}$}
& \colhead{$m_{W2}$}
& \colhead{$E(\bv)$\tablenotemark{a}} \\
\colhead{Name}
& \colhead{Name}
& \colhead{(J2000.0)}
& \colhead{(J2000.0)}
& \colhead{}
& \colhead{(arcsec)}
& \colhead{(kpc)}
& \colhead{(mag)}
& \colhead{(mag)}
& \colhead{(mag)}
& \colhead{(mag)}
& \colhead{(mag)}
& \colhead{(mag)}
& \colhead{(mag)}
& \colhead{(mag)}
}
\startdata
ZTF18aabssth & PSO J165.1878+22.2877 & $11{:}00{:}45.07$ & $+22{:}17{:}15.8$ & $0.02291$ & $4.44$ & $2.06$ & $16.89$ & $16.49$ & $16.38$ & $16.38$ & $16.61$ & $15.79$ & $16.38$ & 0.015 \\
ZTF18aabxlsv & PSO J157.4639+09.0106 & $10{:}29{:}51.32$ & $+09{:}00{:}38.1$ & $0.04797$ & $9.53$ & $8.96$ & $16.17$ & $15.48$ & $15.27$ & $15.02$ & $15.02$ & $14.82$ & $15.30$ & 0.025 \\
ZTF18aaemivw & PSO J158.4280+39.4908 & $10{:}33{:}42.72$ & $+39{:}29{:}26.8$ & $0.06807$ & $0.33$ & $0.43$ & $16.66$ & $16.03$ & $15.67$ & $15.53$ & $15.35$ & $14.94$ & $15.08$ & 0.012 \\
ZTF18aagpzjk & PSO J119.8484+16.4214 & $07{:}59{:}23.61$ & $+16{:}25{:}16.9$ & $0.01631$ & $26.83$ & $8.91$ & $14.96$ & $14.49$ & $14.33$ & $14.49$ & $14.87$ & $13.85$ & $14.43$ & 0.031 \\
ZTF18aagrdcs & PSO J218.3331+41.2658 & $14{:}33{:}19.95$ & $+41{:}15{:}56.9$ & $0.01814$ & $5.43$ & $2.00$ & $18.53$ & $17.62$ & $17.45$ & $18.11$ & $18.86$ & $17.32$ & $18.00$ & 0.012 \\
ZTF18aagrtxs & PSO J198.6087+50.9792 & $13{:}14{:}26.08$ & $+50{:}58{:}45.0$ & $0.02966$ & $7.94$ & $4.72$ & $\ldots$ & $\ldots$ & $\ldots$ & $13.77$ & $13.68$ & $13.70$ & $14.36$ & 0.010 \\
ZTF18aagstdc & $\ldots$ & $\ldots$ & $\ldots$ & 
$\ldots$ & $\ldots$ & $\ldots$ & $\ldots$ & $\ldots$ & $\ldots$ & 
$\ldots$ & $\ldots$ & $\ldots$ & $\ldots$ & $\ldots$ \\
ZTF18aagtcxj & PSO J248.0475+42.7139 & $16{:}32{:}11.40$ & $+42{:}42{:}50.0$ & $0.03240$ & $2.40$ & $1.55$ & $16.72$ & $15.70$ & $15.31$ & $15.00$ & $14.82$ & $14.28$ & $14.75$ & 0.011 \\
ZTF18aahesrp & PSO J128.9390+28.2705 & $08{:}35{:}45.35$ & $+28{:}16{:}13.7$ & $\ldots$ & $1.36$ & $1.35$ & $19.43$ & $18.99$ & $18.82$ & $18.69$ & $18.68$ & $19.14$ & $19.86$ & 0.036 \\
ZTF18aahfeiy & PSO J154.3116+43.5219 & $10{:}17{:}14.78$ & $+43{:}31{:}18.8$ & $0.07126$ & $10.09$ & $13.71$ & $17.13$ & $16.69$ & $16.38$ & $16.43$ & $16.50$ & $16.19$ & $16.61$ & 0.010 \\
\enddata
\tablecomments{
This table is available in its entirety in a machine-readable 
form in the online journal. A portion is shown here for guidance 
regarding its form and content. Optical $grizy$ photometry is taken from PS1 Kron magnitude measurements,
while mid-IR $W1$ and $W2$ photometry are from \texttt{Tractor} and \textit{WISE} (see text). All magnitudes 
are reported in the AB system, and no correction for extinction has been applied.}
\tablenotetext{a}{$E(\bv)$ is determined using the \citet{Schlafly11} updates to the 
\citet{Schlegel98} maps.}
\end{deluxetable*}

\begin{figure}
    \centering
    \includegraphics[width=1.0\linewidth]{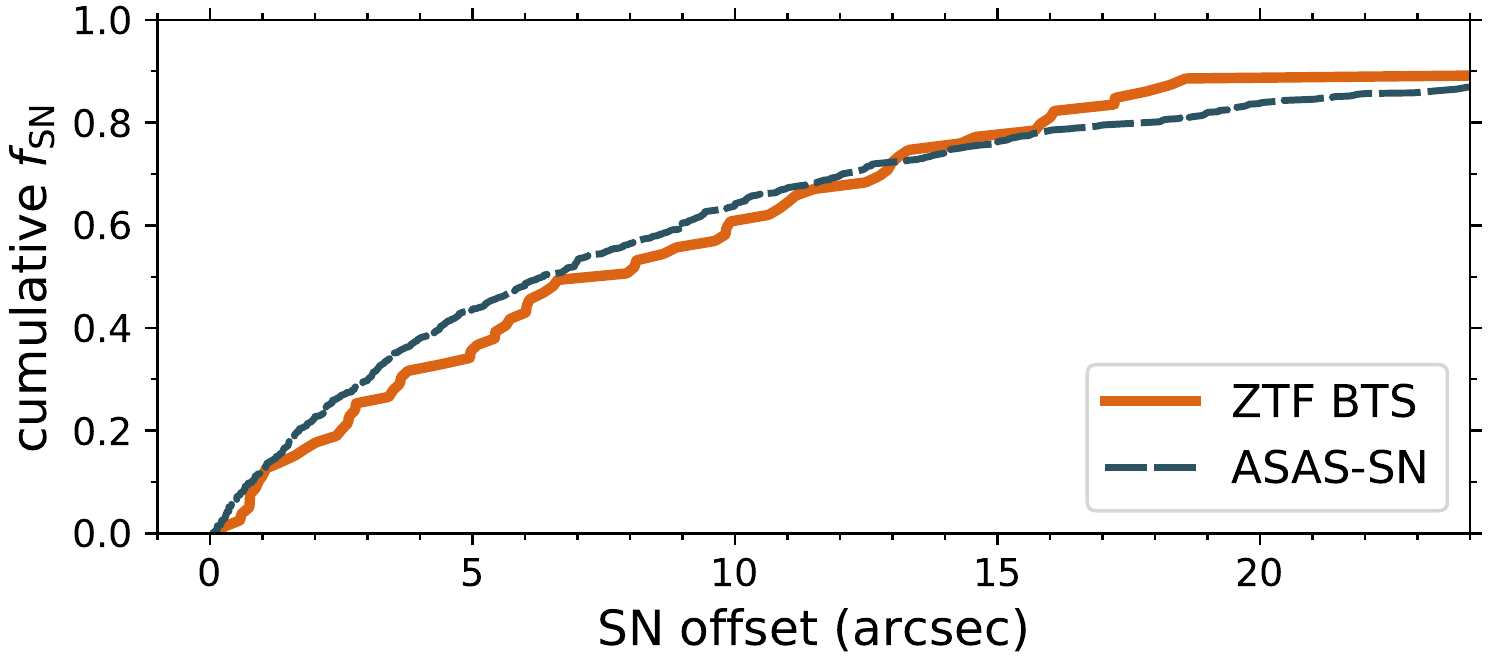}
    \caption{Cumulative distribution of the angular offset between $m_\mathrm{peak} \le 17$\,mag SNe and their host galaxies found by the ZTF BTS (solid line) and ASAS-SN (dashed line). The distributions are generally very similar, with ASAS-SN doing slightly better at small separations ($< 5\arcsec$), however, these differences may simply be due to small number statistics.}
    \label{fig:asassn_offset}
\end{figure}

Using the newly identified host offsets, we can additionally examine whether or not there is a bias against finding nuclear SNe in ZTF, as has been found for other surveys (see \citealt{Holoien19}). Figure~\ref{fig:asassn_offset} compares the cumulative distribution of angular offsets for bright ($m_\mathrm{peak} \le 17$\,mag) BTS SNe and ASAS-SN.\footnote{As shown in Figure~\ref{fig:host_separation} the BTS sample needs to be restricted for a fair comparison to ASAS-SN as high-$z$ SNe have smaller angular offsets.} We find remarkably similar distributions between ZTF and ASAS-SN. Both a two-sample KS test and a $\chi^2$ test for independence do not show a statistically significant difference between the two samples. This suggests that ZTF is not significantly biased against finding nuclear SNe. In contrast, other surveys that detect bright SNe are biased away from galaxy nuclei \citep{Holoien19}. 

Following host identification, we need to determine the absolute magnitude of the host galaxies in order to measure the RCF (see \citealt{Kulkarni2018} and below for further details). In this study we focus on the mid-infrared (mid-IR) flux of the host galaxies, primarily for 2 reasons: (i) a galaxy's mid-IR absolute magnitude serves as a good proxy for the total galactic stellar mass \citep{2013MNRAS.433.2946W}, and (ii) mid-IR photons are mostly transparent to dust in the Milky Way, meaning that significant reddening corrections are not needed to estimate a galaxy's absolute magnitude in the mid-IR. The \textit{Wide-field Infrared Survey Explorer} (\textit{WISE}; \citealt{Wright10}) satellite conducted an all-sky survey in the mid-IR, and we use \textit{WISE} images to determine the brightness of ZTF BTS SN host galaxies at 3.4\,$\mu$m. 

The largest \textit{WISE} source catalogs (e.g., unWISE; \citealt{Schlafly19}) utilize an unresolved (i.e. stellar) point-spread-function (PSF) to measure source flux. This PSF is not appropriate for many of the low-$z$ galaxies in our sample, which are clearly resolved in \textit{WISE} imaging. The recent development of \texttt{The\,Tractor} \citep{Lang16a}, enables ``forced'' \textit{WISE} flux measurements, where the mid-IR apertures are determined via sources detected in (relatively) high-resolution optical images. \texttt{The\,Tractor} was used to measure the mid-IR flux of $\sim$400 million \textit{WISE} sources that were detected by SDSS \citep{Lang16}, and now is also being applied to Legacy Survey \citep{Dey19} images. There are 371 BTS host galaxies that have \texttt{Tractor} forced mid-IR photometry from both Legacy Survey and SDSS images, while an additional 157 hosts have detections in just Legacy Survey images and another 90 hosts have detections in just SDSS\footnote{We only retain galaxies with a signal-to-noise ratio $>$5 in the $W1$ filter from the forced photometry catalogs.} (hereafter we refer to this aperture-matched forced photometry as \texttt{Tractor} photometry). We include \texttt{Tractor} photometry based on both Legacy Survey and SDSS images in our analysis of the RCF below. For the 371 sources detected both in the Legacy Survey and SDSS we compare the \texttt{Tractor} photometry dervied from each set of images and measure a sample standard deviation of 0.008 in $\Delta\, \mathrm{flux}/\mathrm{flux}$. This small difference suggests that there are no systematic effects introduced by combining photometry from the two different optical catalogs. Ultimately, this results in 618 host galaxies with 3.4\,$\mu$m flux measurements that we can use in the analysis of the RCF. Including only the brightest mid-IR galaxies in our sample will bias our final measurement of the RCF, as discussed below.

\section{The Redshift Completeness Factor}\label{sec:rcf}

To calculate the Redshift Completeness Factor (RCF), we follow the methodology originally outlined in \citet{Kulkarni2018}. The RCF captures the probability that a random galaxy will have a catalogued spectroscopic redshift as a function of its redshift and IR luminosity. To estimate the RCF we use only SNe Ia, as they occur in both star-forming and passive galaxies, whereas CC SNe would only trace star-forming galaxies. When a ZTF BTS SN has a previously cataloged host galaxy redshift, we consider that a ``hit'' ($\mathrm{NED}_z$), and when the host does not have a known redshift that is considered a ``miss'' ($!\mathrm{NED}_z$).

By raw number, there are 512 SNe Ia with known hosts in the BTS,\footnote{We exclude Ia-02cx, Ia-csm, and Ia-SC events from the RCF calculations. These ``peculiar'' events account for $\sim$1\% of the BTS SN Ia sample, and would not substantially change our analysis.} and 227 of them are ``hits'' (have known redshifts). Thus, over the redshift range sampled by the BTS, the $\mathrm{RCF} = 44\% \pm 1\%$ (90\% confidence interval).\footnote{As in \citet{Kulkarni2018} we find that the RCF as traced by CC SNe is much higher than that traced by SNe Ia. This suggests that redshift catalogs are more complete for star-forming galaxies than passive galaxies.} This estimate is significantly lower than what was found for a lower-redshift sample (\citealt{Kulkarni2018} estimated $\mathrm{RCF} \approx 75\%$). 
 The difference in RCF estimates can be entirely understood by the differing redshift distributions of the two samples. If we restrict our analysis to SNe with $z \le 0.03$, we find the $\mathrm{RCF} = 69\% \pm 4\%$ (90\% confidence interval), which is consistent with the results reported in \citet{Kulkarni2018}. At face value these results show that the RCF decreases as redshift increases, an unsurprising result. 

\begin{figure*}
    \centering
    \includegraphics[width=1.0\linewidth]{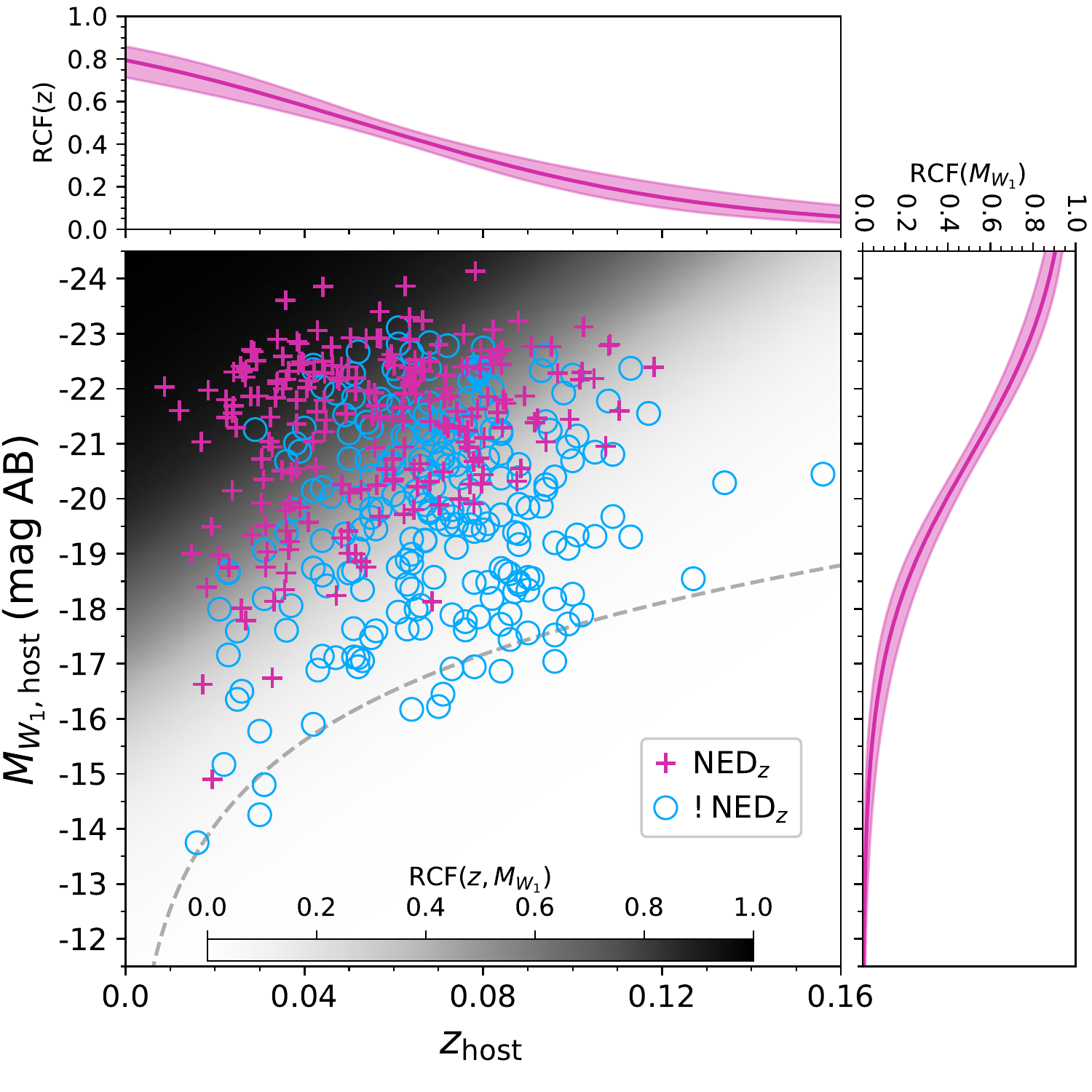}
    \caption{Absolute $W_1$-band magnitude, $M_{W_1,\mathrm{host}}$, vs.\ redshift, $z$, for the host galaxies of SNe Ia in the ZTF BTS. Galaxies with known redshifts (from NED or other databases) prior to SN discovery are shown as magenta pluses, while those lacking redshifts (!NED$_z$) are shown as gold circles. The dashed line roughly corresponds to the \textit{WISE} detection limit $m_{W_1, \mathrm{limit}} \approx 20.629$\,mag \citealt{Schlafly19}. The shaded background shows the probability of a host galaxy having a cataloged redshift given its redshift and $M_{W_1}$ ($\mathrm{RCF}(z, M_{W_1})$), based on 442 galaxies with \textit{WISE} detections. The top and right plots show the probability of a host galaxy having a cataloged redshift given \textit{only} its redshift, $\mathrm{RCF}(z)$, \textit{or} $M_{W_1}$, $\mathrm{RCF}(M_{W_1})$, respectively. In these two plots the solid lines show the median value of the RCF, while the shaded region corresponds to the 90\% credible region on the RCF.}
    \label{fig:RCF}
\end{figure*}

We can further constrain the RCF as a function of redshift and galaxy luminosity by estimating the joint distribution for a galaxy to have a cataloged redshift given its redshift and $M_{W1}$, $\mathrm{RCF}(z, M_{W1})$.\footnote{We convert observed $W1$ magnitudes to absolute magnitude by calculating the distance modulus with \texttt{astropy} \citep{Astropy-Collaboration13}, assuming a concordance cosmology with $\Omega_\Lambda = 0.7$, $\Omega_M = 0.3$, and $H_0 = 70$\,km\,s$^{-1}$\,Mpc$^{-1}$. We also correct for Milky Way extinction, a small effect, using $E(B-V)$ from \citet{Schlafly11}, $R_V=3.1$, and the extinction law from \citet{Fitzpatrick07}.} A detailed summary of the joint probability RCF calculation is included in Appendix~\ref{sec:rcf_calc}. To estimate $\mathrm{RCF}(z, M_{W1})$, we include only those SNe Ia with identified host galaxies that have a measured IR brightness. This reduces the sample to 442 SNe, of which 213 are ``hits.'' An estimate of this joint distribution is shown in Figure~\ref{fig:RCF}, as well as one-dimensional probabilities $\mathrm{RCF}(z)$ and $\mathrm{RCF}(M_{W1})$. All 512 SNe Ia are used to constrain $\mathrm{RCF}(z)$, as a host galaxy identification or brightness measurement is not necessary for that calculation. 

From Figure~\ref{fig:RCF} it is clear that the analysis in \citet{Kulkarni2018} was significantly limited by the lower redshift sample that was available at that time. For example, there are no ``hits'' for $z > 0.12$, and the $\mathrm{RCF}(z)$ tends towards zero at high redshifts, whereas \citet{Kulkarni2018} only found mild evidence that the $\mathrm{RCF}(z)$ decreases with $z$. Unsurprisingly, we still find that low-$z$ and massive galaxies are the most likely to be catalogued. At higher redshifts ($z > 0.1$), only very massive galaxies ($M_{W1} \la -22$\,mag AB, comparable to that of the Milky Way) are catalogued. 

The decline in $\mathrm{RCF}(z)$ as a function of $z$ has important ramifications for the electromagnetic (EM) follow-up of gravitational wave events. The typical localization areas for 2 and 3 detector networks is several hundred to several thousand $\deg^2$ \citep{Kasliwal14}. One strategy to mitigate against these large areas that would be impossible to search even with modest field-of-view instruments, is to target known galaxies within the LIGO localization volume (e.g., \citealt{Gehrels16}). During O3, the horizon distance for binary neutron star (BNS) mergers is $\sim$200\,Mpc (see O3 alerts for S190425z, S190510g, S190901ap, S190910h; \citealp{S190425z,S190510g,S190901ap,S190910h}), roughly corresponding to $z \approx 0.05$. According to the BTS, for $z \le 0.05$ the RCF $\approx 63\%$, while integrating our best model inference from $z = 0$ to $0.05$ yields RCF $\approx 57\%$. Thus, targeted efforts to identify EM radiation from BNS mergers are likely to miss 1/3, or more, of all potential host galaxies for the EM transient. These numbers become significantly worse for events discovered at a distance $z > 0.05$. Any future efforts to quantify the rate of EM counterparts to gravitational wave events should account for the fraction of ``missing'' galaxies that we have identified with the BTS.

\subsection{Catalog Completeness as a Function of Galaxy Brightness}

\begin{figure}
    \centering
    \includegraphics[width=1.0\linewidth]{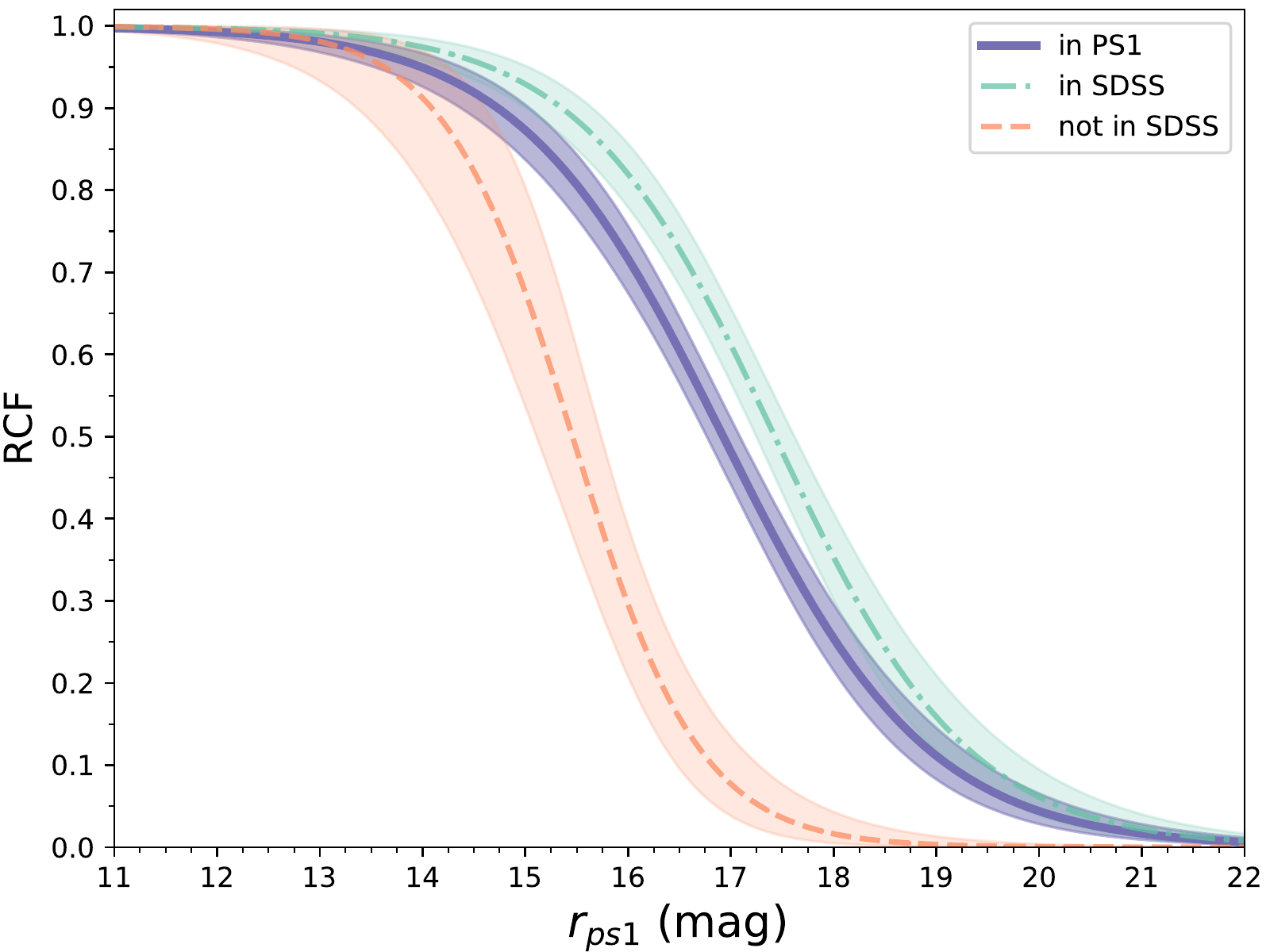}
    \caption{Redshift completeness factor as a function of \textit{apparent} host galaxy brightness, as traced by the  $r_\mathrm{PS1}$ Kron mag measurement. The solid purple curve shows the median value for all hosts in PS1 (95\% of all identified ZTF BTS hosts), while the green, dot-dashed curve shows hosts within the SDSS imaging footprint (72\% of all hosts), and the orange, dashed curve shows hosts outside SDSS (23\% of all hosts). The RCF is higher within the SDSS footprint, likely due to the many SDSS spectroscopic redshift surveys. For all galaxies within PS1, the RCF = 0.5 at $r_\mathrm{PS1} \approx 16.9$\,mag, while within the SDSS footprint, the RCF = 0.5 at $r_\mathrm{PS1} \approx 17.4$\,mag, and outside SDSS the RCF = 0.5 at $r_\mathrm{PS1} \approx 15.5$\,mag.}
    \label{fig:optical_RCF}
\end{figure}

To date SDSS has been the most prolific survey in terms of spectroscopically measuring galaxy redshifts, with $\sim$2.8\,million catalogued galaxies and counting \citep{Aguado19}. At this stage, significant improvements to the RCF will require tens of millions of new redshift measurements. Fortunately, within the next few years we will enter the era of supremely multiplexed spectrographs [e.g., The Dark Energy Spectroscopic Instrument (DESI),\footnote{\url{https://www.desi.lbl.gov/}} the Subaru Prime Focus Spectrograph (PFS),\footnote{\url{https://pfs.ipmu.jp/}} Euclid,\footnote{https://sci.esa.int/web/euclid/} the Wide Field Infrared Survey Telescope (\textit{WFIRST})\footnote{https://wfirst.gsfc.nasa.gov/}], which could dramatically increase the number of galaxies with known spectroscopic redshifts within the local universe. The best strategy to this end is to obtain spectra of bright galaxies with currently unknown redshifts, as is planned as part of the DESI Bright Galaxy Survey \citep{DESI-Collaboration16}. Using the same methodology described above, we can estimate the RCF as a function of galaxy brightness (as traced by the PS1 $r$-band), rather than $z$ or $M_{W1}$. The results from this exercise are summarized in Figure~\ref{fig:optical_RCF}.

From Figure~\ref{fig:optical_RCF} it is clear that the vast majority of extremely bright galaxies ($r_\mathrm{PS1} \la 14$\,mag) have catalogued redshifts. Figure~\ref{fig:optical_RCF} also shows the RCF for galaxies that are and are not within the SDSS imaging footprint,\footnote{This is determined by performing a cross match between BTS host positions and the SDSS imaging catalog. Any hosts with SDSS sources within 1\arcmin are assumed to be within the SDSS imaging footprint.} and a comparison of these two curves highlights the crucial role that SDSS played in terms of identifying galaxies in the local universe and spectroscopically measuring their redshifts. SDSS pushes the completeness of redshift catalogs $\sim$2\,mag fainter than what is observed outside the SDSS footprint. From our analysis, it is possible to estimate the completeness of existing redshift catalogs as a function of depth, and we find that catalogs are currently $\sim$90\%, 50\%, and 10\% complete to a depth of $r_\mathrm{PS1} \approx$14.7\,mag, 16.9\,mag, and 19.1\,mag respectively. Since the DESI bright galaxy survey is an order-of-magnitude increase over SDSS, Figure~\ref{fig:optical_RCF} makes it clear that the DESI bright galaxy survey will dramatically improve our knowledge of which galaxies reside in the local universe. 

\section{Summary and Conclusions}

We have presented early results from the ZTF Bright Transient Survey. The BTS utilizes publicly announced discoveries from the ZTF MSIP surveys to spectroscopically classify all extragalactic transients that peak brighter than 18.5\,mag. Simple filters are applied to the ZTF alert stream with the aim of minimizing false negatives, and we estimate that $\sim$95\% of all $m_\mathrm{peak} \le 18.5$\,mag SNe in the BTS are spectroscopically classified (during the period of the survey in 2018). This effort has resulted in the classification of \nBTS~SNe.

Spectroscopic observations are primarily conducted with SEDM, which is optimized to classify SN with $m < 19$\,mag. For sources that are inaccessible to SEDM, we utilize any available resource in order to obtain a classification. Final BTS classifications utilize spectral template matching, via \texttt{SNID}, and visual inspection of the spectra and (publically available) light curves. Of the \nBTS~classified BTS SNe, the majority are SNe Ia (\nIa), followed by CC SNe, both SNe II (\nII) and SNe Ib/c (\nIbc), with a relatively small number of SLSNe (\nslsne). The relative fraction of each of these types of SNe agrees with what has been found in previous studies \citep{Li11, Holoien19}. 

In this early release paper, we have focused on measuring the redshift completeness of local galaxy catalogs by using SNe Ia as a relatively unbiased tracer of galaxies in the local universe. By raw number, we find that less than half of the BTS SN host galaxies have known spectroscopic redshifts. In more detail, we find that the RCF falls steeply as a function of redshift, with only $\sim$half of all galaxies having known redshifts at $z \approx 0.05$, and $\la$20\% of galaxies having known redshifts at $z \approx 0.1$. The ``missing'' galaxies with unknown redshifts have important ramifications when searching for electromagnetic counterparts to multimessenger astronomical events, and suggest that the most complete method to find the EM counterparts for events at $d \ga 100$\,Mpc is to tile the entire error regions associated with gravitational wave (GW) or neutrino event alerts. However, even if this is done the same problem arises again when deciding which candidates among those found by the wide-field searches to follow up spectroscopically. Given the resources available, typically candidates with known distances, that also coincide with the distance constraints in a GW alert, are heavily prioritized for follow-up (e.g., \citealp{Andreoni2019,Coughlin2019}). Incorporation of redshifts from the Census of the Local Universe narrowband H$\alpha$ catalog \citep{Cook2019} may alleviate this problem to some extent in the near future. 

The combination of ZTF and SEDM has illustrated the power of focused efforts in the era of very large time-domain surveys. Within the next few years the Large Synoptic Survey Telescope (LSST; \citealt{Ivezic08}) will begin full survey operations. LSST will increase the volume of transient discoveries by an order of magnitude relative to on-going surveys in much the same way that PTF, PS1, and others built upon LOSS, and ATLAS and ZTF have built upon those surveys. We have already reached the era where true spectroscopic completeness is impossible for SN surveys, and LSST will greatly exacerbate this problem. Nevertheless, the use of an ultra-low-resolution instrument has allowed us to spectroscopically classify a nearly-complete subset of the discoveries made by ZTF (those with $m_\mathrm{peak} \le 18.5$\,mag). The simple focus of the BTS -- classify all the bright transients -- has resulted in the largest systematic classification of SNe to date. \nBTS~SNe are included in this early release paper, while the inclusion of 2019 results will eventually bring this number to $>$1800. In addition to measuring the RCF, as was done here, our growing sample can be used to: determine volumetric SN rates, measure the expansion of the Universe using low-$z$ SNe Ia, study the demographics of CC SNe, and measure the luminosity function of a wide range of transient phenomena. This, despite the fact that BTS only targets a tiny minority of all transients discovered by ZTF. The BTS demonstrates that a focused triage of an otherwise overwhelming discovery stream can lead to both impactful and novel results.
%\todo{I think this paragraph kinda sucks}

\acknowledgments
\small{
Based on observations obtained with the Samuel Oschin Telescope 48-inch and the 60-inch Telescope at the Palomar Observatory as part of the Zwicky Transient Facility project. ZTF is supported by the National Science Foundation under Grant No. AST-1440341 and a collaboration including Caltech, IPAC, the Weizmann Institute for Science, the Oskar Klein Center at Stockholm University, the University of Maryland, the University of Washington, Deutsches Elektronen- Synchrotron and Humboldt University, Los Alamos National Laboratories, the TANGO Consortium of Taiwan, the University of Wisconsin at Milwaukee, and Lawrence Berkeley National Laboratories. Operations are conducted by COO, IPAC, and UW. 

This work was supported by the GROWTH project funded by the National Science Foundation under PIRE Grant No 1545949. The ZTF forced-photometry service was funded under the Heising-Simons Foundation grant \#12540303 (PI: Graham). A.A.M. is funded by the Large Synoptic Survey Telescope Corporation, the Brinson Foundation, and the Moore Foundation in support of the LSSTC Data Science Fellowship Program, he also receives support as a CIERA Fellow by the CIERA Postdoctoral Fellowship Program (Center for Interdisciplinary Exploration and Research in Astrophysics, Northwestern University). M.R. has received funding from the European Research Council (ERC) under the European Union's Horizon 2020 research and innovation programme (grant agreement No 759194 - USNAC). The Oskar Klein Centre is funded by the Swedish Research Council.
A.Y.Q.H. is supported by a National Science Foundation Graduate Research Fellowship under Grant No.\,DGE‐1144469.

%Part of this research was carried out at the Jet Propulsion Laboratory, California Institute of Technology, under a contract with the National Aeronautics and Space Administration.

Partially based on observations made with the Nordic Optical Telescope, operated by the Nordic Optical Telescope Scientific Association at the Observatorio del Roque de los Muchachos, La Palma, Spain, of the Instituto de Astrofisica de Canarias. Some of the data presented here were obtained with ALFOSC, which is provided by the Instituto de Astrofisica de Andalucia (IAA) under a joint agreement with the University of Copenhagen and NOTSA.

Some of the data presented herein were obtained at the W. M. Keck Observatory, which is operated as a scientific partnership among the California Institute of Technology, the University of California, and NASA; the observatory was made possible by the generous financial support of the W. M. Keck Foundation. 

This paper is partly based on observations made with the Italian Telescopio Nazionale Galileo (TNG) operated on the island of La Palma by the Fundaci{\'o}n Galileo Galilei of the INAF (Istituto Nazionale di Astrofisica) at the Spanish Observatorio del Roque de los Muchachos of the Instituto de Astrofisica de Canarias. 
Partially based on observations obtained with the Apache Point Observatory 3.5-meter telescope, which is owned and operated by the Astrophysical Research Consortium. 
Partially based on observations from the LCOGT network.
Partially based on public observations collected at the WHT, operated on the island of La Palma by the Isaac Newton Group.

The Liverpool Telescope is operated on the island of La Palma by Liverpool John Moores University in the Spanish Observatorio del Roque de los Muchachos of the Instituto de Astrofisica de Canarias with financial support from the UK Science and Technology Facilities Council.

The SED Machine is based upon work supported by the National Science Foundation under Grant No. 1106171. 

We thank students Shaney Sze and Ho Ko for assisting with the manual candidate vetting during the summer of 2018.
}

\software{\texttt{SNID} \citep{Blondin2007},
          \texttt{astropy} \citep{Astropy-Collaboration13}, 
          \texttt{scipy} \citep{Jones01}, 
          \texttt{matplotlib} \citep{Hunter07},
          \texttt{pandas} \citep{McKinney10},
          \texttt{emcee} \citep{Foreman-Mackey13},
          \texttt{corner} \citep{Foreman-Mackey16},
          \texttt{MultinomCI} \citep{Signorell19},
          \texttt{PYSEDM} \citep{Rigault2019},
          \texttt{IRAF} \citep{Tody1986},
          \texttt{PyRAF} \citep{pyraf},
          \texttt{pyraf-dbsp} \citep{Bellm2016},
          \texttt{LPipe} \citep{Perley2019},
          \texttt{pyDIS} \citep{pydis}.
          }

%This work is part of the research programme VENI, with project number 016.192.277, which is (partly) financed by the Netherlands Organisation for Scientific Research (NWO).

\appendix

\section{Measuring the Conditional Probability of the RCF}\label{sec:rcf_calc}

To estimate the conditional probability that a galaxy has a catalogued redshift based on its distance and IR-luminosity, we model the data $X$ with the Bernoulli distribution
\begin{equation}
	X \sim \operatorname{Bern}(p),
\end{equation}
where $p$ is parameterized with a logistic function with dependence on both redshift $z$ and host galaxy luminosity:
\begin{equation}
p(z, M, \theta) = \frac{1}{1 + \exp(az + bM - c)},
\end{equation}
with host-galaxy absolute magnitude $M$, and $\theta$ representing the model parameters: $a$, $b$, and $c$, which need to be determined. The precise analytic dependence of $p$ on $z$ and $M$ may not be logistic, however, the purpose of this exercise is to provide a general sense for how the RCF relies on $z$ and $M$. Given that it smoothly transitions over an exponential length scale from 1 to 0, the logistic function works well for this general purpose.

From here it follows that the probability of a host galaxy having a previously cataloged redshift is:
\begin{equation}
Pr(q) = \left\{
\begin{array}{ll}
      p(z, M, \theta), & \mathrm{if} \, q=\mathrm{NED}_z \\
      1-p(z, M, \theta), & \mathrm{if} \, q=\mathrm{!NED}_z
\end{array}\right.
\label{eqn:bern}
\end{equation}
and the likelihood of the observations given the data and model parameters is:
\begin{equation}
Pr({q_k}\,|\,{z_k}, M_K, \theta) = \prod_{k=1}^K p(z_k, M_k, \theta)^{q_k} \, (1-p(z_k, M_k, \theta))^{1 - q_k},
\label{eqn:likelihood}
\end{equation}
where $k$ represents the individual observations and $q_k = 1$ for $\mathrm{NED}_z$ galaxies and $q_k = 0$ for $\mathrm{!NED}_z$ galaxies.

\begin{figure}
    \centering
    \includegraphics[width=0.45\linewidth]{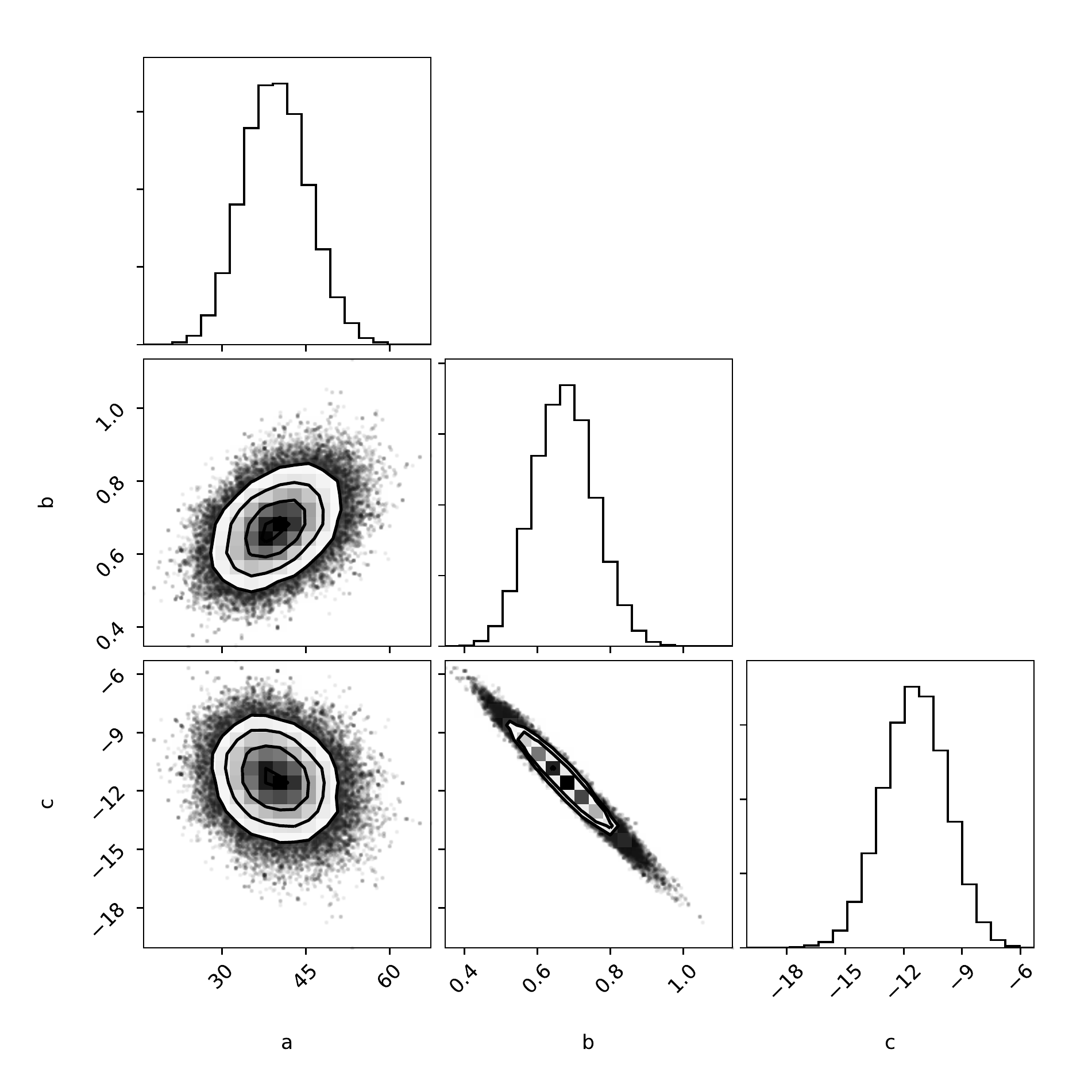}
    \caption{Corner plot showing the posterior distribution of $a$, $b$, and $c$, used to constrain the joint distribution of the RCF on $z$ and $M_{W1}$, $\mathrm{RCF}(z, M_{W_1})$.}
    \label{fig:corner}
\end{figure}

To estimate the model parameters we adopt wide, flat priors, and use the affine-invariant Markov Chain Monte Carlo (MCMC) ensemble sampling technique described by \citet{Goodman10}, as implemented in the \texttt{emcee} software package \citep{Foreman-Mackey13} . For $a$ and $b$ we adopt flat priors boundead between 0 and $10^{6}$. For $c$ we adopt a flat prior between $-100$ and $100$. We use 25 walkers within the sample, and run the ensemble until it has ``converged,'' which we define as $>$100 times longer than the average auto-correlation length of the individual chains from each walker. We find there is a strong covariance between $b$ and $c$, whereas $a$ is relatively independent, as shown in the corner plot in Figure~\ref{fig:corner}. We will happily make our posterior samples available to readers upon request.

%\todo{Keep this figure or no?}. CF: why not keep it

We also constrain the RCF as a function of the host redshift, $z$, or host galaxy luminosity, individually. We do this separately from the analysis above, while using the same MCMC procedure with $p$ in Equations~\ref{eqn:bern} and~\ref{eqn:likelihood} replaced by
\begin{equation}
p(z, \theta) = \frac{1}{1 + \exp(az - c)},
\label{eqn:p_z}
\end{equation}
for redshift, and 
\begin{equation}
p(M, \theta) = \frac{1}{1 + \exp(bM - c)},
\end{equation}
for host galaxy luminosity (where, again, we use absolute magnitude $M_{W1}$ instead of total luminosity). The results of this procedure are shown in the side panels of Figure~\ref{fig:RCF}. In these panels the solid lines show the median value of $p(z)$, $\mathrm{RCF}(z)$, and $p(M)$, $\mathrm{RCF}(M)$, while the shaded region shows the 90\% credible regions for $p(z)$ and $p(M)$.

% \subsection{Tables}

% Our current BTS sample is presented in Table~\ref{tab:sample}. The NED column indicates whether the redshift for the host was known prior to the discovery of the transient. The Host column indicates the associated host galaxy. $M_{K_s}$ and $M_{UV}$ indicate host magnitudes.

% \begin{table}
% \footnotesize
% \centering
% \resizebox{\textwidth}{!}{\begin{tabular}{l|r|r|r|r|r|r|r|r|r|r|r|r|r}
% Survey name & TNS Name & $M_{r,peak}$ & $M_{g,peak}$ & Discovery date & Classification date & Type & RA & DEC & Redshift & NED & Host & $M_{K_s}$ & $M_{UV}$  \\\hline
% ZTF18aahfzea & AT2018xxx & 16.30 & ... & 2018-04-06 & 2018-04-21 & SN Ia & 12:05:26.67 & 
% +20:17:56.6 & 0.02436 & t & NGC 4090 & 10.60 & 17.96 \\
% ZTF18yyyyyyy & AT2018xxx & 0.0 & ... & 2018-xx-xx & 2018-xx-xx & SN Y & 00:00:00.00 & +00:00:00.00 & 0.000 & f & 2MASS-XXX & 0.0 & 0.0 \\
% \end{tabular}}
% \caption{\label{tab:sample}The ZTF BTS Sample.}
% \end{table}

%\bibliographystyle{alpha}
\bibliography{references}

\begin{thebibliography}{}
\expandafter\ifx\csname natexlab\endcsname\relax\def\natexlab#1{#1}\fi
\providecommand{\url}[1]{\href{#1}{#1}}

\bibitem[{{Abolfathi} {et~al.}(2018){Abolfathi}, {Aguado}, {Aguilar}, {Allende
  Prieto}, {Almeida}, {Ananna}, {Anders}, {Anderson}, {Andrews}, {Anguiano},
  {Arag{\'o}n-Salamanca}, {Argudo-Fern{\'a}ndez}, {Armengaud}, {Ata},
  {Aubourg}, {Avila-Reese}, {Badenes}, {Bailey}, {Balland}, {Barger},
  {Barrera-Ballesteros}, {Bartosz}, {Bastien}, {Bates}, {Baumgarten},
  {Bautista}, {Beaton}, {Beers}, {Belfiore}, {Bender}, {Bernardi}, {Bershady},
  {Beutler}, {Bird}, {Bizyaev}, {Blanc}, {Blanton}, {Blomqvist}, {Bolton},
  {Boquien}, {Borissova}, {Bovy}, {Bradna Diaz}, {Brandt}, {Brinkmann},
  {Brownstein}, {Bundy}, {Burgasser}, {Burtin}, {Busca}, {Ca{\~n}as},
  {Cano-D{\'\i}az}, {Cappellari}, {Carrera}, {Casey}, {Cervantes Sodi}, {Chen},
  {Cherinka}, {Chiappini}, {Choi}, {Chojnowski}, {Chuang}, {Chung}, {Clerc},
  {Cohen}, {Comerford}, {Comparat}, {Correa do Nascimento}, {da Costa},
  {Cousinou}, {Covey}, {Crane}, {Cruz-Gonzalez}, {Cunha}, {da Silva Ilha},
  {Damke}, {Darling}, {Davidson}, {Dawson}, {de Icaza Lizaola}, {de la
  Macorra}, {de la Torre}, {De Lee}, {de Sainte Agathe}, {Deconto Machado},
  {Dell'Agli}, {Delubac}, {Diamond-Stanic}, {Donor}, {Downes}, {Drory}, {du Mas
  des Bourboux}, {Duckworth}, {Dwelly}, {Dyer}, {Ebelke}, {Davis Eigenbrot},
  {Eisenstein}, {Elsworth}, {Emsellem}, {Eracleous}, {Erfanianfar},
  {Escoffier}, {Fan}, {Fern{\'a}ndez Alvar}, {Fernandez-Trincado}, {Fernand o
  Cirolini}, {Feuillet}, {Finoguenov}, {Fleming}, {Font-Ribera}, {Freischlad},
  {Frinchaboy}, {Fu}, {G{\'o}mez Maqueo Chew}, {Galbany}, {Garc{\'\i}a
  P{\'e}rez}, {Garcia-Dias}, {Garc{\'\i}a-Hern{\'a}ndez}, {Garma Oehmichen},
  {Gaulme}, {Gelfand }, {Gil-Mar{\'\i}n}, {Gillespie}, {Goddard}, {Gonz{\'a}lez
  Hern{\'a}ndez}, {Gonzalez-Perez}, {Grabowski}, {Green}, {Grier}, {Gueguen},
  {Guo}, {Guy}, {Hagen}, {Hall}, {Harding}, {Hasselquist}, {Hawley}, {Hayes},
  {Hearty}, {Hekker}, {Hernand ez}, {Hernandez Toledo}, {Hogg},
  {Holley-Bockelmann}, {Holtzman}, {Hou}, {Hsieh}, {Hunt}, {Hutchinson},
  {Hwang}, {Jimenez Angel}, {Johnson}, {Jones}, {J{\"o}nsson}, {Jullo}, {Khan},
  {Kinemuchi}, {Kirkby}, {Kirkpatrick}, {Kitaura}, {Knapp}, {Kneib},
  {Kollmeier}, {Lacerna}, {Lane}, {Lang}, {Law}, {Le Goff}, {Lee}, {Li}, {Li},
  {Lian}, {Liang}, {Lima}, {Lin}, {Long}, {Lucatello}, {Lundgren}, {Mackereth},
  {MacLeod}, {Mahadevan}, {Maia}, {Majewski}, {Manchado}, {Maraston},
  {Mariappan}, {Marques-Chaves}, {Masseron}, {Masters}, {McDermid}, {McGreer},
  {Melendez}, {Meneses-Goytia}, {Merloni}, {Merrifield}, {Meszaros}, {Meza},
  {Minchev}, {Minniti}, {Mueller}, {Muller-Sanchez}, {Muna}, {Mu{\~n}oz},
  {Myers}, {Nair}, {Nand ra}, {Ness}, {Newman}, {Nichol}, {Nidever},
  {Nitschelm}, {Noterdaeme}, {O'Connell}, {Oelkers}, {Oravetz}, {Oravetz},
  {Ort{\'\i}z}, {Osorio}, {Pace}, {Padilla}, {Palanque-Delabrouille},
  {Palicio}, {Pan}, {Pan}, {Parikh}, {P{\^a}ris}, {Park}, {Peirani},
  {Pellejero-Ibanez}, {Penny}, {Percival}, {Perez-Fournon}, {Petitjean},
  {Pieri}, {Pinsonneault}, {Pisani}, {Prada}, {Prakash}, {Queiroz}, {Raddick},
  {Raichoor}, {Barboza Rembold}, {Richstein}, {Riffel}, {Riffel}, {Rix},
  {Robin}, {Rodr{\'\i}guez Torres}, {Rom{\'a}n-Z{\'u}{\~n}iga}, {Ross},
  {Rossi}, {Ruan}, {Ruggeri}, {Ruiz}, {Salvato}, {S{\'a}nchez}, {S{\'a}nchez},
  {Sanchez Almeida}, {S{\'a}nchez-Gallego}, {Santana Rojas}, {Santiago},
  {Schiavon}, {Schimoia}, {Schlafly}, {Schlegel}, {Schneider}, {Schuster},
  {Schwope}, {Seo}, {Serenelli}, {Shen}, {Shen}, {Shetrone}, {Shull}, {Silva
  Aguirre}, {Simon}, {Skrutskie}, {Slosar}, {Smethurst}, {Smith}, {Sobeck},
  {Somers}, {Souter}, {Souto}, {Spindler}, {Stark}, {Stassun}, {Steinmetz},
  {Stello}, {Storchi-Bergmann}, {Streblyanska}, {Stringfellow}, {Su{\'a}rez},
  {Sun}, {Szigeti}, {Taghizadeh-Popp}, {Talbot}, {Tang}, {Tao}, {Tayar},
  {Tembe}, {Teske}, {Thakar}, {Thomas}, {Tissera}, {Tojeiro}, {Tremonti},
  {Troup}, {Urry}, {Valenzuela}, {van den Bosch}, {Vargas-Gonz{\'a}lez},
  {Vargas-Maga{\~n}a}, {Vazquez}, {Villanova}, {Vogt}, {Wake}, {Wang},
  {Weaver}, {Weijmans}, {Weinberg}, {Westfall}, {Whelan}, {Wilcots}, {Wild},
  {Williams}, {Wilson}, {Wood-Vasey}, {Wylezalek}, {Xiao}, {Yan}, {Yang},
  {Ybarra}, {Y{\`e}che}, {Zakamska}, {Zamora}, {Zarrouk}, {Zasowski}, {Zhang},
  {Zhao}, {Zhao}, {Zheng}, {Zheng}, {Zhou}, {Zhu}, {Zinn}, \&
  {Zou}}]{Abolfathi18}
{Abolfathi}, B., {Aguado}, D.~S., {Aguilar}, G., {et~al.} 2018, \apjs, 235, 42

\bibitem[{{Aguado} {et~al.}(2019){Aguado}, {Ahumada}, {Almeida}, {Anderson},
  {Andrews}, {Anguiano}, {Aquino Ort{\'\i}z}, {Arag{\'o}n-Salamanca},
  {Argudo-Fern{\'a}ndez}, {Aubert}, {Avila-Reese}, {Badenes}, {Barboza
  Rembold}, {Barger}, {Barrera-Ballesteros}, {Bates}, {Bautista}, {Beaton},
  {Beers}, {Belfiore}, {Bernardi}, {Bershady}, {Beutler}, {Bird}, {Bizyaev},
  {Blanc}, {Blanton}, {Blomqvist}, {Bolton}, {Boquien}, {Borissova}, {Bovy},
  {Brand t}, {Brinkmann}, {Brownstein}, {Bundy}, {Burgasser}, {Byler}, {Cano
  Diaz}, {Cappellari}, {Carrera}, {Cervantes Sodi}, {Chen}, {Cherinka}, {Choi},
  {Chung}, {Coffey}, {Comerford}, {Comparat}, {Covey}, {da Silva Ilha}, {da
  Costa}, {Dai}, {Damke}, {Darling}, {Davies}, {Dawson}, {de Sainte Agathe},
  {Deconto Machado}, {Del Moro}, {De Lee}, {Diamond-Stanic}, {Dom{\'\i}nguez
  S{\'a}nchez}, {Donor}, {Drory}, {du Mas des Bourboux}, {Duckworth}, {Dwelly},
  {Ebelke}, {Emsellem}, {Escoffier}, {Fern{\'a}ndez-Trincado}, {Feuillet},
  {Fischer}, {Fleming}, {Fraser-McKelvie}, {Freischlad}, {Frinchaboy}, {Fu},
  {Galbany}, {Garcia-Dias}, {Garc{\'\i}a-Hern{\'a}ndez}, {Garma Oehmichen},
  {Geimba Maia}, {Gil-Mar{\'\i}n}, {Grabowski}, {Gu}, {Guo}, {Ha},
  {Harrington}, {Hasselquist}, {Hayes}, {Hearty}, {Hernandez Toledo}, {Hicks},
  {Hogg}, {Holley-Bockelmann}, {Holtzman}, {Hsieh}, {Hunt}, {Hwang},
  {Ibarra-Medel}, {Jimenez Angel}, {Johnson}, {Jones}, {J{\"o}nsson},
  {Kinemuchi}, {Kollmeier}, {Krawczyk}, {Kreckel}, {Kruk}, {Lacerna}, {Lan},
  {Lane}, {Law}, {Lee}, {Li}, {Lian}, {Lin}, {Lin}, {Lintott}, {Long},
  {Longa-Pe{\~n}a}, {Mackereth}, {de la Macorra}, {Majewski}, {Malanushenko},
  {Manchado}, {Maraston}, {Mariappan}, {Marinelli}, {Marques-Chaves},
  {Masseron}, {Masters}, {McDermid}, {Medina Pe{\~n}a}, {Meneses-Goytia},
  {Merloni}, {Merrifield}, {Meszaros}, {Minniti}, {Minsley}, {Muna}, {Myers},
  {Nair}, {Correa do Nascimento}, {Newman}, {Nitschelm}, {Olmstead}, {Oravetz},
  {Oravetz}, {Ortega Minakata}, {Pace}, {Padilla}, {Palicio}, {Pan}, {Pan},
  {Parikh}, {Parker}, {Peirani}, {Penny}, {Percival}, {Perez-Fournon},
  {Peterken}, {Pinsonneault}, {Prakash}, {Raddick}, {Raichoor}, {Riffel},
  {Riffel}, {Rix}, {Robin}, {Roman-Lopes}, {Rose}, {Ross}, {Rossi}, {Rowlands},
  {Rubin}, {S{\'a}nchez}, {S{\'a}nchez-Gallego}, {Sayres}, {Schaefer},
  {Schiavon}, {Schimoia}, {Schlafly}, {Schlegel}, {Schneider}, {Schultheis},
  {Seo}, {Shamsi}, {Shao}, {Shen}, {Shetty}, {Simonian}, {Smethurst}, {Sobeck},
  {Souter}, {Spindler}, {Stark}, {Stassun}, {Steinmetz}, {Storchi-Bergmann},
  {Stringfellow}, {Su{\'a}rez}, {Sun}, {Taghizadeh-Popp}, {Talbot}, {Tayar},
  {Thakar}, {Thomas}, {Tissera}, {Tojeiro}, {Troup}, {Unda-Sanzana},
  {Valenzuela}, {Vargas-Maga{\~n}a}, {V{\'a}zquez-Mata}, {Wake}, {Weaver},
  {Weijmans}, {Westfall}, {Wild}, {Wilson}, {Woods}, {Yan}, {Yang}, {Zamora},
  {Zasowski}, {Zhang}, {Zheng}, {Zheng}, {Zhu}, {Zinn}, \& {Zou}}]{Aguado19}
{Aguado}, D.~S., {Ahumada}, R., {Almeida}, A., {et~al.} 2019, \apjs, 240, 23

\bibitem[{{Aldering} {et~al.}(2002){Aldering}, {Adam}, {Antilogus}, {Astier},
  {Bacon}, {Bongard}, {Bonnaud}, {Copin}, {Hardin}, {Henault}, {Howell},
  {Lemonnier}, {Levy}, {Loken}, {Nugent}, {Pain}, {Pecontal}, {Pecontal},
  {Perlmutter}, {Quimby}, {Schahmaneche}, {Smadja}, \&
  {Wood-Vasey}}]{Aldering2002}
{Aldering}, G., {Adam}, G., {Antilogus}, P., {et~al.} 2002, in \procspie, Vol.
  4836, Survey and Other Telescope Technologies and Discoveries, ed. J.~A.
  {Tyson} \& S.~{Wolff}, 61--72

\bibitem[{{Andreoni} {et~al.}(2019){Andreoni}, {Goldstein}, {Anand},
  {Coughlin}, {Singer}, {Ahumada}, {Medford}, {Kool}, {Webb}, {Bulla}, {Bloom},
  {Kasliwal}, {Nugent}, {Bagdasaryan}, {Barnes}, {Cook}, {Cooke}, {Duev},
  {Fremling}, {Gatkine}, {Golkhou}, {Kong}, {Mahabal},
  {Mart{\'{\i}}nez-Palomera}, {Tao}, \& {Zhang}}]{Andreoni2019}
{Andreoni}, I., {Goldstein}, D.~A., {Anand}, S., {et~al.} 2019, \apjl, 881, L16

\bibitem[{{Angus} {et~al.}(2019){Angus}, {Smith}, {Sullivan}, {Inserra},
  {Wiseman}, {D'Andrea}, {Thomas}, {Nichol}, {Galbany}, {Childress}, {Asorey},
  {Brown}, {Casas}, {Castander}, {Curtin}, {Frohmaier}, {Glazebrook}, {Gruen},
  {Gutierrez}, {Kessler}, {Kim}, {Lidman}, {Macaulay}, {Nugent}, {Pursiainen},
  {Sako}, {Soares-Santos}, {Thomas}, {Abbott}, {Avila}, {Bertin}, {Brooks},
  {Buckley-Geer}, {Burke}, {Carnero Rosell}, {Carretero}, {da Costa}, {De
  Vicente}, {Desai}, {Diehl}, {Doel}, {Eifler}, {Flaugher}, {Fosalba},
  {Frieman}, {Garc{\'\i}a-Bellido}, {Gruendl}, {Gschwend}, {Hartley},
  {Hollowood}, {Honscheid}, {Hoyle}, {James}, {Kuehn}, {Kuropatkin}, {Lahav},
  {Lima}, {Maia}, {March}, {Marshall}, {Menanteau}, {Miller}, {Miquel}, {Ogand
  o}, {Plazas}, {Romer}, {Sanchez}, {Schindler}, {Schubnell}, {Sobreira},
  {Suchyta}, {Swanson}, {Tarle}, {Thomas}, {Tucker}, \& {DES
  Collaboration}}]{Angus+2019}
{Angus}, C.~R., {Smith}, M., {Sullivan}, M., {et~al.} 2019, \mnras, 487, 2215

\bibitem[{{Arnett} {et~al.}(1989){Arnett}, {Bahcall}, {Kirshner}, \&
  {Woosley}}]{Arnett89}
{Arnett}, W.~D., {Bahcall}, J.~N., {Kirshner}, R.~P., \& {Woosley}, S.~E. 1989,
  \araa, 27, 629

\bibitem[{{Astier} {et~al.}(2006){Astier}, {Guy}, {Regnault}, {Pain},
  {Aubourg}, {Balam}, {Basa}, {Carlberg}, {Fabbro}, {Fouchez}, {Hook},
  {Howell}, {Lafoux}, {Neill}, {Palanque-Delabrouille}, {Perrett}, {Pritchet},
  {Rich}, {Sullivan}, {Taillet}, {Aldering}, {Antilogus}, {Arsenijevic},
  {Balland}, {Baumont}, {Bronder}, {Courtois}, {Ellis}, {Filiol}, {Gon{\c
  c}alves}, {Goobar}, {Guide}, {Hardin}, {Lusset}, {Lidman}, {McMahon},
  {Mouchet}, {Mourao}, {Perlmutter}, {Ripoche}, {Tao}, \&
  {Walton}}]{Astier2006}
{Astier}, P., {Guy}, J., {Regnault}, N., {et~al.} 2006, \aap, 447, 31

\bibitem[{{Astropy Collaboration} {et~al.}(2013){Astropy Collaboration},
  {Robitaille}, {Tollerud}, {Greenfield}, {Droettboom}, {Bray}, {Aldcroft},
  {Davis}, {Ginsburg}, {Price-Whelan}, {Kerzendorf}, {Conley}, {Crighton},
  {Barbary}, {Muna}, {Ferguson}, {Grollier}, {Parikh}, {Nair}, {Unther},
  {Deil}, {Woillez}, {Conseil}, {Kramer}, {Turner}, {Singer}, {Fox}, {Weaver},
  {Zabalza}, {Edwards}, {Azalee Bostroem}, {Burke}, {Casey}, {Crawford},
  {Dencheva}, {Ely}, {Jenness}, {Labrie}, {Lim}, {Pierfederici}, {Pontzen},
  {Ptak}, {Refsdal}, {Servillat}, \& {Streicher}}]{Astropy-Collaboration13}
{Astropy Collaboration}, {Robitaille}, T.~P., {Tollerud}, E.~J., {et~al.} 2013,
  \aap, 558, A33

\bibitem[{{Baade} \& {Zwicky}(1934)}]{Baade1934}
{Baade}, W., \& {Zwicky}, F. 1934, Physical Review, 46, 76

\bibitem[{{Bellm}(2016)}]{Bellm16a}
{Bellm}, E.~C. 2016, \pasp, 128, 084501

\bibitem[{{Bellm} \& {Sesar}(2016)}]{Bellm2016}
{Bellm}, E.~C., \& {Sesar}, B. 2016, {pyraf-dbsp: Reduction pipeline for the
  Palomar Double Beam Spectrograph}, Astrophysics Source Code Library, , ,
  ascl:1602.002

\bibitem[{{Bellm} {et~al.}(2019{\natexlab{a}}){Bellm}, {Kulkarni}, {Graham},
  {Dekany}, {Smith}, {Riddle}, {Masci}, {Helou}, {Prince}, {Adams},
  {Barbarino}, {Barlow}, {Bauer}, {Beck}, {Belicki}, {Biswas}, {Blagorodnova},
  {Bodewits}, {Bolin}, {Brinnel}, {Brooke}, {Bue}, {Bulla}, {Burruss}, {Cenko},
  {Chang}, {Connolly}, {Coughlin}, {Cromer}, {Cunningham}, {De}, {Delacroix},
  {Desai}, {Duev}, {Eadie}, {Farnham}, {Feeney}, {Feindt}, {Flynn},
  {Franckowiak}, {Frederick}, {Fremling}, {Gal-Yam}, {Gezari}, {Giomi},
  {Goldstein}, {Golkhou}, {Goobar}, {Groom}, {Hacopians}, {Hale}, {Henning},
  {Ho}, {Hover}, {Howell}, {Hung}, {Huppenkothen}, {Imel}, {Ip}, {Ivezi{\'c}},
  {Jackson}, {Jones}, {Juric}, {Kasliwal}, {Kaspi}, {Kaye}, {Kelley},
  {Kowalski}, {Kramer}, {Kupfer}, {Landry}, {Laher}, {Lee}, {Lin}, {Lin},
  {Lunnan}, {Giomi}, {Mahabal}, {Mao}, {Miller}, {Monkewitz}, {Murphy},
  {Ngeow}, {Nordin}, {Nugent}, {Ofek}, {Patterson}, {Penprase}, {Porter},
  {Rauch}, {Rebbapragada}, {Reiley}, {Rigault}, {Rodriguez}, {van Roestel},
  {Rusholme}, {van Santen}, {Schulze}, {Shupe}, {Singer}, {Soumagnac}, {Stein},
  {Surace}, {Sollerman}, {Szkody}, {Taddia}, {Terek}, {Van Sistine}, {van
  Velzen}, {Vestrand}, {Walters}, {Ward}, {Ye}, {Yu}, {Yan}, \&
  {Zolkower}}]{Bellm2019}
{Bellm}, E.~C., {Kulkarni}, S.~R., {Graham}, M.~J., {et~al.}
  2019{\natexlab{a}}, \pasp, 131, 018002

\bibitem[{{Bellm} {et~al.}(2019{\natexlab{b}}){Bellm}, {Kulkarni}, {Barlow},
  {Feindt}, {Graham}, {Goobar}, {Kupfer}, {Ngeow}, {Nugent}, {Ofek}, {Prince},
  {Riddle}, {Walters}, \& {Ye}}]{Bellm2019b}
{Bellm}, E.~C., {Kulkarni}, S.~R., {Barlow}, T., {et~al.} 2019{\natexlab{b}},
  \pasp, 131, 068003

\bibitem[{{Ben-Ami} {et~al.}(2012){Ben-Ami}, {Konidaris}, {Quimby}, {Davis},
  {Ngeow}, {Ritter}, \& {Rudy}}]{Ben-Ami12}
{Ben-Ami}, S., {Konidaris}, N., {Quimby}, R., {et~al.} 2012, in \procspie, Vol.
  8446, Ground-based and Airborne Instrumentation for Astronomy IV, 844686

\bibitem[{{Blagorodnova} {et~al.}(2018){Blagorodnova}, {Neill}, {Walters},
  {Kulkarni}, {Fremling}, {Ben-Ami}, {Dekany}, {Fucik}, {Konidaris}, {Nash},
  {Ngeow}, {Ofek}, {O' Sullivan}, {Quimby}, {Ritter}, \&
  {Vyhmeister}}]{Blagorodnova2018}
{Blagorodnova}, N., {Neill}, J.~D., {Walters}, R., {et~al.} 2018, \pasp, 130,
  035003

\bibitem[{{Blondin} {et~al.}(2015){Blondin}, {Dessart}, \&
  {Hillier}}]{2015MNRAS.448.2766B}
{Blondin}, S., {Dessart}, L., \& {Hillier}, D.~J. 2015, \mnras, 448, 2766

\bibitem[{{Blondin} \& {Tonry}(2007)}]{Blondin2007}
{Blondin}, S., \& {Tonry}, J.~L. 2007, in American Institute of Physics
  Conference Series, Vol. 924, The Multicolored Landscape of Compact Objects
  and Their Explosive Origins, ed. T.~{di Salvo}, G.~L. {Israel},
  L.~{Piersant}, L.~{Burderi}, G.~{Matt}, A.~{Tornambe}, \& M.~T. {Menna},
  312--321

\bibitem[{{Branch} {et~al.}(2006){Branch}, {Dang}, {Hall}, {Ketchum},
  {Melakayil}, {Parrent}, {Troxel}, {Casebeer}, {Jeffery}, \&
  {Baron}}]{Branch06}
{Branch}, D., {Dang}, L.~C., {Hall}, N., {et~al.} 2006, \pasp, 118, 560

\bibitem[{{Cenko} {et~al.}(2006){Cenko}, {Fox}, {Moon}, {Harrison}, {Kulkarni},
  {Henning}, {Guzman}, {Bonati}, {Smith}, {Thicksten}, {Doyle}, {Petrie},
  {Gal-Yam}, {Soderberg}, {Anagnostou}, \& {Laity}}]{Cenko2006}
{Cenko}, S.~B., {Fox}, D.~B., {Moon}, D.-S., {et~al.} 2006, \pasp, 118, 1396

\bibitem[{{Chambers} {et~al.}(2016){Chambers}, {Magnier}, {Metcalfe},
  {Flewelling}, {Huber}, {Waters}, {Denneau}, {Draper}, {Farrow}, {Finkbeiner},
  {Holmberg}, {Koppenhoefer}, {Price}, {Rest}, {Saglia}, {Schlafly}, {Smartt},
  {Sweeney}, {Wainscoat}, {Burgett}, {Chastel}, {Grav}, {Heasley}, {Hodapp},
  {Jedicke}, {Kaiser}, {Kudritzki}, {Luppino}, {Lupton}, {Monet}, {Morgan},
  {Onaka}, {Shiao}, {Stubbs}, {Tonry}, {White}, {Ba{\~n}ados}, {Bell},
  {Bender}, {Bernard}, {Boegner}, {Boffi}, {Botticella}, {Calamida},
  {Casertano}, {Chen}, {Chen}, {Cole}, {Deacon}, {Frenk}, {Fitzsimmons},
  {Gezari}, {Gibbs}, {Goessl}, {Goggia}, {Gourgue}, {Goldman}, {Grant},
  {Grebel}, {Hambly}, {Hasinger}, {Heavens}, {Heckman}, {Henderson}, {Henning},
  {Holman}, {Hopp}, {Ip}, {Isani}, {Jackson}, {Keyes}, {Koekemoer}, {Kotak},
  {Le}, {Liska}, {Long}, {Lucey}, {Liu}, {Martin}, {Masci}, {McLean}, {Mindel},
  {Misra}, {Morganson}, {Murphy}, {Obaika}, {Narayan}, {Nieto-Santisteban},
  {Norberg}, {Peacock}, {Pier}, {Postman}, {Primak}, {Rae}, {Rai}, {Riess},
  {Riffeser}, {Rix}, {R{\"o}ser}, {Russel}, {Rutz}, {Schilbach}, {Schultz},
  {Scolnic}, {Strolger}, {Szalay}, {Seitz}, {Small}, {Smith}, {Soderblom},
  {Taylor}, {Thomson}, {Taylor}, {Thakar}, {Thiel}, {Thilker}, {Unger},
  {Urata}, {Valenti}, {Wagner}, {Walder}, {Walter}, {Watters}, {Werner},
  {Wood-Vasey}, \& {Wyse}}]{Chambers2016}
{Chambers}, K.~C., {Magnier}, E.~A., {Metcalfe}, N., {et~al.} 2016, arXiv
  e-prints, arXiv:1612.05560

\bibitem[{{Cook} {et~al.}(2019){Cook}, {Kasliwal}, {Van Sistine}, {Kaplan},
  {Sutter}, {Kupfer}, {Shupe}, {Laher}, {Masci}, {Dale}, {Sesar}, {Brady},
  {Yan}, {Ofek}, {Reitze}, \& {Kulkarni}}]{Cook2019}
{Cook}, D.~O., {Kasliwal}, M.~M., {Van Sistine}, A., {et~al.} 2019, \apj, 880,
  7

\bibitem[{{Cortini}(2018)}]{Cortini18}
{Cortini}, G. 2018, Transient Name Server Discovery Report, 2018-393, 1

\bibitem[{{Coughlin} {et~al.}(2019){Coughlin}, {Ahumada}, {Anand}, {De},
  {Hankins}, {Kasliwal}, {Singer}, {Bellm}, {Andreoni}, {Cenko}, {Cooke},
  {Copperwheat}, {Dugas}, {Jencson}, {Perley}, {Yu}, {Bhalerao}, {Kumar},
  {Bloom}, {Anupama}, {Ashley}, {Bagdasaryan}, {Biswas}, {Buckley}, {Burdge},
  {Cook}, {Cromer}, {Cunningham}, {D'A{\`i}}, {Dekany}, {Delacroix},
  {Dichiara}, {Duev}, {Dutta}, {Feeney}, {Frederick}, {Gatkine}, {Ghosh},
  {Goldstein}, {Golkhou}, {Goobar}, {Graham}, {Hanayama}, {Horiuchi}, {Hung},
  {Jha}, {Kong}, {Giomi}, {Kaplan}, {Karambelkar}, {Kowalski}, {Kulkarni},
  {Kupfer}, {La Parola}, {Masci}, {Mazzali}, {Moore}, {Mogotsi}, {Neill},
  {Ngeow}, {Mart{\'{\i}}nez-Palomera}, {Pavana}, {Ofek}, {Patil}, {Riddle},
  {Rigault}, {Rusholme}, {Serabyn}, {Shupe}, {Sharma}, {Sollerman}, {Soon},
  {Staats}, {Taggart}, {Tan}, {Travouillon}, {Troja}, {Waratkar}, \&
  {Yatsu}}]{Coughlin2019}
{Coughlin}, M.~W., {Ahumada}, T., {Anand}, S., {et~al.} 2019, arXiv e-prints,
  arXiv:1907.12645

\bibitem[{Davenport {et~al.}(2016)Davenport, de~Val-Borro, \&
  Wilkinson}]{pydis}
Davenport, J., de~Val-Borro, M., \& Wilkinson, T.~D. 2016,
  doi:10.5281/zenodo.58753

\bibitem[{{De Cia} {et~al.}(2018){De Cia}, {Gal-Yam}, {Rubin}, {Leloudas},
  {Vreeswijk}, {Perley}, {Quimby}, {Yan}, {Sullivan}, {Fl{\"o}rs}, {Sollerman},
  {Bersier}, {Cenko}, {Gal-Yam}, {Maguire}, {Ofek}, {Prentice}, {Schulze},
  {Spyromilio}, {Valenti}, {Arcavi}, {Corsi}, {Howell}, {Mazzali}, {Kasliwal},
  {Taddia}, \& {Yaron}}]{DeCia+2018}
{De Cia}, A., {Gal-Yam}, A., {Rubin}, A., {et~al.} 2018, \apj, 860, 100

\bibitem[{{DESI Collaboration} {et~al.}(2016){DESI Collaboration}, {Aghamousa},
  {Aguilar}, {Ahlen}, {Alam}, {Allen}, {Allende Prieto}, {Annis}, {Bailey},
  {Balland}, {Ballester}, {Baltay}, {Beaufore}, {Bebek}, {Beers}, {Bell},
  {Bernal}, {Besuner}, {Beutler}, {Blake}, {Bleuler}, {Blomqvist}, {Blum},
  {Bolton}, {Briceno}, {Brooks}, {Brownstein}, {Buckley-Geer}, {Burden},
  {Burtin}, {Busca}, {Cahn}, {Cai}, {Cardiel-Sas}, {Carlberg}, {Carton},
  {Casas}, {Castand er}, {Cervantes-Cota}, {Claybaugh}, {Close}, {Coker},
  {Cole}, {Comparat}, {Cooper}, {Cousinou}, {Crocce}, {Cuby}, {Cunningham},
  {Davis}, {Dawson}, {de la Macorra}, {De Vicente}, {Delubac}, {Derwent},
  {Dey}, {Dhungana}, {Ding}, {Doel}, {Duan}, {Ealet}, {Edelstein},
  {Eftekharzadeh}, {Eisenstein}, {Elliott}, {Escoffier}, {Evatt}, {Fagrelius},
  {Fan}, {Fanning}, {Farahi}, {Farihi}, {Favole}, {Feng}, {Fernandez},
  {Findlay}, {Finkbeiner}, {Fitzpatrick}, {Flaugher}, {Flender}, {Font-Ribera},
  {Forero-Romero}, {Fosalba}, {Frenk}, {Fumagalli}, {Gaensicke}, {Gallo},
  {Garcia-Bellido}, {Gaztanaga}, {Pietro Gentile Fusillo}, {Gerard},
  {Gershkovich}, {Giannantonio}, {Gillet}, {Gonzalez-de-Rivera},
  {Gonzalez-Perez}, {Gott}, {Graur}, {Gutierrez}, {Guy}, {Habib}, {Heetderks},
  {Heetderks}, {Heitmann}, {Hellwing}, {Herrera}, {Ho}, {Holland}, {Honscheid},
  {Huff}, {Hutchinson}, {Huterer}, {Hwang}, {Illa Laguna}, {Ishikawa},
  {Jacobs}, {Jeffrey}, {Jelinsky}, {Jennings}, {Jiang}, {Jimenez}, {Johnson},
  {Joyce}, {Jullo}, {Juneau}, {Kama}, {Karcher}, {Karkar}, {Kehoe}, {Kennamer},
  {Kent}, {Kilbinger}, {Kim}, {Kirkby}, {Kisner}, {Kitanidis}, {Kneib},
  {Koposov}, {Kovacs}, {Koyama}, {Kremin}, {Kron}, {Kronig}, {Kueter-Young},
  {Lacey}, {Lafever}, {Lahav}, {Lambert}, {Lampton}, {Land riau}, {Lang},
  {Lauer}, {Le Goff}, {Le Guillou}, {Le Van Suu}, {Lee}, {Lee}, {Leitner},
  {Lesser}, {Levi}, {L'Huillier}, {Li}, {Liang}, {Lin}, {Linder}, {Loebman},
  {Luki{\'c}}, {Ma}, {MacCrann}, {Magneville}, {Makarem}, {Manera}, {Manser},
  {Marshall}, {Martini}, {Massey}, {Matheson}, {McCauley}, {McDonald},
  {McGreer}, {Meisner}, {Metcalfe}, {Miller}, {Miquel}, {Moustakas}, {Myers},
  {Naik}, {Newman}, {Nichol}, {Nicola}, {Nicolati da Costa}, {Nie}, {Niz},
  {Norberg}, {Nord}, {Norman}, {Nugent}, {O'Brien}, {Oh}, {Olsen}, {Padilla},
  {Padmanabhan}, {Padmanabhan}, {Palanque-Delabrouille}, {Palmese},
  {Pappalardo}, {P{\^a}ris}, {Park}, {Patej}, {Peacock}, {Peiris}, {Peng},
  {Percival}, {Perruchot}, {Pieri}, {Pogge}, {Pollack}, {Poppett}, {Prada},
  {Prakash}, {Probst}, {Rabinowitz}, {Raichoor}, {Ree}, {Refregier}, {Regal},
  {Reid}, {Reil}, {Rezaie}, {Rockosi}, {Roe}, {Ronayette}, {Roodman}, {Ross},
  {Ross}, {Rossi}, {Rozo}, {Ruhlmann-Kleider}, {Rykoff}, {Sabiu}, {Samushia},
  {Sanchez}, {Sanchez}, {Schlegel}, {Schneider}, {Schubnell}, {Secroun},
  {Seljak}, {Seo}, {Serrano}, {Shafieloo}, {Shan}, {Sharples}, {Sholl},
  {Shourt}, {Silber}, {Silva}, {Sirk}, {Slosar}, {Smith}, {Smoot}, {Som},
  {Song}, {Sprayberry}, {Staten}, {Stefanik}, {Tarle}, {Sien Tie}, {Tinker},
  {Tojeiro}, {Valdes}, {Valenzuela}, {Valluri}, {Vargas-Magana}, {Verde},
  {Walker}, {Wang}, {Wang}, {Weaver}, {Weaverdyck}, {Wechsler}, {Weinberg},
  {White}, {Yang}, {Yeche}, {Zhang}, {Zhao}, {Zheng}, {Zhou}, {Zhou}, {Zhu},
  {Zou}, \& {Zu}}]{DESI-Collaboration16}
{DESI Collaboration}, {Aghamousa}, A., {Aguilar}, J., {et~al.} 2016, arXiv
  e-prints, arXiv:1611.00036

\bibitem[{{Dey} {et~al.}(2019){Dey}, {Schlegel}, {Lang}, {Blum}, {Burleigh},
  {Fan}, {Findlay}, {Finkbeiner}, {Herrera}, {Juneau}, {Landriau}, {Levi},
  {McGreer}, {Meisner}, {Myers}, {Moustakas}, {Nugent}, {Patej}, {Schlafly},
  {Walker}, {Valdes}, {Weaver}, {Y{\`e}che}, {Zou}, {Zhou}, {Abareshi},
  {Abbott}, {Abolfathi}, {Aguilera}, {Alam}, {Allen}, {Alvarez}, {Annis},
  {Ansarinejad}, {Aubert}, {Beechert}, {Bell}, {BenZvi}, {Beutler}, {Bielby},
  {Bolton}, {Brice{\~n}o}, {Buckley-Geer}, {Butler}, {Calamida}, {Carlberg},
  {Carter}, {Casas}, {Castander}, {Choi}, {Comparat}, {Cukanovaite}, {Delubac},
  {DeVries}, {Dey}, {Dhungana}, {Dickinson}, {Ding}, {Donaldson}, {Duan},
  {Duckworth}, {Eftekharzadeh}, {Eisenstein}, {Etourneau}, {Fagrelius},
  {Farihi}, {Fitzpatrick}, {Font-Ribera}, {Fulmer}, {G{\"a}nsicke},
  {Gaztanaga}, {George}, {Gerdes}, {Gontcho}, {Gorgoni}, {Green}, {Guy},
  {Harmer}, {Hernand ez}, {Honscheid}, {Huang}, {James}, {Jannuzi}, {Jiang},
  {Joyce}, {Karcher}, {Karkar}, {Kehoe}, {Kneib}, {Kueter-Young}, {Lan},
  {Lauer}, {Le Guillou}, {Le Van Suu}, {Lee}, {Lesser}, {Perreault Levasseur},
  {Li}, {Mann}, {Marshall}, {Mart{\'\i}nez-V{\'a}zquez}, {Martini}, {du Mas des
  Bourboux}, {McManus}, {Meier}, {M{\'e}nard}, {Metcalfe},
  {Mu{\~n}oz-Guti{\'e}rrez}, {Najita}, {Napier}, {Narayan}, {Newman}, {Nie},
  {Nord}, {Norman}, {Olsen}, {Paat}, {Palanque-Delabrouille}, {Peng},
  {Poppett}, {Poremba}, {Prakash}, {Rabinowitz}, {Raichoor}, {Rezaie},
  {Robertson}, {Roe}, {Ross}, {Ross}, {Rudnick}, {Safonova}, {Saha},
  {S{\'a}nchez}, {Savary}, {Schweiker}, {Scott}, {Seo}, {Shan}, {Silva},
  {Slepian}, {Soto}, {Sprayberry}, {Staten}, {Stillman}, {Stupak}, {Summers},
  {Sien Tie}, {Tirado}, {Vargas-Maga{\~n}a}, {Vivas}, {Wechsler}, {Williams},
  {Yang}, {Yang}, {Yapici}, {Zaritsky}, {Zenteno}, {Zhang}, {Zhang}, {Zhou}, \&
  {Zhou}}]{Dey19}
{Dey}, A., {Schlegel}, D.~J., {Lang}, D., {et~al.} 2019, \aj, 157, 168

\bibitem[{{Duev} {et~al.}(2019){Duev}, {Mahabal}, {Masci}, {Graham},
  {Rusholme}, {Walters}, {Karmarkar}, {Frederick}, {Kasliwal}, {Rebbapragada},
  \& {Ward}}]{Duev2019}
{Duev}, D.~A., {Mahabal}, A., {Masci}, F.~J., {et~al.} 2019, arXiv e-prints,
  arXiv:1907.11259

\bibitem[{{Filippenko}(1997)}]{Filippenko1997}
{Filippenko}, A.~V. 1997, \araa, 35, 309

\bibitem[{{Filippenko} {et~al.}(2001){Filippenko}, {Li}, {Treffers}, \&
  {Modjaz}}]{Filippenko01}
{Filippenko}, A.~V., {Li}, W.~D., {Treffers}, R.~R., \& {Modjaz}, M. 2001, in
  Astronomical Society of the Pacific Conference Series, Vol. 246, IAU Colloq.
  183: Small Telescope Astronomy on Global Scales, ed. B.~{Paczynski}, W.-P.
  {Chen}, \& C.~{Lemme}, 121--+

\bibitem[{{Fitzpatrick} \& {Massa}(2007)}]{Fitzpatrick07}
{Fitzpatrick}, E.~L., \& {Massa}, D. 2007, \apj, 663, 320

\bibitem[{{Flesch}(2015)}]{Flesch2015}
{Flesch}, E.~W. 2015, \pasa, 32, e010

\bibitem[{{Foley} {et~al.}(2006){Foley}, {Li}, {Moore}, {Wong}, {Pooley}, \&
  {Filippenko}}]{Foley06}
{Foley}, R.~J., {Li}, W., {Moore}, M., {et~al.} 2006, Central Bureau Electronic
  Telegrams, 695, 1

\bibitem[{{Foley} {et~al.}(2007){Foley}, {Smith}, {Ganeshalingam}, {Li},
  {Chornock}, \& {Filippenko}}]{foley06jc}
{Foley}, R.~J., {Smith}, N., {Ganeshalingam}, M., {et~al.} 2007, \apjl, 657,
  L105

\bibitem[{Foreman-Mackey(2016)}]{Foreman-Mackey16}
Foreman-Mackey, D. 2016, The Journal of Open Source Software, 24,
  doi:10.21105/joss.00024.
\newblock \url{http://dx.doi.org/10.5281/zenodo.45906}

\bibitem[{{Foreman-Mackey} {et~al.}(2013){Foreman-Mackey}, {Hogg}, {Lang}, \&
  {Goodman}}]{Foreman-Mackey13}
{Foreman-Mackey}, D., {Hogg}, D.~W., {Lang}, D., \& {Goodman}, J. 2013, \pasp,
  125, 306

\bibitem[{{Fox} {et~al.}(2014){Fox}, {Azalee Bostroem}, {Van Dyk},
  {Filippenko}, {Fransson}, {Matheson}, {Cenko}, {Chandra}, {Dwarkadas}, {Li},
  {Parker}, \& {Smith}}]{Fox2014}
{Fox}, O.~D., {Azalee Bostroem}, K., {Van Dyk}, S.~D., {et~al.} 2014, \apj,
  790, 17

\bibitem[{{Frieman} {et~al.}(2008){Frieman}, {Bassett}, {Becker}, {Choi},
  {Cinabro}, {DeJongh}, {Depoy}, {Dilday}, {Doi}, {Garnavich}, {Hogan},
  {Holtzman}, {Im}, {Jha}, {Kessler}, {Konishi}, {Lampeitl}, {Marriner},
  {Marshall}, {McGinnis}, {Miknaitis}, {Nichol}, {Prieto}, {Riess}, {Richmond},
  {Romani}, {Sako}, {Schneider}, {Smith}, {Takanashi}, {Tokita}, {van der
  Heyden}, {Yasuda}, {Zheng}, {Adelman-McCarthy}, {Annis}, {Assef},
  {Barentine}, {Bender}, {Blandford}, {Boroski}, {Bremer}, {Brewington},
  {Collins}, {Crotts}, {Dembicky}, {Eastman}, {Edge}, {Edmondson}, {Elson},
  {Eyler}, {Filippenko}, {Foley}, {Frank}, {Goobar}, {Gueth}, {Gunn},
  {Harvanek}, {Hopp}, {Ihara}, {Ivezi{\'c}}, {Kahn}, {Kaplan}, {Kent},
  {Ketzeback}, {Kleinman}, {Kollatschny}, {Kron}, {Krzesi{\'n}ski}, {Lamenti},
  {Leloudas}, {Lin}, {Long}, {Lucey}, {Lupton}, {Malanushenko}, {Malanushenko},
  {McMillan}, {Mendez}, {Morgan}, {Morokuma}, {Nitta}, {Ostman}, {Pan},
  {Rockosi}, {Romer}, {Ruiz-Lapuente}, {Saurage}, {Schlesinger}, {Snedden},
  {Sollerman}, {Stoughton}, {Stritzinger}, {Subba Rao}, {Tucker}, {Vaisanen},
  {Watson}, {Watters}, {Wheeler}, {Yanny}, \& {York}}]{Frieman2008}
{Frieman}, J.~A., {Bassett}, B., {Becker}, A., {et~al.} 2008, \aj, 135, 338

\bibitem[{{Gaia Collaboration} {et~al.}(2016){Gaia Collaboration}, {Prusti},
  {de Bruijne}, {Brown}, {Vallenari}, {Babusiaux}, {Bailer-Jones}, {Bastian},
  {Biermann}, {Evans}, {Eyer}, {Jansen}, {Jordi}, {Klioner}, {Lammers},
  {Lindegren}, {Luri}, {Mignard}, {Milligan}, {Panem}, {Poinsignon},
  {Pourbaix}, {Randich}, {Sarri}, {Sartoretti}, {Siddiqui}, {Soubiran},
  {Valette}, {van Leeuwen}, {Walton}, {Aerts}, {Arenou}, {Cropper}, {Drimmel},
  {H{\o}g}, {Katz}, {Lattanzi}, {O'Mullane}, {Grebel}, {Holland}, {Huc},
  {Passot}, {Bramante}, {Cacciari}, {Casta{\~n}eda}, {Chaoul}, {Cheek}, {De
  Angeli}, {Fabricius}, {Guerra}, {Hern{\'a}ndez}, {Jean-Antoine-Piccolo},
  {Masana}, {Messineo}, {Mowlavi}, {Nienartowicz}, {Ord{\'o}{\~n}ez- Blanco},
  {Panuzzo}, {Portell}, {Richards}, {Riello}, {Seabroke}, {Tanga},
  {Th{\'e}venin}, {Torra}, {Els}, {Gracia- Abril}, {Comoretto},
  {Garcia-Reinaldos}, {Lock}, {Mercier}, {Altmann}, {Andrae}, {Astraatmadja},
  {Bellas-Velidis}, {Benson}, {Berthier}, {Blomme}, {Busso}, {Carry},
  {Cellino}, {Clementini}, {Cowell}, {Creevey}, {Cuypers}, {Davidson}, {De
  Ridder}, {de Torres}, {Delchambre}, {Dell'Oro}, {Ducourant}, {Fr{\'e}mat},
  {Garc{\'\i}a-Torres}, {Gosset}, {Halbwachs}, {Hambly}, {Harrison}, {Hauser},
  {Hestroffer}, {Hodgkin}, {Huckle}, {Hutton}, {Jasniewicz}, {Jordan},
  {Kontizas}, {Korn}, {Lanzafame}, {Manteiga}, {Moitinho}, {Muinonen},
  {Osinde}, {Pancino}, {Pauwels}, {Petit}, {Recio-Blanco}, {Robin}, {Sarro},
  {Siopis}, {Smith}, {Smith}, {Sozzetti}, {Thuillot}, {van Reeven}, {Viala},
  {Abbas}, {Abreu Aramburu}, {Accart}, {Aguado}, {Allan}, {Allasia},
  {Altavilla}, {{\'A}lvarez}, {Alves}, {Anderson}, {Andrei}, {Anglada Varela},
  {Antiche}, {Antoja}, {Ant{\'o}n}, {Arcay}, {Atzei}, {Ayache}, {Bach},
  {Baker}, {Balaguer-N{\'u}{\~n}ez}, {Barache}, {Barata}, {Barbier}, {Barblan},
  {Baroni}, {Barrado y Navascu{\'e}s}, {Barros}, {Barstow}, {Becciani},
  {Bellazzini}, {Bellei}, {Bello Garc{\'\i}a}, {Belokurov}, {Bendjoya},
  {Berihuete}, {Bianchi}, {Bienaym{\'e}}, {Billebaud}, {Blagorodnova},
  {Blanco-Cuaresma}, {Boch}, {Bombrun}, {Borrachero}, {Bouquillon}, {Bourda},
  {Bouy}, {Bragaglia}, {Breddels}, {Brouillet}, {Br{\"u}semeister},
  {Bucciarelli}, {Budnik}, {Burgess}, {Burgon}, {Burlacu}, {Busonero}, {Buzzi},
  {Caffau}, {Cambras}, {Campbell}, {Cancelliere}, {Cantat-Gaudin}, {Carlucci},
  {Carrasco}, {Castellani}, {Charlot}, {Charnas}, {Charvet}, {Chassat},
  {Chiavassa}, {Clotet}, {Cocozza}, {Collins}, {Collins}, {Costigan}, {Crifo},
  {Cross}, {Crosta}, {Crowley}, {Dafonte}, {Damerdji}, {Dapergolas}, {David},
  {David}, {De Cat}, {de Felice}, {de Laverny}, {De Luise}, {De March}, {de
  Martino}, {de Souza}, {Debosscher}, {del Pozo}, {Delbo}, {Delgado},
  {Delgado}, {di Marco}, {Di Matteo}, {Diakite}, {Distefano}, {Dolding}, {Dos
  Anjos}, {Drazinos}, {Dur{\'a}n}, {Dzigan}, {Ecale}, {Edvardsson}, {Enke},
  {Erdmann}, {Escolar}, {Espina}, {Evans}, {Eynard Bontemps}, {Fabre},
  {Fabrizio}, {Faigler}, {Falc{\~a}o}, {Farr{\`a}s Casas}, {Faye}, {Federici},
  {Fedorets}, {Fern{\'a}ndez-Hern{\'a}ndez}, {Fernique}, {Fienga}, {Figueras},
  {Filippi}, {Findeisen}, {Fonti}, {Fouesneau}, {Fraile}, {Fraser}, {Fuchs},
  {Furnell}, {Gai}, {Galleti}, {Galluccio}, {Garabato}, {Garc{\'\i}a-Sedano},
  {Gar{\'e}}, {Garofalo}, {Garralda}, {Gavras}, {Gerssen}, {Geyer}, {Gilmore},
  {Girona}, {Giuffrida}, {Gomes}, {Gonz{\'a}lez-Marcos},
  {Gonz{\'a}lez-N{\'u}{\~n}ez}, {Gonz{\'a}lez-Vidal}, {Granvik}, {Guerrier},
  {Guillout}, {Guiraud}, {G{\'u}rpide}, {Guti{\'e}rrez-S{\'a}nchez}, {Guy},
  {Haigron}, {Hatzidimitriou}, {Haywood}, {Heiter}, {Helmi}, {Hobbs},
  {Hofmann}, {Holl}, {Holland}, {Hunt}, {Hypki}, {Icardi}, {Irwin}, {Jevardat
  de Fombelle}, {Jofr{\'e}}, {Jonker}, {Jorissen}, {Julbe}, {Karampelas},
  {Kochoska}, {Kohley}, {Kolenberg}, {Kontizas}, {Koposov}, {Kordopatis},
  {Koubsky}, {Kowalczyk}, {Krone-Martins}, {Kudryashova}, {Kull}, {Bachchan},
  {Lacoste-Seris}, {Lanza}, {Lavigne}, {Le Poncin-Lafitte}, {Lebreton},
  {Lebzelter}, {Leccia}, {Leclerc}, {Lecoeur-Taibi}, {Lemaitre}, {Lenhardt},
  {Leroux}, {Liao}, {Licata}, {Lindstr{\o}m}, {Lister}, {Livanou}, {Lobel},
  {L{\"o}ffler}, {L{\'o}pez}, {Lopez-Lozano}, {Lorenz}, {Loureiro},
  {MacDonald}, {Magalh{\~a}es Fernandes}, {Managau}, {Mann}, {Mantelet},
  {Marchal}, {Marchant}, {Marconi}, {Marie}, {Marinoni}, {Marrese},
  {Marschalk{\'o}}, {Marshall}, {Mart{\'\i}n-Fleitas}, {Martino}, {Mary},
  {Matijevi{\v{c}}}, {Mazeh}, {McMillan}, {Messina}, {Mestre}, {Michalik},
  {Millar}, {Miranda}, {Molina}, {Molinaro}, {Molinaro}, {Moln{\'a}r},
  {Moniez}, {Montegriffo}, {Monteiro}, {Mor}, {Mora}, {Morbidelli}, {Morel},
  {Morgenthaler}, {Morley}, {Morris}, {Mulone}, {Muraveva}, {Musella},
  {Narbonne}, {Nelemans}, {Nicastro}, {Noval}, {Ord{\'e}novic},
  {Ordieres-Mer{\'e}}, {Osborne}, {Pagani}, {Pagano}, {Pailler}, {Palacin},
  {Palaversa}, {Parsons}, {Paulsen}, {Pecoraro}, {Pedrosa}, {Pentik{\"a}inen},
  {Pereira}, {Pichon}, {Piersimoni}, {Pineau}, {Plachy}, {Plum}, {Poujoulet},
  {Pr{\v{s}}a}, {Pulone}, {Ragaini}, {Rago}, {Rambaux}, {Ramos-Lerate},
  {Ranalli}, {Rauw}, {Read}, {Regibo}, {Renk}, {Reyl{\'e}}, {Ribeiro},
  {Rimoldini}, {Ripepi}, {Riva}, {Rixon}, {Roelens}, {Romero-G{\'o}mez},
  {Rowell}, {Royer}, {Rudolph}, {Ruiz-Dern}, {Sadowski}, {Sagrist{\`a}
  Sell{\'e}s}, {Sahlmann}, {Salgado}, {Salguero}, {Sarasso}, {Savietto},
  {Schnorhk}, {Schultheis}, {Sciacca}, {Segol}, {Segovia}, {Segransan},
  {Serpell}, {Shih}, {Smareglia}, {Smart}, {Smith}, {Solano}, {Solitro},
  {Sordo}, {Soria Nieto}, {Souchay}, {Spagna}, {Spoto}, {Stampa}, {Steele},
  {Steidelm{\"u}ller}, {Stephenson}, {Stoev}, {Suess}, {S{\"u}veges}, {Surdej},
  {Szabados}, {Szegedi-Elek}, {Tapiador}, {Taris}, {Tauran}, {Taylor},
  {Teixeira}, {Terrett}, {Tingley}, {Trager}, {Turon}, {Ulla}, {Utrilla},
  {Valentini}, {van Elteren}, {Van Hemelryck}, {van Leeuwen}, {Varadi},
  {Vecchiato}, {Veljanoski}, {Via}, {Vicente}, {Vogt}, {Voss}, {Votruba},
  {Voutsinas}, {Walmsley}, {Weiler}, {Weingrill}, {Werner}, {Wevers},
  {Whitehead}, {Wyrzykowski}, {Yoldas}, {{\v{Z}}erjal}, {Zucker}, {Zurbach},
  {Zwitter}, {Alecu}, {Allen}, {Allende Prieto}, {Amorim},
  {Anglada-Escud{\'e}}, {Arsenijevic}, {Azaz}, {Balm}, {Beck}, {Bernstein},
  {Bigot}, {Bijaoui}, {Blasco}, {Bonfigli}, {Bono}, {Boudreault}, {Bressan},
  {Brown}, {Brunet}, {Bunclark}, {Buonanno}, {Butkevich}, {Carret}, {Carrion},
  {Chemin}, {Ch{\'e}reau}, {Corcione}, {Darmigny}, {de Boer}, {de Teodoro}, {de
  Zeeuw}, {Delle Luche}, {Domingues}, {Dubath}, {Fodor}, {Fr{\'e}zouls},
  {Fries}, {Fustes}, {Fyfe}, {Gallardo}, {Gallegos}, {Gardiol}, {Gebran},
  {Gomboc}, {G{\'o}mez}, {Grux}, {Gueguen}, {Heyrovsky}, {Hoar}, {Iannicola},
  {Isasi Parache}, {Janotto}, {Joliet}, {Jonckheere}, {Keil}, {Kim},
  {Klagyivik}, {Klar}, {Knude}, {Kochukhov}, {Kolka}, {Kos}, {Kutka}, {Lainey},
  {LeBouquin}, {Liu}, {Loreggia}, {Makarov}, {Marseille}, {Martayan},
  {Martinez-Rubi}, {Massart}, {Meynadier}, {Mignot}, {Munari}, {Nguyen},
  {Nordlander}, {Ocvirk}, {O'Flaherty}, {Olias Sanz}, {Ortiz}, {Osorio},
  {Oszkiewicz}, {Ouzounis}, {Palmer}, {Park}, {Pasquato}, {Peltzer}, {Peralta},
  {P{\'e}turaud}, {Pieniluoma}, {Pigozzi}, {Poels}, {Prat}, {Prod'homme},
  {Raison}, {Rebordao}, {Risquez}, {Rocca-Volmerange}, {Rosen}, {Ruiz-Fuertes},
  {Russo}, {Sembay}, {Serraller Vizcaino}, {Short}, {Siebert}, {Silva},
  {Sinachopoulos}, {Slezak}, {Soffel}, {Sosnowska}, {Strai{\v{z}}ys}, {ter
  Linden}, {Terrell}, {Theil}, {Tiede}, {Troisi}, {Tsalmantza}, {Tur},
  {Vaccari}, {Vachier}, {Valles}, {Van Hamme}, {Veltz}, {Virtanen}, {Wallut},
  {Wichmann}, {Wilkinson}, {Ziaeepour}, \& {Zschocke}}]{Gaia-Collaboration16}
{Gaia Collaboration}, {Prusti}, T., {de Bruijne}, J.~H.~J., {et~al.} 2016,
  \aap, 595, A1

\bibitem[{{Gal-Yam}(2012)}]{Gal-Yam12}
{Gal-Yam}, A. 2012, Science, 337, 927

\bibitem[{{Gal-Yam} {et~al.}(2009){Gal-Yam}, {Mazzali}, {Ofek}, {Nugent},
  {Kulkarni}, {Kasliwal}, {Quimby}, {Filippenko}, {Cenko}, {Chornock},
  {Waldman}, {Kasen}, {Sullivan}, {Beshore}, {Drake}, {Thomas}, {Bloom},
  {Poznanski}, {Miller}, {Foley}, {Silverman}, {Arcavi}, {Ellis}, \&
  {Deng}}]{Gal-yam2009}
{Gal-Yam}, A., {Mazzali}, P., {Ofek}, E.~O., {et~al.} 2009, \nat, 462, 624

\bibitem[{{Galama} {et~al.}(1998){Galama}, {Vreeswijk}, {van Paradijs},
  {Kouveliotou}, {Augusteijn}, {B{\"o}hnhardt}, {Brewer}, {Doublier},
  {Gonzalez}, {Leibundgut}, {Lidman}, {Hainaut}, {Patat}, {Heise}, {in't Zand},
  {Hurley}, {Groot}, {Strom}, {Mazzali}, {Iwamoto}, {Nomoto}, {Umeda},
  {Nakamura}, {Young}, {Suzuki}, {Shigeyama}, {Koshut}, {Kippen}, {Robinson},
  {de Wildt}, {Wijers}, {Tanvir}, {Greiner}, {Pian}, {Palazzi}, {Frontera},
  {Masetti}, {Nicastro}, {Feroci}, {Costa}, {Piro}, {Peterson}, {Tinney},
  {Boyle}, {Cannon}, {Stathakis}, {Sadler}, {Begam}, \& {Ianna}}]{Galama1998}
{Galama}, T.~J., {Vreeswijk}, P.~M., {van Paradijs}, J., {et~al.} 1998, \nat,
  395, 670

\bibitem[{{Gehrels} {et~al.}(2016){Gehrels}, {Cannizzo}, {Kanner}, {Kasliwal},
  {Nissanke}, \& {Singer}}]{Gehrels16}
{Gehrels}, N., {Cannizzo}, J.~K., {Kanner}, J., {et~al.} 2016, \apj, 820, 136

\bibitem[{{Goobar} \& {Leibundgut}(2011)}]{Goobar2011}
{Goobar}, A., \& {Leibundgut}, B. 2011, Annual Review of Nuclear and Particle
  Science, 61, 251

\bibitem[{{Goodman} \& {Weare}(2010)}]{Goodman10}
{Goodman}, J., \& {Weare}, J. 2010, Communications in Applied Mathematics and
  Computational Science, 5, 65

\bibitem[{Goodman(1965)}]{Goodman65}
Goodman, L.~A. 1965, Technometrics, 7, 247.
\newblock
  \url{https://www.tandfonline.com/doi/abs/10.1080/00401706.1965.10490252}

\bibitem[{{Gorbovskoy} {et~al.}(2013){Gorbovskoy}, {Lipunov}, {Kornilov},
  {Belinski}, {Kuvshinov}, {Tyurina}, {Sankovich}, {Krylov}, {Shatskiy},
  {Balanutsa}, {Chazov}, {Kuznetsov}, {Zimnukhov}, {Shumkov}, {Shurpakov},
  {Senik}, {Gareeva}, {Pruzhinskaya}, {Tlatov}, {Parkhomenko}, {Dormidontov},
  {Krushinsky}, {Punanova}, {Zalozhnyh}, {Popov}, {Burdanov}, {Yazev},
  {Budnev}, {Ivanov}, {Konstantinov}, {Gress}, {Chuvalaev}, {Yurkov},
  {Sergienko}, {Kudelina}, {Sinyakov}, {Karachentsev}, {Moiseev}, \&
  {Fatkhullin}}]{Gorbovskoy13}
{Gorbovskoy}, E.~S., {Lipunov}, V.~M., {Kornilov}, V.~G., {et~al.} 2013,
  Astronomy Reports, 57, 233

\bibitem[{{Graham} {et~al.}(2019){Graham}, {Kulkarni}, {Bellm}, {Adams},
  {Barbarino}, {Blagorodnova}, {Bodewits}, {Bolin}, {Brady}, {Cenko}, {Chang},
  {Coughlin}, {De}, {Eadie}, {Farnham}, {Feindt}, {Franckowiak}, {Fremling},
  {Gezari}, {Ghosh}, {Goldstein}, {Golkhou}, {Goobar}, {Ho}, {Huppenkothen},
  {Ivezi{\'c}}, {Jones}, {Juric}, {Kaplan}, {Kasliwal}, {Kelley}, {Kupfer},
  {Lee}, {Lin}, {Lunnan}, {Mahabal}, {Miller}, {Ngeow}, {Nugent}, {Ofek},
  {Prince}, {Rauch}, {van Roestel}, {Schulze}, {Singer}, {Sollerman}, {Taddia},
  {Yan}, {Ye}, {Yu}, {Barlow}, {Bauer}, {Beck}, {Belicki}, {Biswas}, {Brinnel},
  {Brooke}, {Bue}, {Bulla}, {Burruss}, {Connolly}, {Cromer}, {Cunningham},
  {Dekany}, {Delacroix}, {Desai}, {Duev}, {Feeney}, {Flynn}, {Frederick},
  {Gal-Yam}, {Giomi}, {Groom}, {Hacopians}, {Hale}, {Helou}, {Henning},
  {Hover}, {Hillenbrand}, {Howell}, {Hung}, {Imel}, {Ip}, {Jackson}, {Kaspi},
  {Kaye}, {Kowalski}, {Kramer}, {Kuhn}, {Landry}, {Laher}, {Mao}, {Masci},
  {Monkewitz}, {Murphy}, {Nordin}, {Patterson}, {Penprase}, {Porter},
  {Rebbapragada}, {Reiley}, {Riddle}, {Rigault}, {Rodriguez}, {Rusholme}, {van
  Santen}, {Shupe}, {Smith}, {Soumagnac}, {Stein}, {Surace}, {Szkody}, {Terek},
  {Van Sistine}, {van Velzen}, {Vestrand}, {Walters}, {Ward}, {Zhang}, \&
  {Zolkower}}]{Graham2019}
{Graham}, M.~J., {Kulkarni}, S.~R., {Bellm}, E.~C., {et~al.} 2019, \pasp, 131,
  078001

\bibitem[{{Grzegorzek}(2018)}]{Grzegorzek18}
{Grzegorzek}, J. 2018, Transient Name Server Discovery Report, 2018-582, 1

\bibitem[{{Guti{\'e}rrez} {et~al.}(2017){Guti{\'e}rrez}, {Anderson}, {Hamuy},
  {Morrell}, {Gonz{\'a}lez-Gaitan}, {Stritzinger}, {Phillips}, {Galbany},
  {Folatelli}, {Dessart}, {Contreras}, {Della Valle}, {Freedman}, {Hsiao},
  {Krisciunas}, {Madore}, {Maza}, {Suntzeff}, {Prieto}, {Gonz{\'a}lez},
  {Cappellaro}, {Navarrete}, {Pizzella}, {Ruiz}, {Smith}, \&
  {Turatto}}]{gutirrez}
{Guti{\'e}rrez}, C.~P., {Anderson}, J.~P., {Hamuy}, M., {et~al.} 2017, \apj,
  850, 89

\bibitem[{{Helfand} {et~al.}(2015){Helfand}, {White}, \& {Becker}}]{Helfand15}
{Helfand}, D.~J., {White}, R.~L., \& {Becker}, R.~H. 2015, \apj, 801, 26

\bibitem[{{Hirata} {et~al.}(1987){Hirata}, {Kajita}, {Koshiba}, {Nakahata},
  {Oyama}, {Sato}, {Suzuki}, {Takita}, {Totsuka}, {Kifune}, {Suda},
  {Takahashi}, {Tanimori}, {Miyano}, {Yamada}, {Beier}, {Feldscher}, {Kim},
  {Mann}, {Newcomer}, {van}, {Zhang}, \& {Cortez}}]{Hirata1987}
{Hirata}, K., {Kajita}, T., {Koshiba}, M., {et~al.} 1987, Physical Review
  Letters, 58, 1490

\bibitem[{{Ho} {et~al.}(2019){Ho}, {Phinney}, {Ravi}, {Kulkarni}, {Petitpas},
  {Emonts}, {Bhalerao}, {Blundell}, {Cenko}, {Dobie}, {Howie}, {Kamraj},
  {Kasliwal}, {Murphy}, {Perley}, {Sridharan}, \& {Yoon}}]{Ho2019}
{Ho}, A.~Y.~Q., {Phinney}, E.~S., {Ravi}, V., {et~al.} 2019, \apj, 871, 73

\bibitem[{{Hodgkin} {et~al.}(2013){Hodgkin}, {Wyrzykowski}, {Blagorodnova}, \&
  {Koposov}}]{Hodgkin13}
{Hodgkin}, S.~T., {Wyrzykowski}, L., {Blagorodnova}, N., \& {Koposov}, S. 2013,
  Philosophical Transactions of the Royal Society of London Series A, 371,
  20120239

\bibitem[{{Holoien} {et~al.}(2017{\natexlab{a}}){Holoien}, {Stanek},
  {Kochanek}, {Shappee}, {Prieto}, {Brimacombe}, {Bersier}, {Bishop}, {Dong},
  {Brown}, {Danilet}, {Simonian}, {Basu}, {Beacom}, {Falco}, {Pojmanski},
  {Skowron}, {Wo{\'z}niak}, {{\'A}vila}, {Conseil}, {Contreras}, {Cruz},
  {Fern{\'a}ndez}, {Koff}, {Guo}, {Herczeg}, {Hissong}, {Hsiao}, {Jose},
  {Kiyota}, {Long}, {Monard}, {Nicholls}, {Nicolas}, \& {Wiethoff}}]{Holoien17}
{Holoien}, T.~W.~S., {Stanek}, K.~Z., {Kochanek}, C.~S., {et~al.}
  2017{\natexlab{a}}, \mnras, 464, 2672

\bibitem[{{Holoien} {et~al.}(2017{\natexlab{b}}){Holoien}, {Brown}, {Stanek},
  {Kochanek}, {Shappee}, {Prieto}, {Dong}, {Brimacombe}, {Bishop}, {Basu},
  {Beacom}, {Bersier}, {Chen}, {Danilet}, {Falco}, {Godoy-Rivera}, {Goss},
  {Pojmanski}, {Simonian}, {Skowron}, {Thompson}, {Wo{\'z}niak}, {{\'A}vila},
  {Bock}, {Carballo}, {Conseil}, {Contreras}, {Cruz}, {And{\'u}jar}, {Guo},
  {Hsiao}, {Kiyota}, {Koff}, {Krannich}, {Madore}, {Marples}, {Masi},
  {Morrell}, {Monard}, {Munoz-Mateos}, {Nicholls}, {Nicolas}, {Wagner}, \&
  {Wiethoff}}]{Holoien17a}
{Holoien}, T.~W.~S., {Brown}, J.~S., {Stanek}, K.~Z., {et~al.}
  2017{\natexlab{b}}, \mnras, 467, 1098

\bibitem[{{Holoien} {et~al.}(2017{\natexlab{c}}){Holoien}, {Brown}, {Stanek},
  {Kochanek}, {Shappee}, {Prieto}, {Dong}, {Brimacombe}, {Bishop}, {Bose},
  {Beacom}, {Bersier}, {Chen}, {Chomiuk}, {Falco}, {Godoy-Rivera}, {Morrell},
  {Pojmanski}, {Shields}, {Strader}, {Stritzinger}, {Thompson}, {Wo{\'z}niak},
  {Bock}, {Cacella}, {Conseil}, {Cruz}, {Fernand ez}, {Kiyota}, {Koff},
  {Krannich}, {Marples}, {Masi}, {Monard}, {Nicholls}, {Nicolas}, {Post},
  {Stone}, \& {Wiethoff}}]{Holoien17b}
---. 2017{\natexlab{c}}, \mnras, 471, 4966

\bibitem[{{Holoien} {et~al.}(2019){Holoien}, {Brown}, {Vallely}, {Stanek},
  {Kochanek}, {Shappee}, {Prieto}, {Dong}, {Brimacombe}, {Bishop}, {Bose},
  {Beacom}, {Bersier}, {Chen}, {Chomiuk}, {Falco}, {Holmbo}, {Jayasinghe},
  {Morrell}, {Pojmanski}, {Shields}, {Strader}, {Stritzinger}, {Thompson},
  {Wo{\'z}niak}, {Bock}, {Cacella}, {Carballo}, {Cruz}, {Conseil}, {Farfan},
  {Fernandez}, {Kiyota}, {Koff}, {Krannich}, {Marples}, {Masi}, {Monard},
  {Mu{\~n}oz}, {Nicholls}, {Post}, {Stone}, {Trappett}, \&
  {Wiethoff}}]{Holoien19}
{Holoien}, T.~W.~S., {Brown}, J.~S., {Vallely}, P.~J., {et~al.} 2019, \mnras,
  484, 1899

\bibitem[{{Howell}(2001)}]{Howell2001}
{Howell}, D.~A. 2001, \apjl, 554, L193

\bibitem[{Hunter(2007)}]{Hunter07}
Hunter, J.~D. 2007, Computing In Science \& Engineering, 9, 90

\bibitem[{{Itagaki}(2018{\natexlab{a}})}]{Itagaki18}
{Itagaki}, K. 2018{\natexlab{a}}, Transient Name Server Discovery Report,
  2018-1614, 1

\bibitem[{{Itagaki}(2018{\natexlab{b}})}]{Itagaki18a}
---. 2018{\natexlab{b}}, Transient Name Server Discovery Report, 2018-1766, 1

\bibitem[{{Ivezi{\'c}} {et~al.}(2008){Ivezi{\'c}}, {Tyson}, {Acosta},
  {Allsman}, {Anderson}, {Andrew}, {Angel}, {Axelrod}, {Barr}, {Becker},
  {Becla}, {Beldica}, {Blandford}, {Bloom}, {Borne}, {Brandt}, {Brown},
  {Bullock}, {Burke}, {Chandrasekharan}, {Chesley}, {Claver}, {Connolly},
  {Cook}, {Cooray}, {Covey}, {Cribbs}, {Cutri}, {Daues}, {Delgado}, {Ferguson},
  {Gawiser}, {Geary}, {Gee}, {Geha}, {Gibson}, {Gilmore}, {Gressler}, {Hogan},
  {Huffer}, {Jacoby}, {Jain}, {Jernigan}, {Jones}, {Juric}, {Kahn}, {Kalirai},
  {Kantor}, {Kessler}, {Kirkby}, {Knox}, {Krabbendam}, {Krughoff}, {Kulkarni},
  {Lambert}, {Levine}, {Liang}, {Lim}, {Lupton}, {Marshall}, {Marshall}, {May},
  {Miller}, {Mills}, {Monet}, {Neill}, {Nordby}, {O'Connor}, {Oliver},
  {Olivier}, {Olsen}, {Owen}, {Peterson}, {Petry}, {Pierfederici},
  {Pietrowicz}, {Pike}, {Pinto}, {Plante}, {Radeka}, {Rasmussen}, {Ridgway},
  {Rosing}, {Saha}, {Schalk}, {Schindler}, {Schneider}, {Schumacher}, {Sebag},
  {Seppala}, {Shipsey}, {Silvestri}, {Smith}, {Smith}, {Strauss}, {Stubbs},
  {Sweeney}, {Szalay}, {Thaler}, {Vanden Berk}, {Walkowicz}, {Warner},
  {Willman}, {Wittman}, {Wolff}, {Wood-Vasey}, {Yoachim}, {Zhan}, \& {for the
  LSST Collaboration}}]{Ivezic08}
{Ivezi{\'c}}, {\v Z}., {Tyson}, J.~A., {Acosta}, E., {et~al.} 2008, ArXiv
  e-prints, arXiv:0805.2366

\bibitem[{{Jarrett} {et~al.}(2011){Jarrett}, {Cohen}, {Masci}, {Wright},
  {Stern}, {Benford}, {Blain}, {Carey}, {Cutri}, {Eisenhardt}, {Lonsdale},
  {Mainzer}, {Marsh}, {Padgett}, {Petty}, {Ressler}, {Skrutskie}, {Stanford},
  {Surace}, {Tsai}, {Wheelock}, \& {Yan}}]{Jarrett11}
{Jarrett}, T.~H., {Cohen}, M., {Masci}, F., {et~al.} 2011, \apj, 735, 112

\bibitem[{Jones {et~al.}(2001)Jones, Oliphant, Peterson, {et~al.}}]{Jones01}
Jones, E., Oliphant, T., Peterson, P., {et~al.} 2001, {SciPy}: Open source
  scientific tools for {Python}, , , [Online; accessed <today>].
\newblock \url{http://www.scipy.org/}

\bibitem[{{Kasen}(2006)}]{2006ApJ...649..939K}
{Kasen}, D. 2006, \apj, 649, 939

\bibitem[{{Kasliwal} \& {Nissanke}(2014)}]{Kasliwal14}
{Kasliwal}, M.~M., \& {Nissanke}, S. 2014, \apjl, 789, L5

\bibitem[{{Kasliwal} {et~al.}(2012){Kasliwal}, {Kulkarni}, {Gal-Yam}, {Nugent},
  {Sullivan}, {Bildsten}, {Yaron}, {Perets}, {Arcavi}, {Ben-Ami}, {Bhalerao},
  {Bloom}, {Cenko}, {Filippenko}, {Frail}, {Ganeshalingam}, {Horesh}, {Howell},
  {Law}, {Leonard}, {Li}, {Ofek}, {Polishook}, {Poznanski}, {Quimby},
  {Silverman}, {Sternberg}, \& {Xu}}]{Kasliwal12}
{Kasliwal}, M.~M., {Kulkarni}, S.~R., {Gal-Yam}, A., {et~al.} 2012, \apj, 755,
  161

\bibitem[{{Kasliwal} {et~al.}(2019){Kasliwal}, {Cannella}, {Bagdasaryan},
  {Hung}, {Feindt}, {Singer}, {Coughlin}, {Fremling}, {Walters}, {Duev},
  {Itoh}, \& {Quimby}}]{Kasliwal2019}
{Kasliwal}, M.~M., {Cannella}, C., {Bagdasaryan}, A., {et~al.} 2019, \pasp,
  131, 038003

\bibitem[{{Kulkarni} {et~al.}(2018){Kulkarni}, {Perley}, \&
  {Miller}}]{Kulkarni2018}
{Kulkarni}, S.~R., {Perley}, D.~A., \& {Miller}, A.~A. 2018, \apj, 860, 22

\bibitem[{{Lang} {et~al.}(2016{\natexlab{a}}){Lang}, {Hogg}, \&
  {Mykytyn}}]{Lang16a}
{Lang}, D., {Hogg}, D.~W., \& {Mykytyn}, D. 2016{\natexlab{a}}, {The Tractor:
  Probabilistic astronomical source detection and measurement}, , ,
  ascl:1604.008

\bibitem[{{Lang} {et~al.}(2016{\natexlab{b}}){Lang}, {Hogg}, \&
  {Schlegel}}]{Lang16}
{Lang}, D., {Hogg}, D.~W., \& {Schlegel}, D.~J. 2016{\natexlab{b}}, \aj, 151,
  36

\bibitem[{{Law} {et~al.}(2009){Law}, {Kulkarni}, {Dekany}, {Ofek}, {Quimby},
  {Nugent}, {Surace}, {Grillmair}, {Bloom}, {Kasliwal}, {Bildsten}, {Brown},
  {Cenko}, {Ciardi}, {Croner}, {Djorgovski}, {van Eyken}, {Filippenko}, {Fox},
  {Gal-Yam}, {Hale}, {Hamam}, {Helou}, {Henning}, {Howell}, {Jacobsen},
  {Laher}, {Mattingly}, {McKenna}, {Pickles}, {Poznanski}, {Rahmer}, {Rau},
  {Rosing}, {Shara}, {Smith}, {Starr}, {Sullivan}, {Velur}, {Walters}, \&
  {Zolkower}}]{Law2009}
{Law}, N.~M., {Kulkarni}, S.~R., {Dekany}, R.~G., {et~al.} 2009, \pasp, 121,
  1395

\bibitem[{{Li} {et~al.}(2011){Li}, {Leaman}, {Chornock}, {Filippenko},
  {Poznanski}, {Ganeshalingam}, {Wang}, {Modjaz}, {Jha}, {Foley}, \&
  {Smith}}]{Li11}
{Li}, W., {Leaman}, J., {Chornock}, R., {et~al.} 2011, \mnras, 412, 1441

\bibitem[{{Li} {et~al.}(2000){Li}, {Filippenko}, {Treffers}, {Friedman},
  {Halderson}, {Johnson}, {King}, {Modjaz}, {Papenkova}, {Sato}, \&
  {Shefler}}]{Li2000}
{Li}, W.~D., {Filippenko}, A.~V., {Treffers}, R.~R., {et~al.} 2000, in American
  Institute of Physics Conference Series, Vol. 522, American Institute of
  Physics Conference Series, ed. S.~S. {Holt} \& W.~W. {Zhang}, 103--106

\bibitem[{{Ligo Scientific Collaboration} \& {VIRGO
  Collaboration}(2019{\natexlab{a}})}]{S190425z}
{Ligo Scientific Collaboration}, \& {VIRGO Collaboration}. 2019{\natexlab{a}},
  GRB Coordinates Network, Circular Service, 24168

\bibitem[{{Ligo Scientific Collaboration} \& {VIRGO
  Collaboration}(2019{\natexlab{b}})}]{S190510g}
---. 2019{\natexlab{b}}, GRB Coordinates Network, Circular Service, 24442

\bibitem[{{Ligo Scientific Collaboration} \& {VIRGO
  Collaboration}(2019{\natexlab{c}})}]{S190901ap}
---. 2019{\natexlab{c}}, GRB Coordinates Network, Circular Service, 25606

\bibitem[{{Ligo Scientific Collaboration} \& {VIRGO
  Collaboration}(2019{\natexlab{d}})}]{S190910h}
---. 2019{\natexlab{d}}, GRB Coordinates Network, Circular Service, 25707

\bibitem[{{Liu} {et~al.}(2016){Liu}, {Modjaz}, {Bianco}, \& {Graur}}]{liu2016}
{Liu}, Y.-Q., {Modjaz}, M., {Bianco}, F.~B., \& {Graur}, O. 2016, \apj, 827, 90

\bibitem[{{Lunnan} {et~al.}(2019){Lunnan}, {Yan}, {Perley}, {Schulze},
  {Taggart}, {Gal-Yam}, {Fremling}, {Soumagnac}, {Ofek}, {Adams}, {Barbarino},
  {Bellm}, {De}, {Fransson}, {Frederick}, {Golkhou}, {Graham}, {Hallakoun},
  {Ho}, {Kasliwal}, {Kaspi}, {Kulkarni}, {Laher}, {Masci}, {Pozo Nunez},
  {Rusholme}, {Quimby}, {Shupe}, {Sollerman}, {Taddia}, {van Roestel}, {Yang},
  \& {Yao}}]{Lunnan2019}
{Lunnan}, R., {Yan}, L., {Perley}, D.~A., {et~al.} 2019, arXiv e-prints,
  arXiv:1910.02968

\bibitem[{{Mahabal} {et~al.}(2019){Mahabal}, {Rebbapragada}, {Walters},
  {Masci}, {Blagorodnova}, {van Roestel}, {Ye}, {Biswas}, {Burdge}, {Chang},
  {Duev}, {Golkhou}, {Miller}, {Nordin}, {Ward}, {Adams}, {Bellm}, {Branton},
  {Bue}, {Cannella}, {Connolly}, {Dekany}, {Feindt}, {Hung}, {Fortson},
  {Frederick}, {Fremling}, {Gezari}, {Graham}, {Groom}, {Kasliwal}, {Kulkarni},
  {Kupfer}, {Lin}, {Lintott}, {Lunnan}, {Parejko}, {Prince}, {Riddle},
  {Rusholme}, {Saunders}, {Sedaghat}, {Shupe}, {Singer}, {Soumagnac}, {Szkody},
  {Tachibana}, {Tirumala}, {van Velzen}, \& {Wright}}]{Mahabal2019}
{Mahabal}, A., {Rebbapragada}, U., {Walters}, R., {et~al.} 2019, \pasp, 131,
  038002

\bibitem[{{Maraston} {et~al.}(2013){Maraston}, {Pforr}, {Henriques}, {Thomas},
  {Wake}, {Brownstein}, {Capozzi}, {Tinker}, {Bundy}, {Skibba}, {Beifiori},
  {Nichol}, {Edmondson}, {Schneider}, {Chen}, {Masters}, {Steele}, {Bolton},
  {York}, {Weaver}, {Higgs}, {Bizyaev}, {Brewington}, {Malanushenko},
  {Malanushenko}, {Snedden}, {Oravetz}, {Pan}, {Shelden}, \&
  {Simmons}}]{Maraston13}
{Maraston}, C., {Pforr}, J., {Henriques}, B.~M., {et~al.} 2013, \mnras, 435,
  2764

\bibitem[{{Martin} {et~al.}(2005){Martin}, {Fanson}, {Schiminovich},
  {Morrissey}, {Friedman}, {Barlow}, {Conrow}, {Grange}, {Jelinsky},
  {Milliard}, {Siegmund}, {Bianchi}, {Byun}, {Donas}, {Forster}, {Heckman},
  {Lee}, {Madore}, {Malina}, {Neff}, {Rich}, {Small}, {Surber}, {Szalay},
  {Welsh}, \& {Wyder}}]{Martin05}
{Martin}, D.~C., {Fanson}, J., {Schiminovich}, D., {et~al.} 2005, \apjl, 619,
  L1

\bibitem[{{Masci} {et~al.}(2019){Masci}, {Laher}, {Rusholme}, {Shupe}, {Groom},
  {Surace}, {Jackson}, {Monkewitz}, {Beck}, {Flynn}, {Terek}, {Landry},
  {Hacopians}, {Desai}, {Howell}, {Brooke}, {Imel}, {Wachter}, {Ye}, {Lin},
  {Cenko}, {Cunningham}, {Rebbapragada}, {Bue}, {Miller}, {Mahabal}, {Bellm},
  {Patterson}, {Juri{\'c}}, {Golkhou}, {Ofek}, {Walters}, {Graham}, {Kasliwal},
  {Dekany}, {Kupfer}, {Burdge}, {Cannella}, {Barlow}, {Van Sistine}, {Giomi},
  {Fremling}, {Blagorodnova}, {Levitan}, {Riddle}, {Smith}, {Helou}, {Prince},
  \& {Kulkarni}}]{Masci2019}
{Masci}, F.~J., {Laher}, R.~R., {Rusholme}, B., {et~al.} 2019, \pasp, 131,
  018003

\bibitem[{{McCray}(1993)}]{McCray93}
{McCray}, R. 1993, \araa, 31, 175

\bibitem[{McKinney(2010)}]{McKinney10}
McKinney, W. 2010, in Proceedings of the 9th Python in Science Conference, ed.
  S.~van~der Walt \& J.~Millman, 51 -- 56

\bibitem[{{Modjaz} {et~al.}(2016){Modjaz}, {Liu}, {Bianco}, \&
  {Graur}}]{modjaz2016}
{Modjaz}, M., {Liu}, Y.~Q., {Bianco}, F.~B., \& {Graur}, O. 2016, \apj, 832,
  108

\bibitem[{{Modjaz} {et~al.}(2014){Modjaz}, {Blondin}, {Kirshner}, {Matheson},
  {Berlind}, {Bianco}, {Calkins}, {Challis}, {Garnavich}, {Hicken}, {Jha},
  {Liu}, \& {Marion}}]{modjaz2014}
{Modjaz}, M., {Blondin}, S., {Kirshner}, R.~P., {et~al.} 2014, \aj, 147, 99

\bibitem[{{Nordin} {et~al.}(2019){Nordin}, {Brinnel}, {van Santen}, {Bulla},
  {Feindt}, {Franckowiak}, {Fremling}, {Gal-Yam}, {Giomi}, {Kowalski},
  {Mahabal}, {Miranda}, {Rauch}, {Rigault}, {Schulze}, {Reusch}, {Sollerman},
  {Stein}, {Yaron}, {van Velzen}, \& {Ward}}]{Nordin2019}
{Nordin}, J., {Brinnel}, V., {van Santen}, J., {et~al.} 2019, arXiv e-prints,
  arXiv:1904.05922

\bibitem[{{Norgaard-Nielsen} {et~al.}(1989){Norgaard-Nielsen}, {Hansen},
  {Jorgensen}, {Aragon Salamanca}, \& {Ellis}}]{Norgaard89}
{Norgaard-Nielsen}, H.~U., {Hansen}, L., {Jorgensen}, H.~E., {Aragon
  Salamanca}, A., \& {Ellis}, R.~S. 1989, \nat, 339, 523

\bibitem[{{Oke} \& {Gunn}(1982)}]{Oke1982}
{Oke}, J.~B., \& {Gunn}, J.~E. 1982, \pasp, 94, 586

\bibitem[{{Oke} {et~al.}(1995){Oke}, {Cohen}, {Carr}, {Cromer}, {Dingizian},
  {Harris}, {Labrecque}, {Lucinio}, {Schaal}, {Epps}, \& {Miller}}]{Oke1995}
{Oke}, J.~B., {Cohen}, J.~G., {Carr}, M., {et~al.} 1995, \pasp, 107, 375

\bibitem[{{Papadogiannakis} {et~al.}(2019{\natexlab{a}}){Papadogiannakis},
  {Dhawan}, {Morosin}, \& {Goobar}}]{2019MNRAS.485.2343P}
{Papadogiannakis}, S., {Dhawan}, S., {Morosin}, R., \& {Goobar}, A.
  2019{\natexlab{a}}, \mnras, 485, 2343

\bibitem[{{Papadogiannakis} {et~al.}(2019{\natexlab{b}}){Papadogiannakis},
  {Goobar}, {Amanullah}, {Bulla}, {Dhawan}, {Doran}, {Feindt}, {Ferretti},
  {Hangard}, {Howell}, {Johansson}, {Kasliwal}, {Laher}, {Masci}, {Nyholm},
  {Ofek}, {Sollerman}, \& {Yan}}]{2019MNRAS.483.5045P}
{Papadogiannakis}, S., {Goobar}, A., {Amanullah}, R., {et~al.}
  2019{\natexlab{b}}, \mnras, 483, 5045

\bibitem[{{Pastorello} {et~al.}(2007){Pastorello}, {Smartt}, {Mattila},
  {Eldridge}, {Young}, {Itagaki}, {Yamaoka}, {Navasardyan}, {Valenti}, {Patat},
  {Agnoletto}, {Augusteijn}, {Benetti}, {Cappellaro}, {Boles}, {Bonnet-Bidaud},
  {Botticella}, {Bufano}, {Cao}, {Deng}, {Dennefeld}, {Elias-Rosa},
  {Harutyunyan}, {Keenan}, {Iijima}, {Lorenzi}, {Mazzali}, {Meng}, {Nakano},
  {Nielsen}, {Smoker}, {Stanishev}, {Turatto}, {Xu}, \&
  {Zampieri}}]{pastorello07-06jc}
{Pastorello}, A., {Smartt}, S.~J., {Mattila}, S., {et~al.} 2007, \nat, 447, 829

\bibitem[{{Pastorello} {et~al.}(2010){Pastorello}, {Smartt}, {Botticella},
  {Maguire}, {Fraser}, {Smith}, {Kotak}, {Magill}, {Valenti}, {Young},
  {Gezari}, {Bresolin}, {Kudritzki}, {Howell}, {Rest}, {Metcalfe}, {Mattila},
  {Kankare}, {Huang}, {Urata}, {Burgett}, {Chambers}, {Dombeck}, {Flewelling},
  {Grav}, {Heasley}, {Hodapp}, {Kaiser}, {Luppino}, {Lupton}, {Magnier},
  {Monet}, {Morgan}, {Onaka}, {Price}, {Rhoads}, {Siegmund}, {Stubbs},
  {Sweeney}, {Tonry}, {Wainscoat}, {Waterson}, {Waters}, \&
  {Wynn-Williams}}]{Pastorello2010}
{Pastorello}, A., {Smartt}, S.~J., {Botticella}, M.~T., {et~al.} 2010, \apjl,
  724, L16

\bibitem[{{Patat} {et~al.}(2001){Patat}, {Cappellaro}, {Danziger}, {Mazzali},
  {Sollerman}, {Augusteijn}, {Brewer}, {Doublier}, {Gonzalez}, {Hainaut},
  {Lidman}, {Leibundgut}, {Nomoto}, {Nakamura}, {Spyromilio}, {Rizzi},
  {Turatto}, {Walsh}, {Galama}, {van Paradijs}, {Kouveliotou}, {Vreeswijk},
  {Frontera}, {Masetti}, {Palazzi}, \& {Pian}}]{Patat2001}
{Patat}, F., {Cappellaro}, E., {Danziger}, J., {et~al.} 2001, \apj, 555, 900

\bibitem[{{Patterson} {et~al.}(2019){Patterson}, {Bellm}, {Rusholme}, {Masci},
  {Juric}, {Krughoff}, {Golkhou}, {Graham}, {Kulkarni}, {Helou}, \& {Zwicky
  Transient Facility Collaboration}}]{Patterson2019}
{Patterson}, M.~T., {Bellm}, E.~C., {Rusholme}, B., {et~al.} 2019, \pasp, 131,
  018001

\bibitem[{{Perley}(2019)}]{Perley2019}
{Perley}, D.~A. 2019, \pasp, 131, 084503

\bibitem[{{Perlmutter} {et~al.}(1999){Perlmutter}, {Aldering}, {Goldhaber},
  {Knop}, {Nugent}, {Castro}, {Deustua}, {Fabbro}, {Goobar}, {Groom}, {Hook},
  {Kim}, {Kim}, {Lee}, {Nunes}, {Pain}, {Pennypacker}, {Quimby}, {Lidman},
  {Ellis}, {Irwin}, {McMahon}, {Ruiz-Lapuente}, {Walton}, {Schaefer}, {Boyle},
  {Filippenko}, {Matheson}, {Fruchter}, {Panagia}, {Newberg}, {Couch}, \&
  {Project}}]{Perlmutter1999}
{Perlmutter}, S., {Aldering}, G., {Goldhaber}, G., {et~al.} 1999, \apj, 517,
  565

\bibitem[{{Piascik} {et~al.}(2014){Piascik}, {Steele}, {Bates}, {Mottram},
  {Smith}, {Barnsley}, \& {Bolton}}]{SPRAT}
{Piascik}, A.~S., {Steele}, I.~A., {Bates}, S.~D., {et~al.} 2014, in \procspie,
  Vol. 9147, Ground-based and Airborne Instrumentation for Astronomy V, 91478H

\bibitem[{{Prentice} {et~al.}(2018){Prentice}, {Maguire}, {Smartt}, {Magee},
  {Schady}, {Sim}, {Chen}, {Clark}, {Colin}, {Fulton}, {McBrien}, {O'Neill},
  {Smith}, {Ashall}, {Chambers}, {Denneau}, {Flewelling}, {Heinze}, {Holoien},
  {Huber}, {Kochanek}, {Mazzali}, {Prieto}, {Rest}, {Shappee}, {Stalder},
  {Stanek}, {Stritzinger}, {Thompson}, \& {Tonry}}]{Prentice2018}
{Prentice}, S.~J., {Maguire}, K., {Smartt}, S.~J., {et~al.} 2018, \apjl, 865,
  L3

\bibitem[{{Quimby} {et~al.}(2007){Quimby}, {Aldering}, {Wheeler},
  {H{\"o}flich}, {Akerlof}, \& {Rykoff}}]{Quimby2007}
{Quimby}, R.~M., {Aldering}, G., {Wheeler}, J.~C., {et~al.} 2007, \apjl, 668,
  L99

\bibitem[{{Quimby} {et~al.}(2011){Quimby}, {Kulkarni}, {Kasliwal}, {Gal-Yam},
  {Arcavi}, {Sullivan}, {Nugent}, {Thomas}, {Howell}, {Nakar}, {Bildsten},
  {Theissen}, {Law}, {Dekany}, {Rahmer}, {Hale}, {Smith}, {Ofek}, {Zolkower},
  {Velur}, {Walters}, {Henning}, {Bui}, {McKenna}, {Poznanski}, {Cenko}, \&
  {Levitan}}]{quimby11}
{Quimby}, R.~M., {Kulkarni}, S.~R., {Kasliwal}, M.~M., {et~al.} 2011, \nat,
  474, 487

\bibitem[{{Quimby} {et~al.}(2018){Quimby}, {De Cia}, {Gal-Yam}, {Leloudas},
  {Lunnan}, {Perley}, {Vreeswijk}, {Yan}, {Bloom}, {Cenko}, {Cooke}, {Ellis},
  {Filippenko}, {Kasliwal}, {Kleiser}, {Kulkarni}, {Matheson}, {Nugent}, {Pan},
  {Silverman}, {Sternberg}, {Sullivan}, \& {Yaron}}]{Quimby2018}
{Quimby}, R.~M., {De Cia}, A., {Gal-Yam}, A., {et~al.} 2018, \apj, 855, 2

\bibitem[{{Riess} {et~al.}(1998){Riess}, {Filippenko}, {Challis},
  {Clocchiatti}, {Diercks}, {Garnavich}, {Gilliland}, {Hogan}, {Jha},
  {Kirshner}, {Leibundgut}, {Phillips}, {Reiss}, {Schmidt}, {Schommer},
  {Smith}, {Spyromilio}, {Stubbs}, {Suntzeff}, \& {Tonry}}]{Riess1998}
{Riess}, A.~G., {Filippenko}, A.~V., {Challis}, P., {et~al.} 1998, \aj, 116,
  1009

\bibitem[{{Rigault} {et~al.}(2019){Rigault}, {Neill}, {Blagorodnova}, {Dugas},
  {Feeney}, {Walters}, {Brinnel}, {Copin}, {Fremling}, {Nordin}, \&
  {Sollerman}}]{Rigault2019}
{Rigault}, M., {Neill}, J.~D., {Blagorodnova}, N., {et~al.} 2019, \aap, 627,
  A115

\bibitem[{{Schlafly} \& {Finkbeiner}(2011)}]{Schlafly11}
{Schlafly}, E.~F., \& {Finkbeiner}, D.~P. 2011, \apj, 737, 103

\bibitem[{{Schlafly} {et~al.}(2019){Schlafly}, {Meisner}, \&
  {Green}}]{Schlafly19}
{Schlafly}, E.~F., {Meisner}, A.~M., \& {Green}, G.~M. 2019, \apjs, 240, 30

\bibitem[{{Schlegel} {et~al.}(1998){Schlegel}, {Finkbeiner}, \&
  {Davis}}]{Schlegel98}
{Schlegel}, D.~J., {Finkbeiner}, D.~P., \& {Davis}, M. 1998, \apj, 500, 525

\bibitem[{{Schlegel}(1990)}]{Schlegel90}
{Schlegel}, E.~M. 1990, \mnras, 244, 269

\bibitem[{{Schmidt} {et~al.}(1993){Schmidt}, {Kirshner}, {Eastman}, {Grashuis},
  {dell'Antonio}, {Caldwell}, {Foltz}, {Huchra}, \& {Milone}}]{Schmidt1993}
{Schmidt}, B.~P., {Kirshner}, R.~P., {Eastman}, R.~G., {et~al.} 1993, \nat,
  364, 600

\bibitem[{{Science Software Branch at STScI}(2012)}]{pyraf}
{Science Software Branch at STScI}. 2012, {PyRAF: Python alternative for IRAF},
  Astrophysics Source Code Library, , , ascl:1207.011

\bibitem[{{Secrest} {et~al.}(2015){Secrest}, {Dudik}, {Dorland}, {Zacharias},
  {Makarov}, {Fey}, {Frouard}, \& {Finch}}]{Secrest2015}
{Secrest}, N.~J., {Dudik}, R.~P., {Dorland}, B.~N., {et~al.} 2015, \apjs, 221,
  12

\bibitem[{{Shappee} {et~al.}(2014){Shappee}, {Prieto}, {Stanek}, {Kochanek},
  {Holoien}, {Jencson}, {Basu}, {Beacom}, {Szczygiel}, {Pojmanski},
  {Brimacombe}, {Dubberley}, {Elphick}, {Foale}, {Hawkins}, {Mullins},
  {Rosing}, {Ross}, \& {Walker}}]{Shappee2014}
{Shappee}, B., {Prieto}, J., {Stanek}, K.~Z., {et~al.} 2014, in American
  Astronomical Society Meeting Abstracts, Vol. 223, American Astronomical
  Society Meeting Abstracts \#223, 236.03

\bibitem[{Signorell \& et~al.(2019)}]{Signorell19}
Signorell, A., \& et~al. 2019, DescTools: Tools for Descriptive Statistics, r
  package version 0.99.29.
\newblock \url{https://cran.r-project.org/package=DescTools}

\bibitem[{{Silverman} {et~al.}(2012{\natexlab{a}}){Silverman}, {Foley},
  {Filippenko}, {Ganeshalingam}, {Barth}, {Chornock}, {Griffith}, {Kong},
  {Lee}, {Leonard}, {Matheson}, {Miller}, {Steele}, {Barris}, {Bloom}, {Cobb},
  {Coil}, {Desroches}, {Gates}, {Ho}, {Jha}, {Kandrashoff}, {Li}, {Mandel},
  {Modjaz}, {Moore}, {Mostardi}, {Papenkova}, {Park}, {Perley}, {Poznanski},
  {Reuter}, {Scala}, {Serduke}, {Shields}, {Swift}, {Tonry}, {Van Dyk}, {Wang},
  \& {Wong}}]{bsnip}
{Silverman}, J.~M., {Foley}, R.~J., {Filippenko}, A.~V., {et~al.}
  2012{\natexlab{a}}, \mnras, 425, 1789

\bibitem[{{Silverman} {et~al.}(2012{\natexlab{b}}){Silverman}, {Foley},
  {Filippenko}, {Ganeshalingam}, {Barth}, {Chornock}, {Griffith}, {Kong},
  {Lee}, {Leonard}, {Matheson}, {Miller}, {Steele}, {Barris}, {Bloom}, {Cobb},
  {Coil}, {Desroches}, {Gates}, {Ho}, {Jha}, {Kandrashoff}, {Li}, {Mandel},
  {Modjaz}, {Moore}, {Mostardi}, {Papenkova}, {Park}, {Perley}, {Poznanski},
  {Reuter}, {Scala}, {Serduke}, {Shields}, {Swift}, {Tonry}, {Van Dyk}, {Wang},
  \& {Wong}}]{Silverman12}
---. 2012{\natexlab{b}}, \mnras, 425, 1789

\bibitem[{{Smartt} {et~al.}(2015){Smartt}, {Valenti}, {Fraser}, {Inserra},
  {Young}, {Sullivan}, {Pastorello}, {Benetti}, {Gal-Yam}, {Knapic},
  {Molinaro}, {Smareglia}, {Smith}, {Taubenberger}, {Yaron}, {Anderson},
  {Ashall}, {Balland}, {Baltay}, {Barbarino}, {Bauer}, {Baumont}, {Bersier},
  {Blagorodnova}, {Bongard}, {Botticella}, {Bufano}, {Bulla}, {Cappellaro},
  {Campbell}, {Cellier-Holzem}, {Chen}, {Childress}, {Clocchiatti},
  {Contreras}, {Dall'Ora}, {Danziger}, {de Jaeger}, {De Cia}, {Della Valle},
  {Dennefeld}, {Elias-Rosa}, {Elman}, {Feindt}, {Fleury}, {Gall},
  {Gonzalez-Gaitan}, {Galbany}, {Morales Garoffolo}, {Greggio}, {Guillou},
  {Hachinger}, {Hadjiyska}, {Hage}, {Hillebrandt}, {Hodgkin}, {Hsiao}, {James},
  {Jerkstrand}, {Kangas}, {Kankare}, {Kotak}, {Kromer}, {Kuncarayakti},
  {Leloudas}, {Lundqvist}, {Lyman}, {Hook}, {Maguire}, {Manulis}, {Margheim},
  {Mattila}, {Maund}, {Mazzali}, {McCrum}, {McKinnon}, {Moreno-Raya},
  {Nicholl}, {Nugent}, {Pain}, {Pignata}, {Phillips}, {Polshaw}, {Pumo},
  {Rabinowitz}, {Reilly}, {Romero-Ca{\~n}izales}, {Scalzo}, {Schmidt},
  {Schulze}, {Sim}, {Sollerman}, {Taddia}, {Tartaglia}, {Terreran},
  {Tomasella}, {Turatto}, {Walker}, {Walton}, {Wyrzykowski}, {Yuan}, \&
  {Zampieri}}]{Smartt15}
{Smartt}, S.~J., {Valenti}, S., {Fraser}, M., {et~al.} 2015, \aap, 579, A40

\bibitem[{{Smith} {et~al.}(2007){Smith}, {Li}, {Foley}, {Wheeler}, {Pooley},
  {Chornock}, {Filippenko}, {Silverman}, {Quimby}, {Bloom}, \&
  {Hansen}}]{smith07-2006gy}
{Smith}, N., {Li}, W., {Foley}, R.~J., {et~al.} 2007, \apj, 666, 1116

\bibitem[{{Soumagnac} \& {Ofek}(2018)}]{Soumagnac2018}
{Soumagnac}, M.~T., \& {Ofek}, E.~O. 2018, \pasp, 130, 075002

\bibitem[{{Sun} {et~al.}(2018){Sun}, {Lau}, {Liu}, {Dai}, {Chen}, {Li}, \&
  {Zhao}}]{Sun18}
{Sun}, P., {Lau}, A., {Liu}, J., {et~al.} 2018, Transient Name Server Discovery
  Report, 2018-2007, 1

\bibitem[{{Tachibana} \& {Miller}(2018)}]{Tachibana2018}
{Tachibana}, Y., \& {Miller}, A.~A. 2018, \pasp, 130, 128001

\bibitem[{{Tanaka}(2018)}]{Tanaka18}
{Tanaka}, Y. 2018, Transient Name Server Discovery Report, 2018-469, 1

\bibitem[{{Tartaglia} {et~al.}(2018){Tartaglia}, {Sand}, {Valenti}, {Wyatt},
  {Anderson}, {Arcavi}, {Ashall}, {Botticella}, {Cartier}, {Chen}, {Cikota},
  {Coulter}, {Della Valle}, {Foley}, {Gal-Yam}, {Galbany}, {Gall}, {Haislip},
  {Harmanen}, {Hosseinzadeh}, {Howell}, {Hsiao}, {Inserra}, {Jha}, {Kankare},
  {Kilpatrick}, {Kouprianov}, {Kuncarayakti}, {Maccarone}, {Maguire},
  {Mattila}, {Mazzali}, {McCully}, {Meland ri}, {Morrell}, {Phillips},
  {Pignata}, {Piro}, {Prentice}, {Reichart}, {Rojas-Bravo}, {Smartt}, {Smith},
  {Sollerman}, {Stritzinger}, {Sullivan}, {Taddia}, \& {Young}}]{Tartaglia18}
{Tartaglia}, L., {Sand}, D.~J., {Valenti}, S., {et~al.} 2018, \apj, 853, 62

\bibitem[{{Taubenberger}(2017)}]{Taubenberger17}
{Taubenberger}, S. 2017, {The Extremes of Thermonuclear Supernovae}, 317

\bibitem[{{Tody}(1986)}]{Tody1986}
{Tody}, D. 1986, in \procspie, Vol. 627, Instrumentation in astronomy VI, ed.
  D.~L. {Crawford}, 733

\bibitem[{{Tonry} {et~al.}(2018){Tonry}, {Denneau}, {Heinze}, {Stalder},
  {Smith}, {Smartt}, {Stubbs}, {Weiland}, \& {Rest}}]{Tonry2018}
{Tonry}, J.~L., {Denneau}, L., {Heinze}, A.~N., {et~al.} 2018, \pasp, 130,
  064505

\bibitem[{{Turatto} {et~al.}(1993){Turatto}, {Cappellaro}, {Danziger},
  {Benetti}, {Gouiffes}, \& {della Valle}}]{turatto93}
{Turatto}, M., {Cappellaro}, E., {Danziger}, I.~J., {et~al.} 1993, \mnras, 262,
  128

\bibitem[{{Wen} {et~al.}(2013){Wen}, {Wu}, {Zhu}, {Lam}, {Wu}, {Wicker}, \&
  {Zhao}}]{2013MNRAS.433.2946W}
{Wen}, X.-Q., {Wu}, H., {Zhu}, Y.-N., {et~al.} 2013, \mnras, 433, 2946

\bibitem[{{Williamson} {et~al.}(2019){Williamson}, {Modjaz}, \&
  {Bianco}}]{williamson}
{Williamson}, M., {Modjaz}, M., \& {Bianco}, F.~B. 2019, \apjl, 880, L22

\bibitem[{{Wright} {et~al.}(2010){Wright}, {Eisenhardt}, {Mainzer}, {Ressler},
  {Cutri}, {Jarrett}, {Kirkpatrick}, {Padgett}, {McMillan}, {Skrutskie},
  {Stanford}, {Cohen}, {Walker}, {Mather}, {Leisawitz}, {Gautier}, {McLean},
  {Benford}, {Lonsdale}, {Blain}, {Mendez}, {Irace}, {Duval}, {Liu}, {Royer},
  {Heinrichsen}, {Howard}, {Shannon}, {Kendall}, {Walsh}, {Larsen}, {Cardon},
  {Schick}, {Schwalm}, {Abid}, {Fabinsky}, {Naes}, \& {Tsai}}]{Wright10}
{Wright}, E.~L., {Eisenhardt}, P.~R.~M., {Mainzer}, A.~K., {et~al.} 2010, \aj,
  140, 1868

\bibitem[{{Zackay} {et~al.}(2016){Zackay}, {Ofek}, \& {Gal-Yam}}]{Zackay16}
{Zackay}, B., {Ofek}, E.~O., \& {Gal-Yam}, A. 2016, \apj, 830, 27

\bibitem[{{Zhang} {et~al.}(2018){Zhang}, {Ding}, {Liu}, {Wang}, {Xu}, {Zhang},
  {Cao}, {Zhao}, {Jiang}, {Ruan}, \& {Gao}}]{Zhang18}
{Zhang}, M., {Ding}, Y., {Liu}, S., {et~al.} 2018, Transient Name Server
  Discovery Report, 2018-1393, 1

\bibitem[{{Zhang} {et~al.}(2015){Zhang}, {Wang}, {Chen}, {Zhang}, {Zhou}, {Li},
  {Liu}, {Mo}, {Zhang}, {Yao}, {Zhao}, {Zhou}, {Nie}, {Huang}, {Jiang}, {Ma},
  {Wang}, {Wu}, {Zhou}, {Zou}, \& {Wang}}]{Zhang15}
{Zhang}, T.-M., {Wang}, X.-F., {Chen}, J.-C., {et~al.} 2015, Research in
  Astronomy and Astrophysics, 15, 215

\bibitem[{{Zwicky}(1938{\natexlab{a}})}]{Zwicky1938PASP}
{Zwicky}, F. 1938{\natexlab{a}}, \pasp, 50, 215

\bibitem[{{Zwicky}(1938{\natexlab{b}})}]{Zwicky1938ApJ}
---. 1938{\natexlab{b}}, \apj, 88, 529

\bibitem[{{Zwicky}(1942)}]{Zwicky1942}
---. 1942, \apj, 96, 28

\end{thebibliography}

\end{document}